\documentclass[amsmath,amssymb,11pt]{article}
\usepackage{jheppub2}
\pdfoutput=1
\usepackage{graphicx,epsfig}
\usepackage{float}
\usepackage{subfig}
\usepackage{subfloat}
\usepackage[utf8]{inputenc}
\usepackage{hyperref}
\usepackage{caption}
\usepackage{color}
\usepackage{overpic}
\usepackage[dvipsnames]{xcolor}

\usepackage{physics}

\usepackage{tikz}
\usetikzlibrary{positioning}
\usetikzlibrary{intersections}
\usetikzlibrary{fadings} 
\usetikzlibrary{arrows.meta} 
\usetikzlibrary{arrows}

\tikzfading[name=fade out,
inner color=transparent!0,
outer color=transparent!100]

\definecolor{cherryblossompink}{rgb}{1.0, 0.72, 0.77}
\definecolor{lightblue}{rgb}{0.68, 0.85, 0.9}

\usetikzlibrary{decorations.pathmorphing}
\usetikzlibrary{decorations.pathreplacing,decorations.markings}

\usetikzlibrary{backgrounds,automata}

\newcommand*{\affmark}[1][*]{\textsuperscript{#1}}

\definecolor{pink1}{rgb}{0.858, 0.188, 0.478}
\newcommand{\red}[1]{{\color{red}{#1}}}

\newcommand{\MV}[1]{{\red{MV: #1}}} 

\newcommand{\beq}{\begin{equation}}

\newcommand{\eeq}{\end{equation}}

\title{Quasi-local energy and microcanonical entropy in two-dimensional nearly de Sitter gravity
}

\author{Andrew Svesko\affmark[1],}
\emailAdd{a.svesko@ucl.ac.uk}
\author{Evita Verheijden\affmark[2],}
\emailAdd{e.m.h.verheijden@uva.nl}
\author{Erik P.  Verlinde\affmark[2],}
\emailAdd{e.p.verlinde@uva.nl}
\author{and Manus R. Visser\affmark[3]\\}
\emailAdd{manus.visser@unige.ch}

\affiliation{\affmark[1]Department of Physics and Astronomy, University College London,\\
London, WC1E 6BT, United Kingdom\\\\
\affmark[2]Institute of Physics \& Delta Institute for Theoretical Physics, University of Amsterdam, \\Science Park 904,
1090 GL Amsterdam, The Netherlands\\\\
\affmark[3]Department of Theoretical Physics, University of Geneva\\
24 quai Ernest-Ansermet, 1211 Gen\`{e}ve 4, Switzerland\\}

\abstract{We study the semi-classical thermodynamics of two-dimensional de Sitter space ($\text{dS}_{2}$) in Jackiw-Teitelboim (JT) gravity coupled to conformal matter. We extend the quasi-local formalism of Brown and York to $\text{dS}_{2}$, where  a timelike   boundary  is introduced in the static patch to uniquely define conserved charges, including quasi-local energy. The   boundary  divides the static patch into two systems, a cosmological system and a black hole system, the former  being    unstable under thermal fluctuations while the latter is stable. A semi-classical quasi-local first law  is derived, where the Gibbons--Hawking entropy is replaced by the generalized entropy. In the microcanonical ensemble the generalized entropy is stationary. Further, we show the on-shell Euclidean microcanonical  action of a causal diamond in semi-classical JT gravity equals minus the generalized entropy of the diamond, hence  extremization of the   entropy follows from minimizing the action. Thus, we provide a first principles derivation of the island rule for $U(1)$ symmetric $\text{dS}_{2}$ backgrounds, without invoking the replica trick. We discuss the implications of our findings for static patch de Sitter holography.
}

\begin{document}

\maketitle

\section{Introduction}\label{sec:intro}

Observation suggests our universe is currently in a phase of accelerated expansion. If this growth continues, the measurable universe will asymptotically approach de Sitter (dS) spacetime, a maximally symmetric space with positive cosmological constant describing an exponentially expanding spacetime.
A striking feature of dS space is that, due to the exponential inflation to the future, a static observer only sees a portion of the full spacetime; confined to the static patch, they encounter a cosmological horizon. The dS cosmological horizon and  event horizons surrounding black holes share similar features. Chiefly, both have a temperature and an associated entropy proportional to the area of the horizon due to thermal radiation emitted from their respective horizons \cite{Gibbons:1977mu}.  However, the thermodynamics of the dS horizon, and the subsequent microscopic interpretation, is more mysterious than for their black hole counterparts due to the observer-dependent nature of the cosmological horizon and lack of unbroken supersymmetry in pure dS (see, \emph{e.g.},  \cite{Banks:2000fe,Witten:2001kn,Spradlin:2001pw,Goheer:2002vf,Anninos:2012qw}).

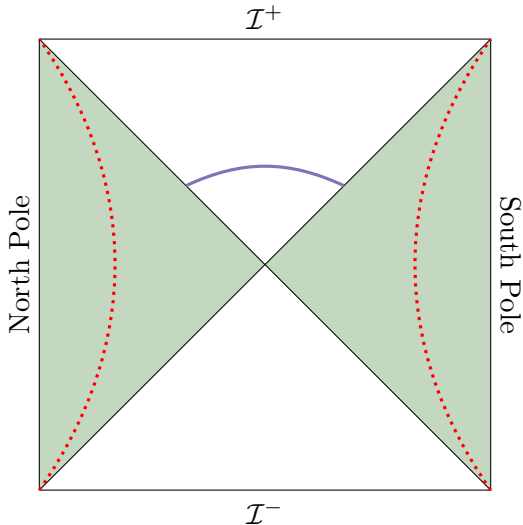
\begin{figure}[t!]
\centering
\begin{tikzpicture}[scale=1.5]
	\fill[fill=OliveGreen, fill opacity=0.3] (0,0) coordinate (a) -- (0,4) coordinate (b) -- (2,2) coordinate (m) -- cycle;
	\fill[fill=OliveGreen, fill opacity=0.3] (4,4) coordinate (c) -- (4,0) coordinate (d) -- (m) -- cycle; 
	\draw (a) -- (b) node[pos=.5, above, sloped] {\text{North Pole}} -- (c) node[pos=.5, above] {$\mathcal{I}^+$} -- (d) node[pos=.5, above, sloped] {\text{South Pole}} -- (a) node[pos=.5, below] {$\mathcal{I}^-$};
	\draw (a) -- (c);
	\draw (b) -- (d);
	\draw[very thick, dotted, red] (a) to[bend right=35] (b); 
	\draw[very thick, dotted, red] (d) to[bend left=35] (c);
	\draw[very thick, Periwinkle] (1.3, 2.7) to[bend left=25] (2.7, 2.7);
\end{tikzpicture}
\caption{Penrose diagram of de Sitter space. The left and right (green) regions describe the static patch of de Sitter space. 
The dotted curves (red) represent anchor curves which we use to define quasi-local thermodynamics. The boundaries of the (blue) bulk spatial surface anchored between the two (stretched) cosmological horizons are extremal surfaces whose area is proposed to compute the entanglement entropy.}
\label{fig:dSanchor} 
\end{figure}

 A promising explanation for the microscopics of dS thermodynamics relies on holography.  In particular,   gravitational entropy in dS may correspond to a fine grained entropy of a dual quantum mechanical theory. However, it is still debated on which boundary the dual  microscopic theory 
should be placed and where the extremal surface $X$ whose area gives the fine grained entropy is located. In the dS/CFT correspondence the dual theory lives on the future conformal boundary $\mathcal I^+$ \cite{Strominger:2001pn,Maldacena:2002vr,Anninos:2011ui,Anninos:2017eib}, whereas in static patch holography it lives on a timelike surface inside the dS static patch \cite{Banks:2005bm,Parikh:2004wh,Anninos:2011af,Anninos:2017hhn,Leuven:2018ejp,Susskind:2021omt}. In this paper we are interested in the static patch and its holographic description, for which there are different proposals in the literature. For example, according to the worldline holography by \cite{Anninos:2011af,Anninos:2017hhn}   the dual quantum theory lives on a screen near the north and south poles in the static patch. Alternatively, it has recently been suggested to place the dual microscopic theory on the (stretched) cosmological horizon, with a bulk  surface $\Sigma$ anchored between the two stretched horizons whose boundaries $\partial \Sigma$ are the extremal surfaces \cite{Susskind:2021dfc,Susskind:2021esx,Shaghoulian:2021cef,Shaghoulian:2022fop}; see Figure~\ref{fig:dSanchor} for a comparison.  As the red timelike curve hugs the south and north poles, one has the worldline holography described in~\cite{Anninos:2011af,Anninos:2017hhn}, while as the curve approaches the horizon one has the holographic description given by~\cite{Susskind:2021dfc,Susskind:2021esx,Shaghoulian:2021cef,Shaghoulian:2022fop}. 
The two proposals for dS static patch holography can be made  consistent with each other if the stretched horizon describes the IR of the underlying microscopic theory, while the worldline at the poles corresponds to the UV of the theory. This would imply there exists a family of  timelike surfaces in between the poles and the stretched horizons which interpolate between the UV and IR of the dual quantum theory. Moreover, note that in this unifying picture large distances (IR) in the bulk correspond to low energies (IR) in the boundary theory, inverting the standard UV/IR correspondence in AdS/CFT \cite{Leuven:2018ejp}.

In this article, we address the aforementioned puzzles regarding both the thermodynamic and microscopic aspects of de Sitter space. To do so, we consider Jackiw-Teitelboim (JT) gravity \cite{Jackiw:1984je,Teitelboim:1983ux} with a positive cosmological constant. There are in fact two distinct versions of this type of JT gravity
depending on the higher-dimensional geometry one spherically reduces: the half reduction of pure three-dimensional dS or the full reduction of the four-dimensional Schwarschild-de Sitter black hole in the near-Nariai limit \cite{Sybesma:2020fxg,Aalsma:2021bit}. Both versions of JT gravity admit two-dimensional de Sitter space as the background, though the global $\text{dS}_{2}$ geometry is different in each version. Specifically, $\text{dS}_{2}$ in the full reduction inherits a black hole horizon, while the half reduction is more reminiscent of higher-dimensional pure de Sitter space (see Figures \ref{fig:dS2halfred} and \ref{fig:dS2fullred} below).

In either model, following \cite{York:1986it,PhysRevLett.61.1336,Brown:1989fa,Brown:1992br}, we enclose the horizons in a box by introducing finite timelike anchor curves between the poles and cosmological horizons (Figure \ref{fig:dSanchor}). Doing so allows us to study the thermodynamics of de Sitter space more carefully in the canonical ensemble, where the dilaton and the local (Tolman) temperature are fixed on this timelike boundary $B$. Using both covariant phase space techniques and a Euclidean path integral we derive a quasi-local first law, cf. Eq. \eqref{eq:quasilocalfirstlawyeah},
\beq d E=T d S_{H}-\sigma d\phi_{B}\;,\label{eq:quasilocfirstlawintro}\eeq
where $E$ is the quasi-local energy, $T$ the Tolman temperature, $S_{H}=\frac{\phi_H}{4G_2}$ the entropy of the bifurcate horizon $H$, $\sigma$ is a ``surface pressure", and $\phi_{B}$ is the value of the dilaton evaluated at~$B$. In the limit the timelike boundary is placed such that the thermodynamic system fills the full static patch, we recover the  2D analog of the global first law of a Schwarzschild-de Sitter black hole \cite{Gibbons:1977mu}. An appealing feature of the quasi-local approach is that the timelike anchor curves we introduce interpolate between the boundaries where presumably a dual microscopic theory lives, namely, the stretched horizons and the poles. Further, the anchor curve naturally divides the spacetime into two systems: a ``black hole system" between the black hole horizon and the anchor curve, and a ``cosmological system" between the boundary $B$ and the cosmological horizon. 
We find that the black hole system in the full reduction  model has positive heat capacity, while the  cosmological system has negative heat capacity (see Figure \ref{fig:EandCvsrB}).

An advantage of working with JT gravity is that we have full analytic control of quantum backreaction. This is because in two dimensions semi-classical effects are fully captured by the 1-loop Polyakov action \cite{Polyakov:1981rd}. In such toy models many conceptual issues of horizon thermodynamics can be resolved. Recently, for example, the authors in \cite{Pedraza:2021cvx} showed that for conformal matter in an eternal $\text{AdS}_{2}$ black hole, the Wald entropy is equal to the generalized entropy $S_{\text{gen}}$ \cite{Bekenstein:1974ax}, the sum of the classical gravitational entropy $S_{\text{BH}}$ and von Neumann entropy $S_{\text{vN}}$ of quantum matter, 
\beq S_{\text{gen}}=S_{\text{BH}}+S_{\text{vN}}\;.\label{eq:genentintro}\eeq
When semi-classical effects are included, the classical Bekenstein--Hawking entropy appearing in the first law is supplanted by $S_{\text{gen}}$, and where the area of the black hole horizon is replaced with the area of a quantum extremal surface (QES), a codimension-2 surface extremizing $S_{\text{gen}}$, also denoted by $X$ \cite{Engelhardt:2014gca}. Likewise, upon including semi-classical effects, we will derive a semi-classical generalization of the quasi-local first law, where, particularly, the classical entropy $S_{H}$ in (\ref{eq:quasilocfirstlawintro}) is replaced by the generalized entropy. 
Further, we find that in the microcanonical ensemble the generalized entropy obeys the stationarity condition
\beq \delta S_{\text{gen}}=0\;,\label{eq:statent}\eeq
a central result of this article, cf. Eq.  \eqref{eq:stationarityentropysc}.

Crucially, this observation offers another way to think about how to compute fine grained entropies in de Sitter space. Indeed, the fact that the entropy is stationary in the microcanonical ensemble is consistent with the extremization of the generalized entropy in the QES formula \cite{Faulkner:2013ana,Engelhardt:2014gca}.
The QES formula is a generalization of the (classical) Ryu--Takayanagi formula \cite{Ryu:2006bv,Hubeny:2007xt}, which says that 
the von Neumann entropy in quantum gravity $S_{\text{vN}}(\Sigma_X)$ of a codimension-1 slice $\Sigma_{X}$ bounded by a QES $X$ may be 
computed in the semi-classical approximation using the following extremization prescription
\beq S_{\text{vN}}(\Sigma_X)=\underset{X}{\text{min}}\,\underset{X}{\text{ext}}\left[\frac{\text{Area}(X)}{4G_{d}}+S^{\text{sc}}_{\text{vN}}(\Sigma_{X})\right] .\label{eq:QESformula}\eeq
On the right-hand side $S^{\text{sc}}_{\text{vN}}$ is the von Neumann entropy of quantum fields in the semi-classical approximation. The term in brackets is thus the generalized entropy $S_{\text{gen}}(\Sigma_X)$ (\ref{eq:genentintro}). The QES formula (\ref{eq:QESformula}) also holds for the von Neumann entropy of Hawking radiation $S_{\text{vN}}^{\text{rad}}$, where it is known as the ``island formula'' \cite{Almheiri:2019hni}. In this case $\Sigma_{X}$ may be disconnected, $\Sigma_{X}=\Sigma_{\text{rad}}\cup I$, where $\Sigma_{\text{rad}}$ is the region outside of the black hole bounded by a cutoff surface and a region at infinity containing the radiation, and $I$ is an ``island'' with $X=\partial I$. 
Applying  (\ref{eq:QESformula}) to black holes in AdS$_{2}$ reveals a Page curve \cite{Penington:2019npb,Almheiri:2019psf,Almheiri:2019yqk}, arguably resolving the black hole information paradox: while the semi-classical fine grained matter entropy may exceed the coarse grained thermodynamic entropy, thus violating the Bekenstein entropy bound \cite{Bekenstein:1980jp}, the total fine grained entropy in quantum gravity does not.

In cosmology one encounters a puzzle similar to the black hole information paradox, such that fine grained matter entropies violate the Bekenstein entropy bound \cite{Fischler:1998st} (see also \cite{Arkani-Hamed:2007ryv}). Consequently, the QES and island formulae (\ref{eq:QESformula}) have been employed to analyze fine grained entropies in de Sitter space in different settings \cite{Hartman:2020khs,Balasubramanian:2020coy,Balasubramanian:2020xqf,Sybesma:2020fxg,Aalsma:2021bit,Kames-King:2021etp,Teresi:2021qff}, \emph{e.g.}, in the full or half reduction model, and for radiation collected inside the static patch or at future infinity. Most relevant to our discussion here is the distinction between the full and half reduction model of de Sitter JT gravity. 
In particular, in the full reduction model, and for radiation collected at future infinity, 
the only non-trivial island is located in the interior of the black hole near the singularity, and the full quantum gravity fine grained entropy obeys a Page-like curve. On the other hand, in the half reduction model there are no non-trivial islands.

Motivated by \cite{Lewkowycz:2013nqa,Dong:2017xht}, the island formula  has been derived using the ``replica trick'' in the context of JT gravity in AdS \cite{Almheiri:2019qdq,Penington:2019kki}. The Page curve  arises from a  competition between two saddle point geometries dominating the Euclidean gravitational path integral, where ``replica wormholes'' dominate over the standard Euclidean black hole solution at late times. Thus far, however, the replica trick derivation of the island formula has not yet been accomplished   in de Sitter space.

Our equilibrium thermodynamic result (\ref{eq:statent}) leads us to provide a first principles derivation of $S_{\text{gen}}$ and its extremization, as in the QES formula, in de Sitter JT gravity without invoking the replica trick. We work in the microcanonical ensemble \cite{Brown:1992bq,Banados:1993qp}, defined using a Euclidean gravitational $\text{dS}_{2}$ path integral, and show the on-shell microcanonical action of $\text{dS}_{2}$ causal diamonds is equal to (minus) the generalized entropy. Minimizing the action with respect to the background corresponds to extremizing $S_{\text{gen}}$ with respect to the location of a QES, analogous to the $\text{AdS}_{2}$ result in \cite{Pedraza:2021ssc}. As an application, we find islands -- only in the full reduction de Sitter JT model -- from which we can compute the fine grained entropy of thermal radiation in dS quantum gravity. Our derivation thus justifies the use of the island formula in $\text{dS}_{2}$ spacetimes. 

To summarize, after detailing the differences between the half and full reductions of de Sitter JT gravity in Section \ref{sec:red2DdS}, we study the quasi-local thermodynamics of $\text{dS}_{2}$  found in both JT models in Section \ref{sec:quasithermo}. We provide a complete analysis of semi-classical de Sitter JT gravity, where we show the semi-classical Wald entropy is equal to $S_{\text{gen}}$, and appears in the semi-classical extension of the quasi-local first law. In Section \ref{sec:islandsfrommicact}, we derive the  microcanonical action of Euclidean causal diamonds in $\text{dS}_{2}$ in semi-classical de Sitter JT gravity, and show that the extremization   of   generalized entropy as in the QES and island formulae follows from the minimization of the action.

To keep the article self contained  we include a number of appendices. In Appendix \ref{app:reductionscoords} we derive the two versions of de Sitter JT gravity via a spherical reduction of the $d$-dimensional Einstein--Hilbert action. We also list some useful coordinate systems of $\text{dS}_{2}$. Appendix \ref{app:nariaigeom} details the geometry of Schwarzschild-de Sitter black hole in the near-Nariai limit in arbitrary dimensions. Appendix \ref{app:noetherchargesumm} summarizes the Noether charge formalism for arbitrary theories of two-dimensional dilaton gravity, and in Appendix \ref{app:diamondcoordinates} we describe the geometry of causal diamonds in Lorentzian and Euclidean  $\text{dS}_{2}$.

\section{Two roads to de Sitter JT gravity} \label{sec:red2DdS}

Two-dimensional dilaton gravity is well known to describe the low-energy dynamics of a wide class of charged, near-extremal black holes and branes in higher dimensions. A popular such model is classical JT gravity in $\text{AdS}_{2}$ \cite{Jackiw:1984je,Teitelboim:1983ux}, following from a spherical reduction of the Einstein--Hilbert action describing near-extremal black holes with near-horizon geometry $\text{AdS}_{2}\times X$  \cite{Achucarro:1993fd,Fabbri:2000xh,Nayak:2018qej,Sachdev:2019bjn}, where $X$ is the transverse space whose size is controlled by the dilaton. Solutions to the theory are ``nearly'' $\text{AdS}_{2}$ in that the spacetime is asymptotically $\text{AdS}_{2}$, and the dilaton encodes deviations from extremality.

Here we review the derivation of  de Sitter JT gravity, which is expected to describe the low-energy physics of near-extremal solutions with a near-horizon geometry of the form $\text{dS}_{2}\times X$. Unlike AdS JT, subtleties arise when performing a spherical reduction of the higher-dimensional theory. In particular, there are two versions of de Sitter JT gravity:\footnote{There are also two distinct versions of AdS JT gravity, obtained by dimensional reduction of higher-dimensional extremal black holes, and by dimensional reduction of $\text{AdS}_{3}$. Similar to the half and full reduction models of de Sitter JT gravity, these two versions of AdS JT gravity differ in that no topological contribution appears when reducing $\text{AdS}_{3}$ (see \cite{Achucarro:1993fd,Verheijden:2021yrb}).} one following from the spherical reduction  of three-dimensional pure de Sitter space ($\text{dS}_{3}$), and another from a spherical reduction of the four-dimensional Schwarzschild-de Sitter (SdS$_4$) black hole in the near-Nariai limit. Both versions of de Sitter JT have ``nearly'' $\text{dS}_{2}$ solutions, however,  we will see the geometry and the  thermodynamics for each will be different. Our discussion largely follows the spirit of \cite{Cotler:2019nbi,Maldacena:2019cbz,Maxfield:2020ale,Sybesma:2020fxg,Kames-King:2021etp}.

\subsection{Half reduction from pure de Sitter}

We first review the derivation of the classical de Sitter JT action via a spherical reduction of pure de Sitter space in three dimensions. Consider the Lorentzian Einstein--Hilbert action with positive cosmological constant $\Lambda$ in $d$ spacetime dimensions,
\beq I_{d}=\frac{1}{16\pi G_{d}}\int_{\hat{M}} \hspace{-2mm}d^{d}X\sqrt{-\hat{g}}[\hat{R}-2\Lambda]+\frac{1}{8\pi G_{d}}\int _{\partial\hat{M}}\hspace{-3mm}d^{d-1}Y\sqrt{-\hat{h}}\hat{K}\;,\quad \Lambda=+\frac{(d-1)(d-2)}{2L_{d}^{2}}\;. \label{eq:ddimaction}\eeq
Here  $\hat{g}_{MN}$ is the $d$-dimensional metric and  $L_{d}$ is the curvature radius of dS$_d$. We have included a $(d-1)$-dimensional Gibbons--Hawking--York (GHY) boundary term, where $\hat{h}_{MN}$ is the induced metric of the boundary with  $\hat{K}$ being the trace of its extrinsic curvature.  

De Sitter space (dS$_d$) is the maximally symmetric spacetime with positive cosmological constant. In static patch coordinates the de Sitter line element is
\beq d\ell^{2}=-f(r)dt^{2}+f^{-1}(r)dr^{2}+r^{2}d\Omega_{d-2}^{2}\;,\quad f(r)=1-\frac{r^{2}}{L^{2}_{d}}\;.\label{eq:statpatchd}\eeq
The positive root $r_{\text{c}}=L_d$ of $f(r)$ gives the location of the observer-dependent cosmological horizon. For an inertial observer moving along any timelike geodesic, the cosmological horizon appears to emit thermal radiation at the Gibbons--Hawking temperature \cite{Gibbons:1977mu}
\beq T_{\text{GH}}=\frac{\kappa_{\text{c}}}{2\pi}\;,\eeq
where $\kappa_{c}$ is the surface gravity of the horizon, defined by $\xi^{a}\nabla_{a}\xi^{b}=\kappa_{\text{c}}\xi^{b}$, and $\xi^{a}$ is the time translation Killing vector. The horizon also has a thermodynamic entropy proportional to the horizon area $A_{\text{c}}$,
\beq S_{\text{GH}}=\frac{A_{\text{c}}}{4G_{d}}\;,\eeq
analogous to the Bekenstein--Hawking area formula for black holes.  In the static patch, moreover, the horizon obeys a first law, 
\beq -\delta H_\xi=T_{\text{GH}}\delta S_{\text{GH}}\;, \qquad \text{where} \qquad \delta H_\xi \equiv \int_\Sigma \delta {T_a}^b \xi^a u_b dV  \label{eq:standardfirstlawpuredS}\eeq
is the variation of the matter Killing energy on a spatial section $\Sigma$ of the static patch with future-pointing unit normal $u^b$.  
The minus sign in front indicates an increase in the matter stress energy inside the static patch leads to a decrease in the cosmological horizon and its associated entropy. 

JT gravity arises from a spherical reduction  of the Einstein--Hilbert action (\ref{eq:ddimaction}) using the metric Ansatz
\beq d\ell^{2}=\hat{g}_{MN}dX^{M}dX^{N}=g_{\mu\nu}(x)dx^{\mu}dx^{\nu}+L^{2}_{d}\Phi^{2/(d-2)}(x)d\Omega_{d-2}^{2}\;.\label{eq:dimansatz}\eeq
Here $M,N=0,1,...,d-1$, $\mu,\nu=0,1$, and $\Phi(x)$ is the dilaton. In $d=3$ we find the following two-dimensional JT action (see Appendix \ref{app:reductionscoords} for details)
\beq I_{\text{JT}}=\frac{1}{16\pi G_{2}}\int_{\mathcal{M}}\hspace{-2mm}d^{2}x\sqrt{-g}\,\Phi\!\left(R-\frac{2}{L_{3}^{2}}\right)+\frac{1}{8\pi G_{2}}\int_{\partial\mathcal{M}}\hspace{-3mm}dy\sqrt{-h}\Phi K\;,\label{eq:JTactv1}\eeq
where we introduced the two-dimensional Newton's constant $2\pi L_{3}/G_{3}=1/G_{2}$. The above action is   the JT action in de Sitter space, which   at this stage we recognize as the Wick rotated ($L_{\text{AdS}}\to  iL_{\text{dS}}$) version of the standard JT action in $\text{AdS}_{2}$. It is worth emphasizing that here we have not explicitly introduced the usual purely topological term.
While the additional topological term does not alter the equations of motion, it does influence the boundary dynamics of the theory and the Euclidean gravitational path integral \cite{Cotler:2019nbi}.  Whether we include the topological term is one of the essential differences between the two versions of JT gravity we mentioned before. 

The gravitational and dilaton equations of motion of the JT action are, respectively, 
\beq T^{\Phi}_{\mu\nu}=0\;,\quad T^{\Phi}_{\mu\nu}\equiv-\frac{2}{\sqrt{-g}}\frac{\delta I_{\text{JT}}}{\delta g^{\mu\nu}}=-\frac{1}{8\pi G_{2}}\left(g_{\mu\nu}\Box-\nabla_{\mu}\nabla_{\nu}+\frac{1}{L_{3}^{2}}g_{\mu\nu}\right)\Phi\;,\label{eq:graveom}\eeq
\beq R-\frac{2}{L_{3}^{2}}=0\;.\label{eq:dilaeom}\eeq 
Thus, the dilaton equation of motion fixes the background to be $\text{dS}_{2}$.  To find explicit expressions for the metric or the dilaton we can solve the field equations outright. From (\ref{eq:dilaeom}), we may write the 2D geometry in static coordinates 
\beq
d \ell^2 = - \left ( 1- \frac{r^{2}}{L^{2}}\right) dt^2 +  \left ( 1- \frac{r^{2}}{L^{2}}\right)^{-1}dr^2\;,\label{eq:statpatch}\eeq
where $L\equiv L_{3}$. The range of coordinates defining the static patch is $0\leq r\leq L$, where at $r=L$ the 2D geometry has a cosmological  horizon. 

Generally, the dilaton may be time-dependent; here, we restrict to a time-independent solution, in which 
\beq \Phi(r)=\Phi_{r}\frac{r}{L}  \label{eq:dilatoneom3d}\eeq
solves the gravitational equations of motion (\ref{eq:graveom}). Here $\Phi_{r}>0$ is some positive constant chosen to normalize the entropy as we see below. When we normalize the timelike Killing vector such that $\xi^{2}=-1$ at the origin $r=0$, \emph{i.e.}, $\xi = \partial_t$, we have that the surface gravities of the 2D and 3D cosmological  horizons are given by $\kappa=1/L=1/L_{3}=\kappa_{\text{c}}$. Therefore, the Gibbons--Hawking temperatures of the 2D and 3D cosmological horizons are both equal to
\beq T_{\text{GH}}=\frac{1}{2\pi L}\;. \label{eq:gibbonshawkingtemperature2d}\eeq
 The entropies in 2D and 3D likewise coincide, when we choose $\Phi_{r}=1$. This can be easily checked using the Wald entropy functional \cite{Wald:1993nt}
\beq S_{\text{JT}}=-2\pi\int_{\mathcal H}dA\frac{\partial\mathcal{L}_{\text{JT}}}{\partial R_{\mu\nu\rho\sigma}}\epsilon_{\mu\nu}\epsilon_{\rho\sigma}=\frac{\Phi_{\mathcal H}}{4G_{2}}=\frac{2\pi L_{3}}{4G_{3}}=S_{\text{GH}}\;, \label{eq:Waldentropy}\eeq
where $\epsilon_{\mu\nu}$ is the binormal to the horizon satisfying $\epsilon_{\mu\nu}\epsilon^{\mu\nu}=-2$, $dA$ is the infinitesimal area element of the bifurcation codimension-2 surface $\mathcal H$ of the Killing horizon $H$, and $\mathcal{L}_{\text{JT}}$ is the Lagrangian density defining the theory. Selecting $\Phi_{r}=1$ is also natural from comparing the 2D reduction to the $\text{dS}_{3}$ geometry (see, \emph{e.g.}, \cite{Sybesma:2020fxg}), but in the following we will keep the constant general.  Further, the first law relating the matter Killing energy $H_\xi$, temperature $T_{\text{GH}}$ and horizon entropy $S_{\text{JT}}$ for the  JT model is given by, cf. Eq. \eqref{eq:global1stlawhalfredffff},
\beq
\delta E = T_{\text{GH}} \delta S_{\text{JT}} + \delta H_\xi\,,
\eeq
where we introduced a new form of energy $E= \pm \frac{\Phi_r}{8\pi G_2 L}$,    the quasi-local energy \eqref{eq:BYE} evaluated at $r_B =0$, which vanishes in higher dimensions but is nonzero in 2D.


The de Sitter JT model found from the reduction of pure $\text{dS}_{3}$ is known as a ``half reduction'', a name inherited from a similar partial reduction of AdS JT gravity \cite{Achucarro:1993fd,Verheijden:2021yrb}. The name follows from the fact that for $r=L_{3}\cos\theta$, the two-dimensional de Sitter line element becomes
\beq d\ell^{2}=-\sin^{2}\theta dt^{2}+L_{3}^{2}d\theta^{2}\;.\label{eq:2dlineelement}\eeq
A constant time slice of $\text{dS}_{3}$ corresponds to a  circle parametrized by $\theta$. The three-dimensional parent geometry demands $\cos\theta\geq0$, \emph{i.e.}, $\theta\in[-\pi/2,\pi/2]$, where $\theta=0$ corresponds to the cosmological horizon. Consequently, the coordinate $\theta$ only covers a semi-circle with endpoints fixed at the north and south poles. The dilaton (\ref{eq:dilatoneom3d}) is never allowed to take negative values, $\Phi\geq0$. 

\subsection{Full reduction from Schwarzschild-de Sitter}

Another solution to the $d$-dimensional Einstein--Hilbert action (\ref{eq:ddimaction}) is the Schwarzschild-de Sitter (SdS) geometry, describing a neutral, non-rotating black hole in de Sitter space. In static coordinates the line element takes the form
\beq d\ell^2=-f(r)dt^{2}+f^{-1}(r)dr^{2}+r^{2}d\Omega^{2}_{d-2}\;,\quad f(r)=1-\frac{r^{2}}{L_{d}^{2}}-\frac{16\pi G_{d}M}{(d-2)\Omega_{d-2}r^{d-3}}\;,\label{eq:SdSmet}\eeq
where $M$ is the mass parameter of the black hole and $\Omega_{d-2}=2\pi^{(d-1)/2}/\Gamma[(d-1)/2]$ is the volume of the unit $(d-2)$-sphere. When $M=0$, the SdS solution (\ref{eq:SdSmet}) reduces to  pure $\text{dS}_{d}$ in static patch coordinates (\ref{eq:statpatchd}). For $d>3$ and $0<M<M_{\text{N}}$, the factor $f(r)$ has two positive roots associated with the locations of the black hole and cosmological horizons, $r_{\text{h}}$ and $r_{\text{c}}$, respectively, with $r_{\text{h}}<r_{\text{c}}$. 


The upper bound $M_{\text{N}}$ corresponds to   the  Nariai solution  \cite{Nariai},   when $r_{\text{h}}=r_{\text{c}}=r_{\text{N}}$ with
\beq r_{\text{N}}=\sqrt{\frac{d-3}{d-1}}L_{d}\;,\quad M_{\text{N}}=\frac{d-2}{d-1}\frac{\Omega_{d-2}}{8\pi G_{d}}r_{\text{N}}^{d-3}\;.\label{eq:Nariailim}\eeq
For masses $M>M_{\text{N}}$ the SdS has a naked singularity, hence the Nariai black hole is the largest physical black hole that fits inside the cosmological horizon. Moreover, the sum of the black hole and cosmological horizon areas is less than the area of the pure de Sitter cosmological horizon, obeying the bound $A(r_{\text{h}})+A(r_{\text{c}})\leq A(L)$, \emph{i.e.}, putting a black hole inside de Sitter only leads to a decrease in entropy. 

The Smarr formula and first law for Schwarzschild-de Sitter are given by \cite{Gibbons:1977mu,Sekiwa:2006qj,Dolan:2013ft} 
\begin{equation} \label{eq:smarrfirstsds}
    0 = \frac{\kappa_{\text h} A_{\text h}}{8 \pi G_d} + \frac{\kappa_{\text c} A_{\text c}}{8 \pi G_d} - \frac{\Theta_\xi \Lambda}{(d-2)4 \pi G_d}\,, \qquad -\delta H_\xi = \frac{\kappa_{\text h} }{8 \pi G_d} \delta  A_{\text h}+ \frac{\kappa_{\text c} }{8 \pi G_d} \delta A_{\text c}\,,
\end{equation}
where $\kappa_{\text{h,c}}$ are the surface gravities associated to the black hole and cosmological horizon,   $A_{\text{h,c}}$ are the respective horizon areas, and $\delta H_\xi$ is the matter Killing energy variation in \eqref{eq:standardfirstlawpuredS}. Further, $\Theta_\xi$ is the quantity conjugate to the cosmological constant in an extended version of the first law where $\Lambda$ is allowed to vary. It can     be defined as a surface integral  of the Killing potential \cite{Kastor:2009wy}, or equivalently as the ``Killing volume''   $\Theta_\xi = \int_\Sigma |\xi| dV$ \cite{Jacobson:2018ahi}, where $|\xi|=\sqrt{-\xi \cdot \xi}$  is the norm of the time translation Killing vector $\xi$, and $dV$ is the proper volume element of the spatial section $\Sigma$. In the limit $r_{\text h} \to 0$ the Smarr formula reduces to the one for pure de Sitter: $ 0 =  \frac{\kappa_{\text c} A_{\text c}}{8 \pi G_d} - \frac{\Theta_\xi \Lambda}{(d-2)4 \pi G_d}$.  Below we will see what form the Smarr relation and   first law  will take   after a dimensional reduction of the SdS black hole.  

We will be interested in the near-Nariai limit of the SdS black hole. 
In this limit the coordinates describing the SdS solution (\ref{eq:SdSmet}) are inappropriate because the function $f(r)\to0$ in between the black hole and cosmological horizons. Instead, following \cite{Nariai,Ginsparg:1982rs}, the Nariai metric may be cast as a $\text{dS}_{2}\times S^{d-2}$ geometry (see Appendix \ref{app:nariaigeom} for details),
\beq d\ell^2=-\left(1-\frac{\tilde{\rho}^{2}}{\hat{L}_{d}^{2}}\right)d\tilde{\tau}^{2}+\left(1-\frac{\tilde{\rho}^{2}}{\hat{L}_{d}^{2}}\right)^{-1}d\tilde{\rho}^{2}+r_{\text{N}}^{2}d\Omega^{2}_{d-2}\;,\label{eq:Nariaimet}\eeq
with $\hat{L}_{d}=L_{d}/\sqrt{d-1}$. In this geometry, the black hole and cosmological horizons are at $\tilde \rho = - \hat L_d$ and $\tilde \rho = \hat L_d$, respectively. They are a finite proper distance apart in a single static patch and are in thermal equilibrium with each other at the Nariai temperature,
\begin{equation} \label{eq:nariaittemp}
T_{\text{N}}=\frac{\tilde \kappa_{\text N}}{2\pi}=\frac{1}{2\pi \hat{L}_{d}}~,
\end{equation}
because the surface gravities are the same for the two horizons in the Nariai limit $\tilde \kappa_{\text N} =    1/ \hat L_d$, cf. Eq. \eqref{eq:nariaisurfacegravity}. 
Moreover, since a single static patch has both the black hole and cosmological horizons, the total entropy of the Nariai solution $S_{\text{N}}$ in this patch is given by the sum of the black hole and cosmological entropies $S_{\text{h}}$ and $S_{\text{c}}$:
\beq S_{\text{N}}=S_{\text{h}}+S_{\text{c}}=\frac{2\Omega_{d-2}r_{\text{N}}^{d-2}}{4G_{d}}\;.\label{eq:entropynariai}\eeq
The dimensional reduction of the near-Nariai limit of the $\text{SdS}_{d}$ solution for $d>3$ leads to another version of de Sitter JT gravity, described by the action (see Appendix \ref{app:reductionscoords} for details)
\beq 
\begin{split}
I_{\text{JT}}&=\frac{1}{16\pi G_{2}}\int_{\mathcal{M}}\hspace{-2mm}d^{2}x\sqrt{-g}\left((\phi_{0}+\phi)R- \frac{2}{L_{d}^{2}} \phi\right)+\frac{1}{8\pi G_{2}}\int_{\partial\mathcal{M}}\hspace{-3mm}dy\sqrt{-h}(\phi_{0}+\phi)K\,,
\end{split}
\label{eq:JTactv2}\eeq
where we have identified the dimensionless two-dimensional Newton's constant $G_{2}$ as
\beq \frac{\Omega_{d-2}r_{\text{N}}^{d-2}}{G_{d}}\equiv\frac{1}{G_{2}}\;.\eeq
The dilaton $\phi$ is related to $\Phi$ via the expansion $\Phi\approx\phi_0+\phi$, where $\Phi=\phi_{0}$ corresponds to the metric Ansatz reducing to the Nariai geometry, and $\phi$ represents a deviation away from the Nariai (``extremal'') solution, analogous to the case of AdS JT gravity. Notice that $\phi_{0}$ is proportional to the entropy of the Nariai black hole 
\beq \frac{\phi_{0}}{4G_{2}}=\frac{\Omega_{d-2}r_{\text{N}}^{d-2}}{4G_{d}}=\frac{1}{2}S_{\text{N}}\;, \label{eq:nariaijtentropy}\eeq
and hence we restrict to positive values $\phi_0 > 0$. 
Since $\phi_0$ in the action is just a topological term, the equations of motion are identical to \eqref{eq:graveom} and \eqref{eq:dilaeom}. The 2D de Sitter geometry in static coordinates is still given by \eqref{eq:statpatch}, but now the radial coordinate ranges from $r_{\text h}=-L$ (black hole horizon) to $r_{\text c}=L$ (cosmological horizon) in the static patch. Furthermore, in this paper we consider the static dilaton solution   $\phi=\phi_r \frac{r}{L},$ with $\phi_r>0$, and the     Gibbons--Hawking temperature is again given by $T_{\text{GH}}=1/2\pi L$ \eqref{eq:gibbonshawkingtemperature2d}.

The total entropy of the near-Nariai solution can be computed using the Wald entropy functional \eqref{eq:Waldentropy}. It includes both the entropy of the Nariai black hole and the dilaton correction, and for each horizon it is given by
\beq
S_{{\text{h,c}} } = \frac{\Phi_{\text{h,c}}}{4G_2} = \frac{\phi_0}{4G_2} +  \frac{\phi_{\text{h,c}}}{4G_2}= S_{\phi_0} + S_{\phi_{\text{h,c}}}\,. \label{eq:JTentropyfullreduction}
\eeq
Here $S_{\phi_0}= \frac{\phi_0}{4G_2}$ is the entropy for each horizon in the extremal Nariai solution and the term $S_{\phi_{\text{h,c}}}=\pm\frac{\phi_r}{4G_2}$ is the non-extremal dilaton correction to the Nariai horizon entropy, where the plus sign corresponds to the cosmological horizon $r_{\text{c}}=L$ and the minus sign to the black hole horizon $r_{\text h}=-L$. Hence, if we add the entropies of the black hole and cosmological horizons, the dilaton corrections cancel in the total entropy, and the sum $2\frac{  \phi_0}{4 G_2}$ matches with the total entropy $S_{\text N}$ \eqref{eq:entropynariai} of the higher-dimensional Nariai black hole. 

Moreover, the Gibbons--Hawking temperature   $T_{\text{GH}}$ \eqref{eq:gibbonshawkingtemperature2d}, the  horizon entropies   $S_{\phi_{\text h,c}} $    \eqref{eq:JTentropyfullreduction} and the matter Killing energy $H_\xi$ \eqref{eq:standardfirstlawpuredS} are related via the  Smarr formula and first law for the dimensional reduction of the near-Nariai solution 
\begin{equation} \label{eq:smarrfirstlaw2dfullred}
    0=T_{\text{GH} } S_{\phi_{\text h}} +  T_{\text{GH} } S_{\phi_{\text c}}\, , \qquad-\delta H_\xi =  T_{\text{GH}} \delta  S_{\phi_{\text h}}+  T_{\text{GH}} \delta  S_{\phi_{\text c}}\,.
\end{equation}
Above we have left out the entropy of the Nariai black hole $S_{\phi_0}$ in both relations. In particular, $\phi_0$  has been held fixed in the first law, and we will do so in the rest of the paper.\footnote{In \cite{Rosso:2020zkk,Pedraza:2021cvx}  an ``extended'' first law for AdS$_2$ was derived, where $\phi_0$, $\Lambda$ and $G$ were allowed to vary.} It is possible to include $S_{\phi_0}$  in the Smarr formula, however, one must then also add a term  proportional to the cosmological constant $\Lambda$ and the Killing volume $\Theta_\xi=\int_\Sigma |\xi| dV$, as in the Smarr formula \eqref{eq:smarrfirstsds} for Schwarzschild-de Sitter space. Indeed, in Section \ref{sec:quasilocaleulerfirst} we derive the following Smarr relation for de Sitter JT gravity, cf. Eq. \eqref{eq:smarrfullstaticpatchh}, 
\begin{equation}
    0=T_{\text{GH} } S_{{\text h}} +  T_{\text{GH} } S_{{\text c}} - \frac{\phi_0 \Lambda}{8 \pi G_2} \Theta_\xi\,. \label{eq:secondsmarrstatic}
\end{equation}
One of the main goals of this paper is to extend this Smarr relation and the first law for dS$_2$ \eqref{eq:smarrfirstlaw2dfullred}  to quasi-local boundaries and to include semi-classical corrections.

Finally, the de Sitter JT action found from reducing $d$-dimensional SdS in the near-Nariai limit is known as the ``full reduction'' model. This is because now the coordinate $\theta$ in (\ref{eq:2dlineelement}) ranges from $0$ to $2\pi$, along the full circle. Consequently,   one relaxes $\phi\geq0$ to $\phi_{0}+\phi\geq0$, such that $\phi$ may take on negative values \cite{Sybesma:2020fxg}. Further, we emphasize spherical reduction of the $d$-dimensional Nariai black hole gives rise to a topological term in the action (\ref{eq:JTactv2}), whose importance was analyzed in   \cite{Hartman:2020khs,Cotler:2019nbi}. Lastly, without loss of generality, for the remainder of the article we work with the JT action following from the full reduction (\ref{eq:JTactv2}), dropping the subscript from $L_{d}$. The half reduction model can be obtained by setting $\phi_0 =0$ and restricting the range of coordinates for dS$_2$.

\subsection{Geometry of dS$_2$}

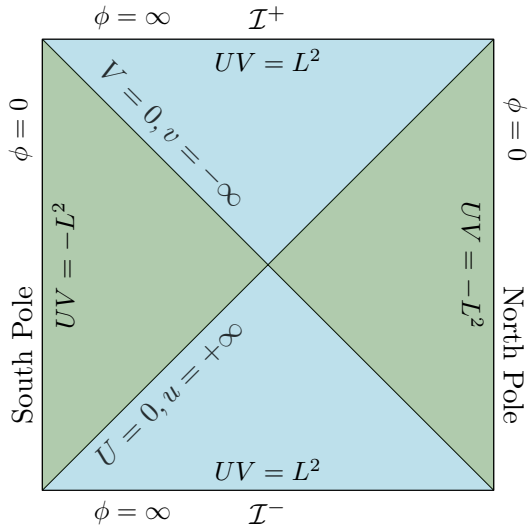
\begin{figure}[t!]
\centering
\begin{tikzpicture}[scale=1.5]
	\pgfmathsetmacro\myunit{4}
	\draw	(0,0)			coordinate (a)
		--++(90:\myunit)	coordinate (b)
							node[pos=.3, above, sloped] {\text{South Pole}}
							node[pos=.8, above, sloped] {{\small $\phi = 0$} }
		--++(0:\myunit)		coordinate (c)
		                    node[pos=.2, above] {{\small $\phi = \infty$}}
							node[pos=.5, above] {$\mathcal{I}^+$}
		--++(-90:\myunit)	coordinate (d)	
							node[pos=.2, sloped, above] {{\small $\phi = 0$}}
							node[pos=.7, sloped, above] {\text{North Pole}}
		--cycle 			node[pos=.5, below] {$\mathcal{I}^-$}
		                    node[pos=.8, below] {{\small $\phi = \infty$}};
    \draw[fill=lightblue, fill opacity=0.8] (a) -- node[pos=.5, below, sloped] {$U = 0, u = + \infty$} (2,2) --  (d);
    \draw[fill=lightblue, fill opacity=0.8] (b) -- node[pos=.5, above, sloped] {$V=0, v=-\infty$} (2,2) -- (c);
    \fill[fill=OliveGreen, fill opacity=0.4] (a) -- (2,2) -- (b);
    \fill[fill=OliveGreen, fill opacity=0.4] (c) -- (2,2) -- (d);
    \draw (b) -- (c) node[pos=.5, below] {{\small $UV = L^2$}};
    \draw (a) -- (d) node[pos=.5, above] {{\small $UV = L^2$}};
    \draw (a) -- (b) node[pos=.5, below, sloped] {{\small $UV = -L^2$}};
    \draw (c) -- (d) node[pos=.5, below, sloped] {{\small $UV = -L^2$}};
\end{tikzpicture}
\caption{Two-dimensional de Sitter space in the half reduction model. The left and right (green) triangles represent the two static patches. In the half reduction model, the dilaton $\phi\geq0$, where it formally diverges at past and future infinity $\mathcal{I}^{\pm}$ and vanishes at the poles.}
\label{fig:dS2halfred} 
\end{figure}

\begin{figure}[t!]
\centering
\begin{tikzpicture}[scale=1.5]
	\pgfmathsetmacro\myunit{4}
	\draw[dashed]	(0,0)			coordinate (a)
		--++(90:\myunit)	coordinate (b);
	\draw (b) --++(0:\myunit)		coordinate (c)
		                    node[pos=.2, above] {{\small $\phi = \infty$}}
							node[pos=.5, above] {$\mathcal{I}^+$}
							node[pos=.8, above] {{\small $r = \infty$}};
	\draw[dashed] (c) --++(-90:\myunit)	coordinate (d);
	\draw (d) -- (a) 		node[pos=.2, below] {{\small $r = \infty$}}
                        	node[pos=.5, below] {$\mathcal{I}^-$}
		                    node[pos=.8, below] {{\small $\phi = \infty$}};
		                  
    \draw[fill=lightblue, fill opacity=0.8] (a) -- node[pos=.5, below, sloped]  {{\small $U = 0, r = L$}} (2,2) --  (d);
    \draw[fill=lightblue, fill opacity=0.8] (b) -- node[pos=.5, above, sloped] {{\small $V=0, r = L$}} (2,2) -- (c);
    \fill[fill=OliveGreen, fill opacity=0.4] (a) -- (2,2) -- (b);
    \fill[fill=OliveGreen, fill opacity=0.4] (c) -- (2,2) -- (d);
    \draw (b) -- (c) node[pos=.5, below] {{\small $UV = L^2$}};
    \draw (a) -- (d) node[pos=.5, above] {{\small $UV = L^2$}};
    \draw[dashed] (a) -- (b) node[pos=.5, below, sloped] {{\small $UV = -L^2$}};
    \draw[dashed] (c) -- (d) node[pos=.5, below, sloped] {{\small $UV = -L^2$}};
    \fill[fill=OliveGreen, fill opacity=0.4] (a) -- (-2,2) -- (b) ;
    \fill[fill=OliveGreen, fill opacity=0.4] (c) -- (6,2)  -- (d);
    \draw (a) -- (-2,2) -- (b) node[pos=.5, above, sloped] {{\small $U = 0, r = -L$}};
    \draw (c) -- (6,2) node[pos=.5, above, sloped] {{\small $V = 0, r = - L$}} -- (d);
    \draw[decorate, decoration={snake, amplitude=0.5mm, segment length=2.5mm}] (a) -- (-2,0) node[pos=.5, below] {{\small $\phi = - \infty$}} coordinate (e);
    \draw (e) -- (-2,4) coordinate (f);
    \draw[decorate, decoration={snake, amplitude=0.5mm, segment length=2.5mm}] (f) -- (b) node[pos=.5, above] {{\small $\phi = - \infty$}}; 
    \draw[decorate, decoration={snake, amplitude=0.5mm, segment length=2.5mm}] (c) -- (6,4) node[pos=.5, above] {{\small $\phi = - \infty$}} coordinate (g); 
    \draw (g) -- (6,0) coordinate (h);
    \draw[decorate, decoration={snake, amplitude=0.5mm, segment length=2.5mm}] (h) -- (d) node[pos=.5, below] {{\small $\phi = - \infty$}};
\end{tikzpicture}
\caption{Two-dimensional de Sitter space in the full reduction model. The left and right green causal diamonds are the static patches, while both the blue shaded and unshaded white regions are referred to as hyperbolic patches. The near-Nariai black hole geometry is imprinted in the two-dimensional geometry via the ``black hole" interiors, the white regions, with past and future singularities residing inside, where  the dilaton takes arbitrarily large negative values. The left and right edges are identified.}
\label{fig:dS2fullred}  
\end{figure}
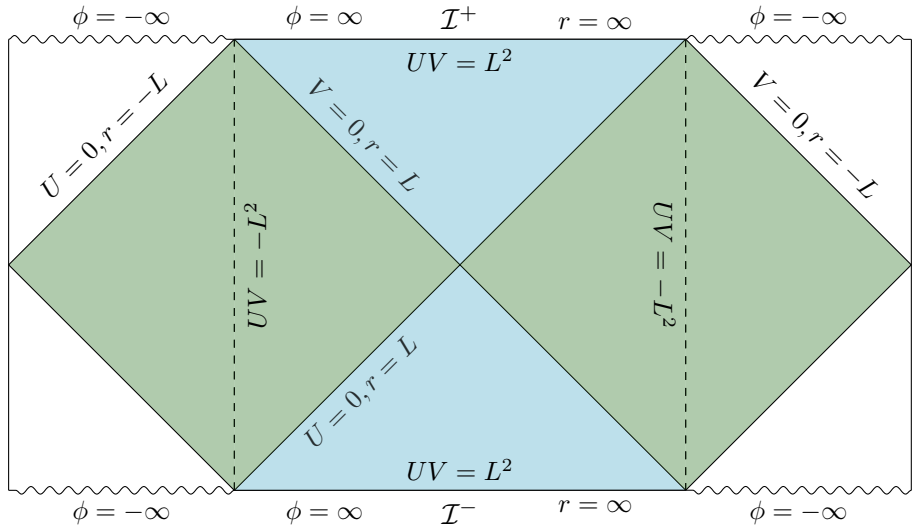

 While $\text{dS}_{2}$ is the fixed geometry in either the full or half reduction versions of JT gravity, the global structure of the two-dimensional space in either model is different due to the higher-dimensional solution from whence they came. In the half reduction model, spatial sections of dS$_2$ are semi-circles where the polar angle runs from $-\pi/2$ to $+\pi/2$, whereas in the full reduction model spatial sections are entire $S^1$'s where the polar angle runs from $-\pi$ to $\pi$ \cite{Sybesma:2020fxg,Aalsma:2021bit}. This explains why the Penrose diagram of the full reduction de Sitter space (Figure~\ref{fig:dS2fullred}) is twice as wide as the Penrose diagram of the half reduction de Sitter space (Figure~\ref{fig:dS2halfred}). Equivalently, the Penrose diagram of two-dimensional de Sitter in the full reduction is a rectangle whereas the Penrose diagram of higher-dimensional de Sitter is a square.\footnote{Note that two-dimensional de Sitter can in principle be infinitely extended, just as the Penrose diagram of Schwarzschild-de Sitter, \emph{i.e.}, there is no requirement from the field equations on the periodicity of the global spatial coordinate ($\varphi$ in Eq. \eqref{eq:globalcoordmetapp}). Only if one demands that $\text{dS}_{2}$ arises as a hyperboloid in the embedding space $\mathbb{R}^{1,2}$ is the global coordinate restricted to be periodic ($\varphi\sim\varphi + 2\pi$). We assume this periodicity here, as represented in Figure \ref{fig:dS2fullred}. We thank Jan Pieter van der Schaar for stressing this. \label{footnoteJanPieter}} This is simply a consequence of the fact that $\text{dS}_{2}$ in the full reduction follows from dimensionally reducing a Schwarzschild-dS black hole.


To illustrate this point, and since it will benefit us when we discuss different de Sitter vacua, let us briefly introduce two sets of coordinate systems for $\text{dS}_{2}$ (see also Appendix \ref{app:reductionscoords}).
First, let $(v,u)$ denote advanced and retarded time coordinates for the static patch (\ref{eq:statpatch}), defined respectively by
 \beq v=t+r_{\ast}\;,\quad u=t-r_{\ast}\;.\eeq
 Here $r_{\ast}=L\text{arctanh} (r/L)$ is a tortoise coordinate, which ranges from $r_\ast = -\infty$ (black hole horizon) to $r_\ast = +\infty$ (cosmological horizon). 
 In these null coordinates the static patch line element (\ref{eq:statpatch}) becomes
 \beq d\ell^2=-\text{sech}^{2}\left(\frac{v-u}{2L}\right)dvdu\;.\label{eq:statnullmet}\eeq
Static patch coordinates (\ref{eq:statnullmet}) only cover a part of de Sitter space. To describe the regions to the future and past of the cosmological horizons one may consider coordinate ranges $r\in(L,\pm\infty)$, however, we can instead cover the full space of the half reduction model by introducing global Kruskal-like coordinates $(V,U)$:
 \beq V=Le^{v/L}\;,\quad U=-Le^{-u/L}\;. \label{eq:kruskalnullcoord}\eeq
 The line element in Kruskal coordinates is
 \beq d\ell^2=-\frac{4}{\left(1-\frac{UV}{L^{2}}\right)^{2}}dVdU\;.\label{eq:kruskalmet}\eeq
 In these coordinates, $UV=-L^{2}$ corresponds to the location of the poles $r=0$, while $UV=+L^{2}$ yields $r=+\infty$, corresponding to the future and past conformal boundary $\mathcal{I}^{\pm}$.  Moreover, the  cosmological horizons are located at ($V=0$, $U=0$).  Both the static patches and global structure of $\text{dS}_{2}$ are depicted in Figures \ref{fig:dS2halfred} and \ref{fig:dS2fullred}. 
 

 Again, the key difference between the half and full reduced models is that in the former the dilaton is strictly non-negative, while the latter allows for $\phi$ to be infinitely negative. The consequence of this is that the  global geometry of $\text{dS}_{2}$ arising from the half reduction resembles that of pure $\text{dS}_{3}$. In this case, the dilaton vanishes at the poles and grows infinite at $\mathcal{I}^{\pm}$, as displayed in Figure \ref{fig:dS2halfred}. Alternatively, the $\text{dS}_{2}$ geometry coming from the full reduction is simply the full dS$_2$  spacetime, which has a different Penrose diagram than the higher-dimensional~dS$_d$, as illustrated in Figure \ref{fig:dS2fullred} (and it can be infinitely extended, see footnote \ref{footnoteJanPieter}). Moreover, it includes features of the higher-dimensional Nariai black hole. Specifically, 
 the $\text{dS}_{2}$ geometry contains the black hole interiors, hiding past and future singularities where the dilaton diverges to negative infinity. 
 


\section{Quasi-local thermodynamics and generalized entropy}\label{sec:quasithermo}
Spatial sections of de Sitter space are compact and hence there is no asymptotic boundary where conserved charges, such as the energy, can be defined. One way to circumvent this is to define conserved charges at future infinity $\mathcal I^+$, as done for instance in \cite{Balasubramanian:2001nb,Anninos:2010zf}, but a static observer does not have access to this region (although a meta-observer does).   Alternatively, one may introduce a timelike boundary $B$ at a radius $r=r_B$, where quasi-local conserved charges can be defined \cite{Brown:1992br} (see Figure \ref{fig:dS2bdryB}). One benefit of the quasi-local method is that the charges, especially the energy, are well defined in the static patch. However, 
the main advantage of this second approach is that by fixing the temperature at the timelike boundary   the canonical thermodynamic ensemble is well defined. More precisely, in a Euclidean path integral description of the canonical ensemble, one has to fix  the temperature at a certain boundary. Since there is no asymptotic boundary in Euclidean de Sitter, one has to introduce an auxiliary boundary where the temperature is uniquely specified. The naive evaluation of the on-shell Euclidean action for de Sitter space \emph{\`a la} Gibbons and Hawking \cite{Gibbons:1976ue,Gibbons:1977mu} indeed gives the correct entropy for de Sitter space, but it is not clear how this entropy follows from a canonical partition function, given that Euclidean dS has no asymptotic boundary where a temperature may be fixed. Thus, the Brown-York quasi-local method seems necessary to properly define the canonical Euclidean path integral for asymptotically de Sitter space.\footnote{We acknowledge Ted Jacobson for pointing this out to us. See also \cite{Banihashemi:2022jys}.} 

\subsection{Tolman temperature and  quasi-local energy}

For de Sitter JT gravity, we define the canonical ensemble by fixing the dilaton and the (local) temperature   at the  timelike boundary $B$ located at $r=r_B$. The boundary $B$   is equivalently defined as a flow line of the Killing vector $\xi=\partial_t$ generating time translations in the static patch, along which the norm of $\xi$ is constant,
\begin{equation} \label{eq:normkilling}
    N  \equiv \sqrt{- \xi^\mu \xi_\mu }=  \sqrt{1 - \frac{r_B^2}{L^2}}\,.
\end{equation}
The temperature at the boundary is uniform and is given by the redshifted Gibbons--Hawking temperature, also known as the Tolman temperature,
\begin{equation} \label{eq:tolmantemp}
    T (r_B)= \frac{\kappa}{2\pi N}  = \frac{1}{2\pi L} \frac{1}{\sqrt{1-\frac{r_B^2}{L^2}}} \, . 
\end{equation}
Notice the Tolman temperature attains its minimum value at the origin, $T(r_B=0)=1/2\pi L$, and diverges at the two horizons $r_{B}=\pm L$.

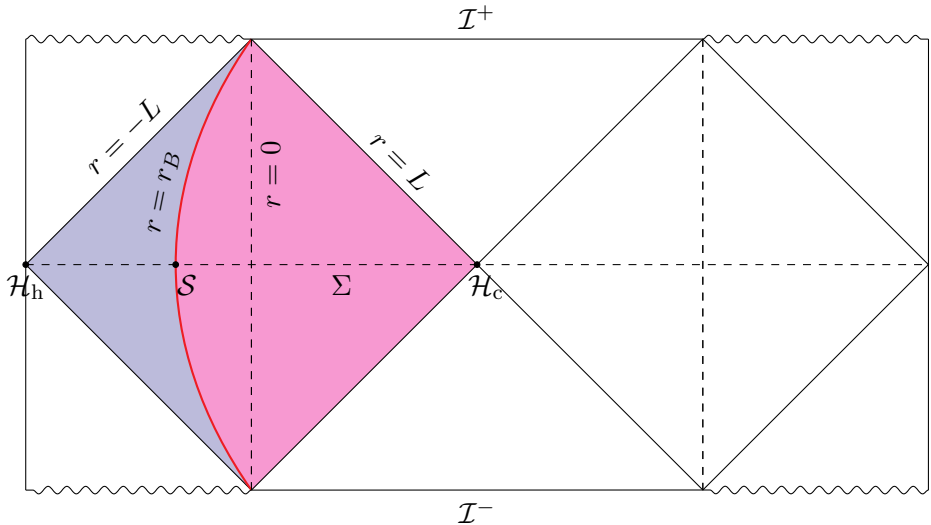
\begin{figure}[t!]
\centering
\begin{tikzpicture}[scale=1.5]
	\pgfmathsetmacro\myunit{4}
	\draw[dashed]	(0,0)			coordinate (a)
		--++(90:\myunit)	coordinate (b);
	\draw (b) --++(0:\myunit)		coordinate (c)
							node[pos=.5, above] {$\mathcal{I}^+$};
	\draw[dashed] (c) --++(-90:\myunit)	coordinate (d);
    \fill[fill=Periwinkle, fill opacity=0.5] (a) to[bend left=35] (b) -- (-2,2) -- (a);
    \fill[fill=Magenta, fill opacity=.4] (a) to[bend left=35] (b) -- (2,2) -- (a);
    \draw (a) -- (-2,2) -- (b) node[pos=.5, above, sloped] {$r = - L$};
    \draw[thick, Red, name path=rB] (a) to[bend left=35] (b);
    \path (a) to[bend left=35] node[pos=.65,above,sloped] {$r=r_B$} (b);
	\draw (d) -- (a) 		node[pos=.5, below] {$\mathcal{I}^-$};
	\draw (b) -- (d) node[pos=.3, above, sloped] {$r = L$} -- (6,2) -- (c) -- (a);
    \draw[dashed] (a) -- (b) node[pos=.7, below, sloped] {$r = 0$};
    \draw[dashed] (c) -- (d);
    \draw[decorate, decoration={snake, amplitude=0.5mm, segment length=2.5mm}] (a) -- (-2,0) coordinate (e);
    \draw (e) -- (-2,4) coordinate (f);
    \draw[decorate, decoration={snake, amplitude=0.5mm, segment length=2.5mm}] (f) -- (b); 
    \draw[decorate, decoration={snake, amplitude=0.5mm, segment length=2.5mm}] (c) -- (6,4) coordinate (g); 
    \draw (g) -- (6,0) coordinate (h);
    \draw[decorate, decoration={snake, amplitude=0.5mm, segment length=2.5mm}] (h) -- (d);
    \draw[dashed, name path=Sigma] (-2,2) -- (6,2) node[pos=.35, below] {$\Sigma$} node[pos=0, below] {$\mathcal{H}_{\text{h}}$}  node[pos=.51, below] {$\mathcal{H}_{\text{c}}$};
    \path[name intersections={of=rB and Sigma, by={int}}] (int) --++(0:1) node[pos=0.1, below] {$\mathcal{S}$};
    \filldraw (int) circle (0.025cm);
    \filldraw (2,2) circle (0.025cm);
    \filldraw (-2,2) circle (0.025cm);
\end{tikzpicture}
\caption{Introducing a Brown-York timelike boundary $B$ (red) at radius $r=r_{B}$ in dS$_2$. We define quasi-local charges with respect to this boundary, a surface with a fixed Tolman temperature. In the full reduction model   $B$ rests somewhere between $r=-L$ and $r=L$ in the static patch. 
The shaded blue region refers to black hole system, while the shaded magenta region describes the cosmological system. The constant-$t$ slice $\Sigma$ has boundary $\partial\Sigma=\mathcal{S}\cup\mathcal{H}$ with $\mathcal{S}$ being the intersection of $\Sigma$ and $B$, and $\mathcal{H}$ is the bifurcation point of the Killing horizon located at $r_{\text{h}}$ or $r_{\text{c}}$ for the black hole  or cosmological system, respectively.}
\label{fig:dS2bdryB}  
\end{figure}

In the following we  consider two different thermodynamic systems: (1) the ``black hole system'' between the black hole horizon and the boundary at radius $r=r_B$, and (2) the ``cosmological system'' between the boundary $B$ and the cosmological horizon (see  Figure~\ref{fig:dS2bdryB}). We derive the thermodynamic  variables  using the  canonical Euclidean path integral for these two systems. A similar analysis was performed, for instance, for Schwarzschild black holes in~\cite{York:1986it},   for two-dimensional black holes \cite{Creighton:1995uj,Lemos:1996bq}, and for SdS black holes in \cite{Banihashemi:2022jys}.



Before we study the path integral, it is worth pointing out that the quasi-local energy can   be directly computed  from the Brown-York stress-energy tensor \cite{Brown:1992br}. In particular, for JT gravity the stress tensor is   (see   Appendix \ref{app:noetherchargesumm})
\begin{equation}
    \tau^{\mu \nu} \equiv \frac{2}{\sqrt{-\gamma}} \frac{\delta I_{\text{JT}}}{\delta \gamma_{\mu \nu}} = \frac{1}{8 \pi G_2} \gamma^{\mu \nu} n^\alpha \nabla_\alpha \phi~,
\end{equation}
where $n^{\mu}$ is an outward-pointing spacelike unit normal to the boundary $B$ of the system under consideration, with induced metric $\gamma_{\mu\nu}=g_{\mu\nu}-n_{\mu}n_{\nu}$. The quasi-local energy $E$ is then 
\begin{equation}
    E =  u_\mu u_\nu \tau^{\mu \nu} = -  \frac{1}{8 \pi G_2}   n^\alpha \nabla_\alpha \phi = \pm \frac{\phi_r}{8 \pi G_2 L} \sqrt{1- \frac{r_B^2}{L^2}}\;, 
\label{eq:BYE}\end{equation}
with $u_{\mu}$ being a future-pointing timelike unit normal to a Cauchy surface $\Sigma$ with induced metric $h_{\mu\nu}=g_{\mu\nu}+u_{\mu}u_{\nu}$.
The plus/minus signs in the last expression for $E$ correspond to the cosmological/black hole systems, since the outward pointing unit vectors normal to $B$ are $n = \mp \sqrt{1- \frac{r_B^2}{L^2}}\partial_r$, respectively.
Notice further  in the full reduction model ($\phi_0 \neq 0$ and $r_B \in [-L,L]$)  the total energy of the  static patch of two-dimensional de Sitter space  vanishes, since  for $r_B = \pm L$ we have $E=0$. However, in the half reduction model ($\phi_0 =0$ and $r_B \in [0,\pm L]$) the total energy of the static patch is non-zero, since for $r_B=0$ we have $E   = \pm\frac{\phi_r}{8 \pi G_{2} L} .$ 


\subsection{On-shell Euclidean action,  free  energy and  heat capacity}

Next, we compute the quasi-local thermodynamic quantities for the black hole and cosmological systems separately by evaluating the Euclidean action on-shell. 
To compute the on-shell Euclidean action, we Euclideanize the Lorentzian $\text{dS}_{2}$ static patch geometry (\ref{eq:statpatch}) by analytically continuing  $t\to-i\tau$,
 \begin{equation}
    d \ell^2 = f(r)d \tau^2 + f^{-1}(r) dr^2 , \qquad f(r) = 1 - \frac{r^2}{L^2} \, .
 \end{equation}
 Removing the conical singularity at the horizons yields a periodicity of the Euclidean time circle, equal to the inverse Gibbons--Hawking  temperature $\beta_{\text{GH}} =  2 \pi /\kappa=2\pi L $. When the Euclidean time has periodicity $\tau \sim \tau + \beta_{\text{GH}}$, the line element of Euclidean dS$_2$ describes a round two-sphere, cf. Eq. \eqref{eq:euclideandsstatic}.  
The proper length of the boundary at radius $r=r_B$ is equal to
\begin{equation}
    \beta (r_B) = \int_0^{\beta_{\text GH}}d \tau  \sqrt{f(r_B)} = \beta_{\text{GH}} \sqrt{f(r_B)} = 2 \pi L \sqrt{1- \frac{r_B^2}{L^2}} \, , \label{eq:inversetolman}
\end{equation}
which we recognize as the inverse of the Tolman temperature \eqref{eq:tolmantemp}.  

Following Gibbons and Hawking \cite{Gibbons:1976ue}, we express the gravitational canonical partition function   as a Euclidean path integral, which can be computed by a saddle-point approximation
\begin{equation}
    Z (\beta) =\text{Tr} \, e^{-\beta H} = \int \mathcal D \psi e^{-I^{\text E}_{\text{JT}}} \approx  e^{-I^{\text{E}}_{\text{JT}}} \,, 
\end{equation}
where $\psi$ denotes the set of dynamical fields, namely the metric $g_{\mu\nu}$ and dilaton $\phi$. We emphasize that here the canonical ensemble is defined with respect to the Tolman temperature, not the Gibbons--Hawking temperature.
The total off-shell Euclidean action of de Sitter JT gravity $I^{\text{E}}_{\text{JT}}$ in the full reduction model is
\begin{equation}
    I_{\text{JT}}^{\text{E}} =-\frac{1}{16\pi G_{2}}\int_{\mathcal{M}_{\text{E}}}\hspace{-2mm}d^{2}x\sqrt{ g} \left [\phi_{0} R+ \phi \left ( R-\frac{2}{L^{2}} \right)\right]-\frac{1}{8\pi G_{2}}\int_{B} d\tau \sqrt{h}(\phi_{0}+\phi)K\,.
\end{equation}
Note that on-shell the bulk contribution proportional to $\phi$ vanishes due to the dilaton equation of motion, $R = 2 / L^2$.  We now compute the on-shell Euclidean JT action   for the two thermodynamic systems, starting with the cosmological system. 

\paragraph{Cosmological system.} The bulk term in the action proportional to $\phi_0$ is
\begin{equation}
    -\frac{1}{16\pi G_{2}}\int_{\mathcal{M}_{\text{E}}}\hspace{-2mm}d^{2}x\sqrt{ g}  \phi_{0} R = - \frac{\beta_{\text{GH}} \phi_0}{8 \pi G_2 L^2} (L- r_B) = \frac{  \phi_0}{4 G_2} \frac{r_B}{L } - \frac{\phi_0}{4 G_2} \,, \label{eq:actionterm1}
\end{equation}
where we integrate from the timelike boundary  $r=r_{B}$ to the cosmological horizon $r_{\text c}=L$. Note that the second term on the right-hand side is (half) the entropy of the Nariai solution, cf. Eq. \eqref{eq:nariaijtentropy}. The  first term actually cancels against the GHY term proportional to $\phi_0$,
\begin{equation}
    -\frac{1}{8\pi G_{2}}\int_{B} d\tau \sqrt{h} \phi_{0} K = - \frac{\beta_{\text{GH}}\phi_0}{8 \pi G_2} \frac{r_B}{L^2}=-\frac{  \phi_0}{4 G_2} \frac{r_B}{L }\,, \label{eq:actionterm2}
\end{equation}
where we inserted the trace of the extrinsic curvature of the boundary $B$  
\begin{equation}
    K = -\frac{1}{2} \frac{f'(r_B)}{\sqrt{f(r_B)}} = \frac{r_B/L^2}{\sqrt{1- \frac{r_B^2}{L^2}}}~. 
\end{equation}
The remaining GHY term contains the essential information about the quasi-local thermodynamics,
\begin{equation}
        -\frac{1}{8\pi G_{2}}\int_{B} d\tau \sqrt{h} \phi  K = - \frac{\beta (r_B)\phi_r}{8 \pi G_2 L} \frac{r_B^2/L^2}{\sqrt{1 - \frac{r_B^2}{L^2}}} = \beta  \frac{\phi_r}{8 \pi G_2 L} \sqrt{1 - \frac{r_B^2}{L^2}}  -  \frac{\phi_r}{4 G_2}\;,   \label{eq:actionterm3}
\end{equation}
where we used the inverse temperature relation (\ref{eq:inversetolman}).

\begin{figure}[t]
\begin{center}
\includegraphics[width=7.2cm]{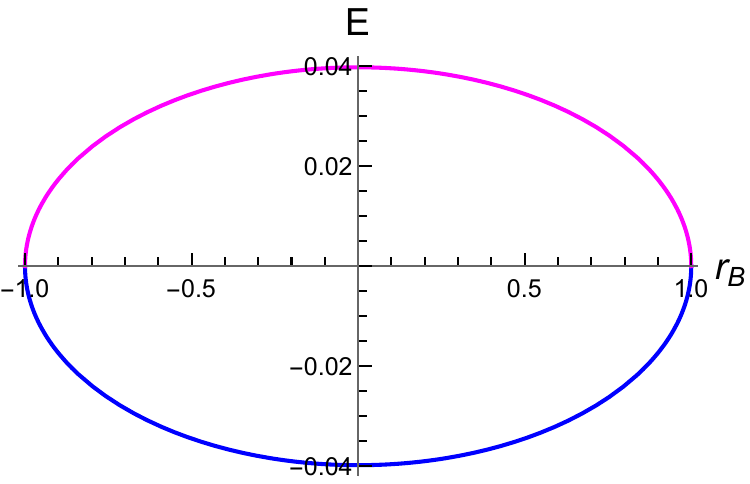}$\qquad\quad$\includegraphics[width=7.2cm]{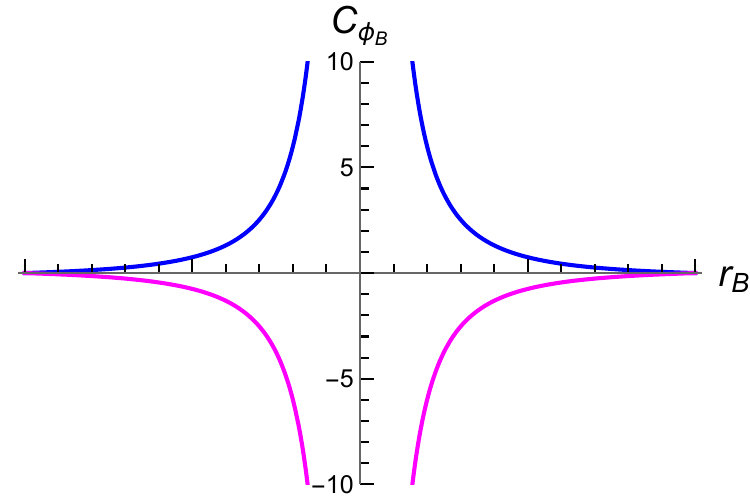}
\end{center}
\caption{Plot of $E$ (left) and $C_{\phi_{B}}$ (right) as a function of radius $r_{B}$ for both cosmological (violet) and black hole (blue) systems. We have set $\phi_{r}=L=G_{2}=1$.}
\label{fig:EandCvsrB} 
\end{figure}

In total, combining the $\phi_0$ and $\phi$ terms \eqref{eq:actionterm1}, \eqref{eq:actionterm2} and \eqref{eq:actionterm3},  the full on-shell Euclidean JT action for the cosmological system is
\begin{equation}
      I_{\text{JT,c}}^{\text{E}}=   - \frac{ \phi_r}{4   G_2 } \frac{r_B^2 }{L^2}   -\frac{\phi_{0}}{4 G_{2} } \;.
\end{equation}
When evaluating at either horizon, we find the total action is minus the cosmological horizon entropy: $I_{\text{JT,c}}^{\text{E}} (r_B= \pm L )= - S_{\text c}$.
From \eqref{eq:actionterm3} we see the on-shell action  can also be expressed as
\begin{equation}
    I_{\text{JT,c}}^{\text{E}} = \beta E_{\text c} - S_{\text c}\,,
\end{equation}
where $\beta$ is the inverse Tolman temperature \eqref{eq:inversetolman}, and the quasi-local energy $E_{\text{c}}$ and  cosmological horizon entropy $S_{\text{c}}$ are
\begin{equation}
    E_{\text c} = \frac{\phi_r}{8 \pi G_2 L} \sqrt{1 - \frac{r_B^2}{L^2}}\,, \qquad \qquad S_{\text c} = \frac{\phi_0 + \phi_r}{4G_2}\,.
\end{equation}
Note that these expressions agree with the quasi-local energy   in \eqref{eq:BYE} and the JT entropy in \eqref{eq:JTentropyfullreduction}. More precisely, these quantities can be obtained from the standard definitions
\begin{equation}
    E = \left ( \frac{\partial   I_{\text{JT}}^{\text{E}}  }{\partial \beta} \right)_{\phi_B}\,,\qquad S = \beta \left ( \frac{\partial   I_{\text{JT}}^{\text{E}}  }{\partial \beta} \right)_{\phi_B} -   I_{\text{JT}}^{\text{E}}\;,
\end{equation}
where instead of the surface area of the boundary, the dilaton $\phi_B$ at $r=r_B$ is kept fixed. See Figure \ref{fig:EandCvsrB} for a plot of $E_{\text c}$ as a function of $r_{B}$ (violet curve).

\begin{figure}[t]
\begin{center}
\includegraphics[width=7.2cm]{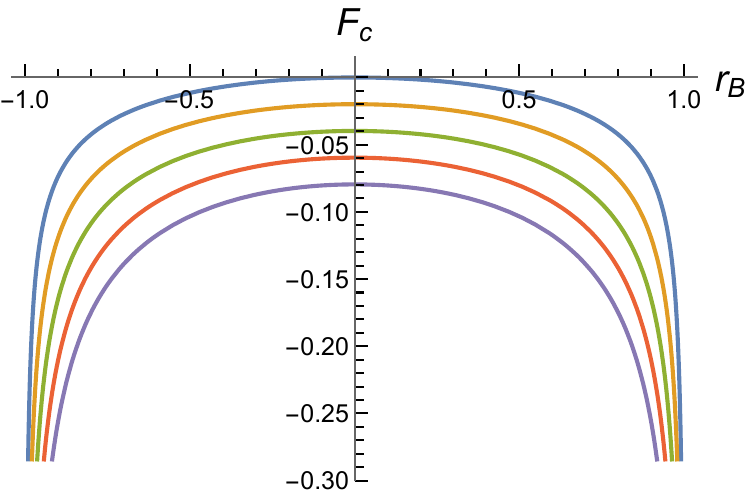}$\qquad\quad$\includegraphics[width=7.2cm]{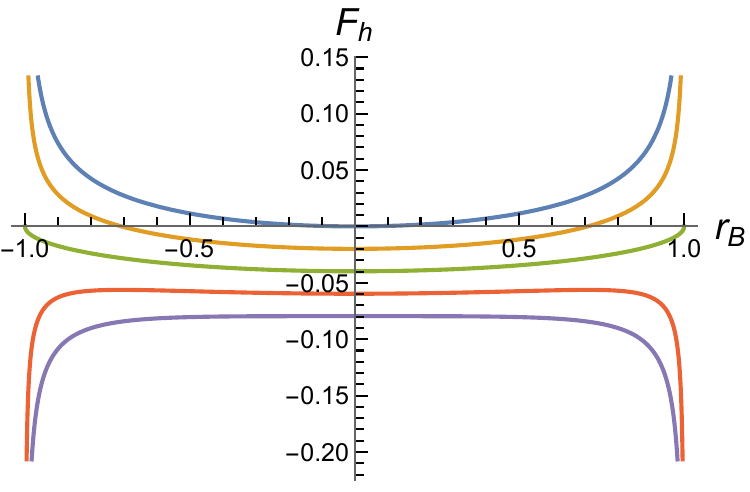}
\end{center}
\caption{Plot of the free energy $F=E-TS$ as a function of radius $r_{B}$ in cosmological (left) and black hole (right) systems, for various values of $\phi_{0}$: $\{0,1/2,1,3/2,2\}=\{\text{blue},\text{orange},\text{green},\text{red},\text{purple}\}$. 
Here $\phi_{r}=L=G_{2}=1$.}
\label{fig:FCandFBHvsrB} 
\end{figure}

The free energy follows directly from the on-shell Euclidean action
\begin{equation}
    F_{\text c} = T I_{\text{JT,c}}^{\text{E}}= E_{\text c} - T S_{\text c} = - \frac{ \phi_r}{8 \pi G_2 L} \frac{r_B^2/L^2}{\sqrt{1 - \frac{r_B^2}{L^2}}} -\frac{\phi_{0}}{8\pi G_{2}L}\frac{1}{\sqrt{1-\frac{r_{B}^{2}}{L^{2}}}}\;.
\label{eq:freeencosmo}\end{equation}
The free energy diverges at the two horizons. Further, recall that $\phi_0 >0$ in the full reduction model and $\phi_0=0$ in the half reduction model, such that the free energy is always less than or equal to zero,  $F_{\text c} \leq 0$. Subtracting the free energy of pure dS$_2$ with a constant dilaton, $F_{0}=- T S_{\phi_0}$, the difference in free energies is always non-positive, $F_{\text c} - F_{0}\leq 0$, implying the nearly dS$_2$ geometry  ($\phi_r \neq 0$) dominates the canonical ensemble over the pure dS$_2$ spacetime ($\phi_r =0$).   
 Since the $F_{\text c}(r_B)$ plot only has a single branch, there is no phase transition for the cosmological system in nearly  dS$_2$ (see Figure \ref{fig:FCandFBHvsrB}).
 
Moreover, the   heat  capacity $C_{\phi_{B},\text{c}}$ for the cosmological system at constant $\phi_B$ is
\begin{equation}
    C_{\phi_B, \text{c}} = T \left ( \frac{\partial S_{\text c}}{\partial T} \right)_{\phi_B} = \left ( \frac{\partial E_{\text c}}{\partial T} \right)_{\phi_B}=  - \frac{\phi_r}{4 G_2} \left ( \frac{L^2}{r_B^2} -1\right) <0\;.
\label{eq:heatcapcosmo}\end{equation}
 The   heat capacity  is  negative everywhere between the two horizons, $r_{\text{h}}=-L$ and $r_{\text c} = L$, and vanishes precisely at the   horizons (see Figure \ref{fig:EandCvsrB}). Hence, the cosmological system  is always  unstable for thermal fluctuations, which seems to be a general  feature of cosmological horizons. For instance,  in higher-dimensional SdS space the cosmological system defined between a boundary at radius $r=r_B$ and the cosmological horizon also has a negative heat capacity.\footnote{This result was thoroughly analyzed in \cite{Banihashemi:2022jys}. See also the recent article \cite{Draper:2022ofa}.} This result appears to be in contradiction with  \cite{Anninos:2017hhn},  where the authors find a positive heat capacity for the cosmological horizon. However, the difference can be attributed to the choice of sign of the dilaton:   they assume the dilaton is negative in dS$_2$, whereas we have taken it to be positive in the vicinity of the cosmological horizon.

\paragraph{Black hole system.} For completeness, consider the black hole system, where $r\in[-L,r_{B}]$,    and $K=- \frac{r_B}{ L^2}/\sqrt{1- \frac{r_B^2}{L^2}} $. The above analysis goes through similarly, where the on-shell Euclidean action is again
\beq I^{\text{E}}_{\text{JT,h}}=\beta E_{\text h}-S_{\text h}\;,\eeq
except now the energy $E_{\text h}$ and entropy $S_{\text h}$  are given by 
\beq E_{\text h}=-\frac{\phi_{r}}{8\pi G_{2}L}\sqrt{1-\frac{r_{B}^{2}}{L^{2}}}\;,\quad S_{\text h}=\frac{\phi_{0}-\phi_{r}}{4G_{2}}\;.\eeq
At the two horizons, the total action is minus the black hole entropy: $I_{\text{JT,c}}^{\text{E}} (r_B= \pm L )= - S_{\text h}$.  
Consequently, the free energy $F_{\text h}$ and heat capacity $C_{\phi_{B},\text{h}}$ are 
\beq F_{\text h}=\frac{ \phi_r}{8 \pi G_2 L} \frac{r_B^2/L^2}{\sqrt{1 - \frac{r_B^2}{L^2}}} -\frac{\phi_{0}}{8\pi G_{2}L}\frac{1}{\sqrt{1-\frac{r_{B}^{2}}{L^{2}}}}\;,
\label{eq:freeenBH}\end{equation}
and
\beq C_{\phi_{B},\text{h}}= \frac{\phi_r}{4 G_2} \left (  \frac{L^2}{r_B^2} -1\right) \label{eq:heatcapBH}>0\eeq
Note that $C_{\phi_{B},\text{h}}$ is positive everywhere between the horizons at $r_{B}=\pm L$, and is zero at the horizons. Thus, the black hole system  is stable with respect to thermal fluctuations (see blue curve on the right side in Figure~\ref{fig:EandCvsrB}). A similar result was obtained for AdS$_2$ black holes in \cite{Lemos:1996bq}, where the system between the black hole horizon and the timelike boundary always has a positive heat capacity.

\begin{figure}[t]
\begin{center}
\includegraphics[width=7cm]{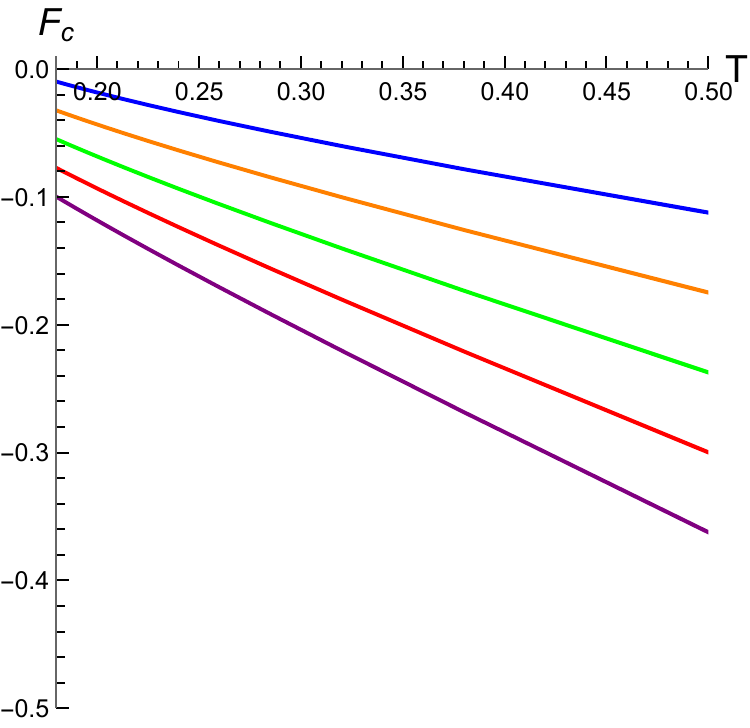}$\qquad\quad$\includegraphics[width=7cm]{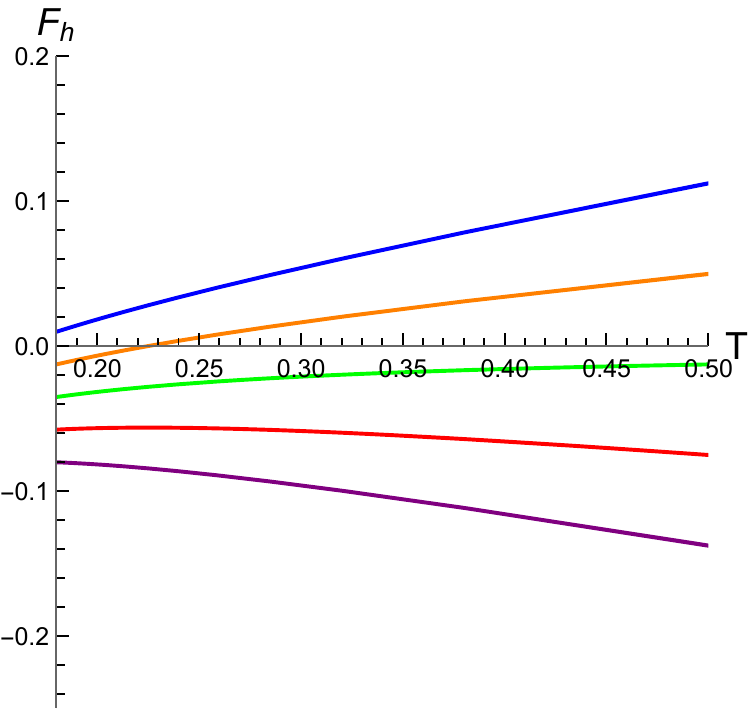}
\end{center}
\caption{Plot of the free energy $F=E-TS$ as a function of the Tolman temperature for the cosmological (left) and black hole (right) systems, for various values of   $\phi_{0}$: $\{0,1/2,1,3/2,2\}=\{\text{blue},\text{orange},\text{green},\text{red},\text{purple}\}$. We have set $\phi_{r}=L=G_{2}=1$.}
\label{fig:FCandFBHvsT} 
\end{figure}

Further, in the half reduction model for $\phi_{0}=0$ we observe that the free energy is non-negative $F_{\text h}\geq 0$ everywhere, approaching positive infinity as one asymptotes to the horizons. In the full reduction model the free energy obeys $F_{\text h}\leq0$ when $\phi_{0}\geq\frac{\phi_{r} r_{B}^{2}}{L^{2}}$. Subtracting the free energy of the pure dS$_2$ solution, $F_{0}=- T S_{\phi_0}$, the difference in free energies is always non-negative, $F_{\text h} - F_{0}\geq 0$, which means the pure dS$_2$ solution  dominates the canonical ensemble for the black hole system. Finally, from Figure  \ref{fig:FCandFBHvsrB} for the $F_{\text h}(r_B)$ plot and Figure \ref{fig:FCandFBHvsT} for the $F_{\text h} (T)$ plot we see there are no phase transitions for the black hole system.    

\subsection{Quasi-local  Euler relation and first law}
\label{sec:quasilocaleulerfirst}

The quasi-local thermodynamic quantities are related to each other by the   Euler   equation (or Smarr formula) and obey a first law, as shown by York   for a Schwarzschild black hole in  \cite{York:1986it}. In this section we derive both of these relations for de Sitter JT gravity using the Noether charge formalism \cite{Wald:1993nt,Iyer:1994ys} (see \cite{Pedraza:2021cvx} or Appendix \ref{app:noetherchargesumm} for a summary). 

\subsubsection*{Quasi-local Euler relation}

In \cite{Liberati:2015xcp,Jacobson:2018ahi,Pedraza:2021cvx} the  Smarr formula  was derived from the following integral identity
 \begin{equation}
     \int_\Sigma j_\xi =  \oint_{\partial \Sigma} Q_\xi= \oint_{\mathcal S} Q_\xi +\oint_{\mathcal H} Q_\xi \;,
 \label{eq:Smarrid}\end{equation}
 where $j_{\xi}$ is the Noether current 1-form associated with the Killing symmetry generated by~$\xi$, and $Q_{\xi}$ is the associated Noether charge $0$-form, obeying the on-shell identity $j_{\xi}=dQ_{\xi}$. In our set-up, $\Sigma$ is a constant-$t$ surface in the static patch, and  its boundary is given by $\partial\Sigma= \mathcal S \cup \mathcal H$, with $\mathcal H$ being the location of the bifurcation point of a   Killing horizon (either the black hole or the cosmological horizon) and $\mathcal S$ is the intersection of $\Sigma$ and the timelike boundary $B$. The first equality in   (\ref{eq:Smarrid}) is an application of Stokes' theorem, and in the second equality   the orientation of the Noether charge integral  at $\mathcal H$ and $\mathcal S$ is taken to be outward. We now compute both sides of the integral  identity explicitly for de Sitter JT gravity.
 
To evaluate the left-hand side we need the definition of the Noether current 1-form $j_{\xi}$
 \begin{equation}
     j_\xi \equiv \theta (\psi, \mathcal L_\xi \psi) - \xi \cdot L_{\text{JT}}\,.
 \end{equation}
The  symplectic potential 1-form of classical JT gravity vanishes when evaluated on the Lie derivative along the Killing vector $\xi$, \emph{i.e.}, $\theta (\psi, \mathcal L_\xi \psi)=0$. The (Lorentzian) JT Lagrangian 2-form $L_{\text{JT}}$ is on shell given by 
 \begin{equation}
     L_{\text{JT}} = \frac{\epsilon}{16 \pi G_2} \left [ (\phi_0 + \phi) R - \frac{2}{L^2}\phi \right]= \frac{\epsilon \phi_0}{8\pi G_2 L^2}\,,
 \end{equation}
 with $\epsilon$ being the spacetime volume form, and    we inserted the dilaton equation of motion $R=2/L^2$ in the last equality. Therefore, the left-hand side of the integral relation is
 \begin{equation}
    \int_\Sigma  j_\xi = -   \frac{\phi_0}{8 \pi G_2 L^2} \int_\Sigma \xi \cdot \epsilon =-\frac{\phi_{0}\Lambda}{8\pi G_{2}}\Theta_{\xi}\, , \label{eq:LHSsmarr} 
 \end{equation}
 Following \cite{Jacobson:2018ahi,Pedraza:2021cvx}, we introduced the ``Killing volume'' $\Theta_{\xi}$, which is defined as the proper volume (length in 2D) of $\Sigma$ locally weighted by the norm of  $\xi$,
 \begin{equation}
     \Theta_\xi = \int_\Sigma |\xi| d \ell\;. \label{eq:killingvolumyeah}
 \end{equation}
  Here we have written $\xi \cdot \epsilon\, |_\Sigma = |\xi| d \ell$, where $d\ell$ is the infinitesimal proper length $d\ell=dr/\sqrt{f(r)}$ and the norm is $|\xi|= \sqrt{f(r)}$. Thus, for the cosmological system  the Killing volume is given by $\Theta_{\xi, \text{c}}=L - r_B$, whereas for the black hole system we have $\Theta_{\xi, \text{h}}= L+r_B.$
 
On the right-hand side of the integral identity (\ref{eq:Smarrid})  we use the expression for the Noether charge in JT gravity, cf. Eq. \eqref{eq:Noethercharge}, 
 \begin{equation}
     Q_\xi^{\text{JT}} =- \frac{1}{16 \pi G_2} \epsilon_{\mu\nu} \left[ (\phi+ \phi_0) \nabla^\mu \xi^\nu + 2 \xi^\mu \nabla^\nu (\phi + \phi_0)\right]\;,
 \label{eq:NoetherclassJT}\end{equation}
 where $\epsilon_{\mu\nu}|_{\partial\Sigma}=(n_{\mu}u_{\nu}-n_{\nu}u_{\mu})$ is the binormal of $\partial\Sigma$, satisfying $\epsilon_{\mu\nu}\epsilon^{\mu \nu}=-2$, and we used that the volume form is  $\epsilon_{\partial \Sigma}=1$ in 2D. At the bifurcation point $\mathcal H$ we have $\xi |_{\mathcal H}=0$ and $\nabla_{\mu}\xi_{\nu} |_{\mathcal H}=-\kappa\epsilon_{\mu\nu}$. Hence,
 \begin{equation}
     \oint_{\mathcal H} Q_\xi=- \frac{\kappa}{8\pi G_2} (\phi_0 + \phi_{\mathcal H})\,,
\label{eq:QxiatH} \end{equation}
 which is equal to minus the  Gibbons--Hawking  temperature $T_{\text{GH}}=\kappa/2\pi$ times the   horizon entropy $S_{\mathcal H}$ (which is the same everywhere on the Killing horizon $H$).\footnote{The minus sign arises here since we have chosen the orientation of the Noether charge integral to be outward  away from the origin (which follows from Stokes' theorem), whereas for black holes the orientation is usually chosen to be  towards spatial infinity, such that $\oint_{\partial \Sigma} Q_\xi = \oint_\infty Q_\xi - \oint_{\mathcal H} Q_\xi$ (see also footnote 8 in \cite{Jacobson:2018ahi}).}
   Meanwhile,  since the boundary $ {\mathcal S}$ is defined as the intersection of $\Sigma$ (with unit normal   $u^\mu=\xi^\mu/N$) and $B$ (with unit normal $n^\mu$) we have  $n\cdot u|_{\mathcal S}=0$ or $n \cdot \xi |_{\mathcal S}=0$, hence the Noether charge at $\mathcal S$ is
 \beq
 \begin{aligned}
      Q_\xi \big |_{\mathcal S} 
      &=- \frac{1}{8 \pi G_2}  \left[ \frac{1}{2}(\phi+ \phi_0)  (-N K  -    n^{\mu}\nabla_\mu  N ) +N n^{\nu}   \nabla_\nu (\phi + \phi_0)\right]\Big  |_{\mathcal S} \\
      &= -\frac{1}{8\pi G_{2}}  \Big[-Nn^\mu a_\mu (\phi_{0}+\phi)+Nn^{\nu}\nabla_{\nu}(\phi_{0}+\phi)  \Big]\Big|_{\mathcal S}\;. 
      \label{eq:QxiatB}
 \end{aligned}
 \eeq
In the first equality we used $- u\cdot \xi = N$ and $- u_\nu \nabla^\mu \xi^\nu = \nabla^\mu N$,  where $ N=|\xi|$ is the norm \eqref{eq:normkilling} of the Killing vector $\xi$. Further, we inserted the     extrinsic curvature of $B$, $K_{\mu \nu} = 2 \nabla_{(\mu} n_{\nu)}$, and the relation $K_{\mu \nu}=K \gamma_{\mu \nu}$ in two dimensions. In the second equality we employed  the trace of the extrinsic curvature $  K=\frac{1}{N}n^\mu \nabla_\mu N =n^\mu a_\mu $, which is equal to the normal component of the acceleration vector $a^{\mu}=u^{\nu}\nabla_{\nu}\xi^{\mu}=\frac{1}{N} \nabla^{\mu} N$.

 Thus, inserting  (\ref{eq:LHSsmarr}), (\ref{eq:QxiatH}) and (\ref{eq:QxiatB}) into the integral identity (\ref{eq:Smarrid}) we arrive to the following   relation
 \begin{equation}
     - \frac{\phi_0 \Lambda}{8 \pi G_2 } \Theta_\xi = - \frac{\kappa}{8\pi G_2} (\phi_0 + \phi_{H}) + \frac{Nn^\mu a_\mu}{8 \pi G_2 } (\phi_0 + \phi_B)  - \frac{N}{8 \pi G_2} n^\mu \nabla_\mu \phi \;.
\label{eq:classELrel} \end{equation}
Our notation above reflects that $\phi_{\mathcal{S}}=\phi(r_{B})=\phi_{B}$, and similarly   $\phi_{\mathcal{H}}=\phi(r_H)=\phi_H$. Dividing by $N$ yields the quasi-local Euler relation
\begin{equation}
     E  = TS_H   - \sigma  (\phi_{0}+\phi_{B}) - \frac{\phi_0 \Lambda}{8 \pi G_2 N} \Theta_\xi\;,
 \label{eq:Smarrclassv2}\end{equation}
  where $E$ is the quasi-local energy \eqref{eq:BYE}, $T$ is the Tolman temperature (\ref{eq:tolmantemp}), $S_H$ is the horizon entropy \eqref{eq:JTentropyfullreduction}, and we introduced the ``surface pressure'' $\sigma$
\begin{equation}
    \sigma  = \frac{n^\mu a_\mu}{8\pi G_{2}}= \pm  \frac{1}{8\pi G_2} \frac{r_B/L^2}{ \sqrt{1-\frac{r_B^2}{L^2}}} \,.
\label{eq:linpress}\end{equation}
The plus sign applies to the cosmological system, while the minus sign is associated to the black hole system. We emphasize that the quasi-local Euler relation  holds for both thermodynamic systems. We can compare the definition of the surface pressure in JT gravity to the standard definition of surface pressure in $d$-dimensional Einstein gravity: $\sigma =\frac{1}{(d-2)8\pi G_d} (-k^{\alpha \beta} + (n^\mu a_\mu +k )\sigma^{\alpha \beta}) \sigma_{\alpha \beta}$, where $\sigma_{\alpha \beta}$ is the induced metric on $\mathcal S$ \cite{Brown:1992br}. Setting the extrinsic curvature $k^{\alpha \beta}$ of the codimension-two surface  $\mathcal S$ to zero, since the surface is just a point in 2D, and using $\sigma^{\alpha \beta}\sigma_{\alpha \beta}=d-2$, we recover the definition in \eqref{eq:linpress}.  

Notice that in the half reduction model the Killing volume term in the Euler relation \eqref{eq:Smarrclassv2} vanishes since $\phi_{0}=0$. In fact, the Euler relation splits into two separate equations
  \begin{equation}
     E  = TS_{\phi_H}   - \sigma   \phi_{B}  \;, \qquad 0 = T S_{\phi_0} - \sigma \phi_0 - \frac{\phi_0 \Lambda}{8 \pi G_2 N} \Theta_\xi\,. \label{eq:eulerblabha}
 \end{equation}
 Hence, in the half reduction model the second equation is trivial and the Euler relation reduces to the first expression. The first equation can be interpreted as the Euler relation for nearly dS$_2$ with $\phi_0=0$ but   $\phi_r \neq 0$, while the second equation is the Euler relation for dS$_2$ with a constant dilaton $\phi_0 \neq 0$ but $\phi_r =0$. 
  The  two relations in \eqref{eq:eulerblabha} can be verified explicitly using the expressions for the thermodynamic variables in static patch coordinates.  
  
In the limit that the thermodynamic systems become the full static patch, \emph{i.e.}, $r_B \to \pm L$ for the respective systems,\footnote{For example, suppose the cosmological system is under consideration, such that $\phi_{H}=\phi_{\text c}$. Then  in the limit the system becomes the full patch we have $r_{B}\to-L$, such that $\frac{1}{4G_{2}}\phi_{B}\to-\frac{1}{4G_2}\phi_{r}=S_{\text h}$.} the product $N n^\mu a_\mu$ in Eq. \eqref{eq:classELrel}
is equal to minus the surface gravity
  \begin{equation}
      N n^\mu a_\mu \to - \kappa \qquad \text{as} \qquad r_B \to r_{\text{h,c}} \,. \label{eq:limitNsig}\end{equation}
This   is 
equivalent to the standard definition of surface gravity $\kappa = \lim_H (Na)$ where the magnitude of the acceleration is defined as $ a =\sqrt{a^\mu a_\mu} $. 
Further, we have $N\to 0$,   $E\to 0$,  and   $\frac{1}{4G_2}(\phi_0 + \phi_B)\to S_{\text{h,c}}$ as $r_B \to r_{\text{h,c}}.$  Hence, if we take the limit of \eqref{eq:classELrel} to the full static patch, then the Euler relation becomes  
  \begin{equation}
       0 = T_{\text{GH}} S_{\text{h}} +T_{\text{GH}} S_{\text{c}} - \frac{\phi_0 \Lambda}{8 \pi G_2 } \Theta_\xi\,, \label{eq:smarrfullstaticpatchh}
  \end{equation}
  where the Killing volume \eqref{eq:killingvolumyeah} is now defined between the cosmological and black hole horizon, $\Theta_\xi = \int_{-L}^L dr =2L$. 
Equivalently, for the full static patch the two separate equations in \eqref{eq:eulerblabha} become 
\begin{equation}
    0 = T_{\text{GH}} S_{\phi_{\text h}} + T_{\text{GH}} S_{\phi_{\text c}} \, , \quad \ \quad 0 = 2 T_{\text{GH}} S_{\phi_0} - \frac{\phi_0 \Lambda}{8 \pi G_2 } \Theta_\xi\,.
\end{equation}
We  anticipated the first relation in Eq.  \eqref{eq:smarrfirstlaw2dfullred}, and the second equation is the Euler relation for the dimensionally reduced extremal Nariai solution.

 \subsubsection*{Quasi-local first law}
The quasi-local first law for both the cosmological and black hole system  is
\begin{equation}
    dE = T dS_H - \sigma d \phi_B\,. \label{eq:quasilocalfirstlawyeah}
\end{equation}
This follows from the coordinate expressions for the relevant thermodynamic quantities. In particular, it can be checked that the Tolman temperature (\ref{eq:tolmantemp}) and surface pressure \eqref{eq:linpress} satisfy
\begin{equation}
    T  = \left ( \frac{\partial E}{\partial S_H} \right)_{\phi_B}\, , \qquad \sigma = - \left ( \frac{\partial E}{\partial \phi_B} \right)_{S_H}\,. \label{eq:derivativesdeftands}
\end{equation}
Ultimately, these relations contain the same content as the quasi-local first law \eqref{eq:quasilocalfirstlawyeah}. Further, we point out that the first law follows from the dimensional reduction of the quasi-local Euler relation for Schwarzschild-de Sitter,  $dE = T dS_H - \sigma dA$, because the area of $\mathcal S$ becomes equal to $\phi_B$ after a spherical reduction.

In addition, although the relations \eqref{eq:derivativesdeftands} can be checked in terms of   static patch coordinates,  a covariant derivation of the quasi-local first law is desired.  In fact, the first law follows also from varying the Smarr relation (\ref{eq:Smarrid}), which leads to the fundamental variational integral identity  \cite{Wald:1993nt,Iyer:1994ys} 
\beq \int_{\Sigma}\omega(\psi,\delta\psi,\mathcal{L}_{\xi}\psi)=\oint_{\partial\Sigma}[\delta Q_{\xi}-\xi\cdot \theta(\psi,\delta\psi)]\;,\label{eq:varSmarr}\eeq
where $\omega(\psi,\delta_{1}\psi,\delta_{2}\psi)\equiv\delta_{1}\theta(\psi,\delta_{2}\psi)-\delta_{2}\theta(\psi,\delta_{1}\psi)$ is the symplectic current 1-form, cf. Eq. \eqref{eq:sympcurrentgen} for an explicit expression in dilaton gravity. Since 
$\mathcal L_\xi \psi =0$, and the symplectic current is linear in $\mathcal L_\xi \psi$, the left-hand side of \eqref{eq:varSmarr} is zero. 
The right-hand side of \eqref{eq:varSmarr} splits into an integral at the bifurcation point $\mathcal{H}$ of the Killing horizon  (either the black hole
or the cosmological horizon)
\begin{equation}
    \oint_\mathcal{H} [\delta Q_\xi - \xi \cdot \theta (\psi, \delta \psi)] =- \frac{\kappa}{8\pi G_2} \delta \phi_H\,,
\label{eq:varyQH}\end{equation}
where we used $\xi |_\mathcal{H} =0$ and Eq. \eqref{eq:QxiatH}, 
and an integral at the intersection point $\mathcal{S}=\Sigma \cap B $. The latter can  be computed by evaluating the variation of the Noether charge    \eqref{eq:QxiatB}
\begin{equation}
     \delta Q_\xi  |_\mathcal{S}= \frac{1}{8\pi G_2} \delta \Big ( N n^\mu a_\mu (\phi_0 + \phi_B)   \Big ) +  \delta ( E N  )\,, \label{eq:varQatBB}
\end{equation}
and the symplectic potential at $B$, cf. Eq. (C.6) of \cite{Pedraza:2021cvx} (ignoring the $dC$ contribution since $\xi \cdot dC = \mathcal L_\xi C - d (\xi \cdot C)$ vanishes after integration over $\mathcal S$),
\begin{equation}\label{eq:thetaatBB}
\begin{aligned}
   \theta (\psi, \delta \psi) |_{B} &= \frac{\epsilon_B}{8\pi G_2} \Big (     K \delta \phi_B  + \frac{1}{2} \gamma^{\mu \nu} n^\alpha \nabla_\alpha \phi \delta \gamma_{\mu \nu} \Big )  -\frac{1}{8\pi G_2}  \delta \Big  ( \epsilon_B K (\phi_0 + \phi_B) \Big )   \\
   &=- \epsilon_B \frac{E}{N} \delta N   - \frac{\epsilon_B}{8\pi G_2}
   \left [          (\phi_0 + \phi_B) \delta    ( n^\mu a_\mu   )    + \frac{1}{N}n^\mu a_\mu (\phi_0 + \phi_B) \delta N \right]  .
    \end{aligned}
\end{equation}
Here, we inserted $K = n^\mu a_\mu $ , $\frac{1}{2} \gamma^{\mu \nu} \delta \gamma_{\mu \nu} = \delta N/N $, $ E = - \frac{1}{8\pi G_{2}} n^\alpha \nabla_\alpha \phi$, and $\delta\epsilon_{B}=(\frac{1}{2}\gamma^{\mu\nu}\delta\gamma_{\mu\nu})\epsilon_{B}$ in the second equality. Using $\xi \cdot \epsilon_B = - N$   
and combining \eqref{eq:varQatBB} and  \eqref{eq:thetaatBB} we find
\begin{equation}
 \oint_{\mathcal S} \left [ \delta Q_\xi - \xi \cdot \theta (\psi, \delta \psi)\right]   =  N \delta E + \frac{N}{8\pi G_2} n^\mu a_\mu \delta \phi_B   \;.
\label{eq:RHSfirstlaw}\end{equation}
Substituting   (\ref{eq:varyQH}), and (\ref{eq:RHSfirstlaw}) into (\ref{eq:varSmarr}) yields the covariant relation
\beq \delta E=\frac{\kappa}{8\pi G_2 N}\delta \phi_{H}-\sigma\delta\phi_{B}\;,\label{eq:quasilocalclasscov}\eeq
recovering the quasi-local first law (\ref{eq:quasilocalfirstlawyeah}).

Multiplying the quasi-local first law by the norm $N$ and taking the limit $r_{B}\to\pm L$, such that the thermodynamic system  becomes the full static patch, we find
\beq 0=T_{\text{GH}}\delta S_{\text{c}}+T_{\text{GH}}\delta S_{\text{h}}\;,\label{eq:global1stlawfullred}\eeq
since $NT=T_{\text{GH}}$, and $N\delta E\to0$ and $N \sigma \to - \frac{1}{4G_2} T_{\text{GH}}$  in this limit.  We recognize this as the 2D  analog of the global first law 
\eqref{eq:smarrfirstsds} for Schwarzschild-de Sitter black holes. Note that this global first law only holds in the full reduction model of JT gravity. In contrast, in the half reduction model, when we take the limit where the thermodynamic system becomes the full patch, such that $r_{B}\to0$ and $N\to1$, the energy contribution is non-vanishing while the surface pressure \eqref{eq:linpress} tends to zero. Consequently, in the half reduction model we attain the following ``global'' first law
\beq \delta E=T_{\text{GH}}\delta S_{H}\;,\label{eq:global1stlawhalfredffff}\eeq
valid for either the cosmological or black hole system. In this case, for $r_B=0$ the energy is $E =\pm\frac{\phi_r}{8 \pi G_{2} L} =T_{\text{GH}}S_{H}$. 

The covariant derivation of the first law (\ref{eq:quasilocalclasscov}) may be generalized by including classical matter contributions,  where   the variation of the matter Hamiltonian $H_{\xi}$ is characterized by the matter energy-momentum tensor $T_{\mu\nu} $ and can be cast as $\delta H_{\xi}=-\int_{\Sigma}\delta( {T_{\mu}}^\nu)  \xi^{\mu}\epsilon_{\nu}$ \cite{Jacobson:2018ahi}. The quasi-local first law with a matter Hamiltonian variation reads
 \beq \delta E=\frac{\kappa}{8\pi G_2 N}\delta \phi_{H}-\sigma\delta\phi_{B} + \delta H_\xi\;,  \eeq
In the limit $r_{B}\to\pm L$, the global first laws (\ref{eq:global1stlawfullred}) and (\ref{eq:global1stlawhalfredffff}) are appropriately modified.

 \subsection{Including semi-classical backreaction} \label{subsec:scbackreact}



A notable feature of JT gravity is that the effects of backreaction are  fixed by the two-dimensional Polyakov action capturing the contributions of the conformal anomaly \cite{Polyakov:1981rd}, in the semi-classical limit. 
 Here we solve the problem of backreaction in de Sitter JT gravity and derive the semi-classical extension of the first law. In particular, we will find that the classical entropy is replaced by the semi-classical Wald entropy, which is equal to the generalized  entropy, as we will discuss. Our treatment here largely follows the recent work \cite{Pedraza:2021cvx}.
 
 


 \subsubsection{Vacuum states and generalized entropy}
 \label{sec:quantummatter}
 Semi-classical JT gravity in de Sitter space is described by minimally coupling the classical JT action (found from the full reduction) (\ref{eq:JTactv2}) to a dynamical two-dimensional conformal field theory $I_{\text{CFT}}$ of central charge $c$. Adding $I_{\text{CFT}}$ makes the semi-classical model an effective theory; unlike the classical action, the 2D CFT action does not follow from a dimensional reduction. Including $I_{\text{CFT}}$ modifies the classical equations of motion (\ref{eq:graveom}) by semi-classical effects
 \beq \frac{1}{8\pi G_{2}}\left(g_{\mu\nu}\Box-\nabla_{\mu}\nabla_{\nu}+\frac{1}{L^{2}}g_{\mu\nu}\right)\phi=\langle T_{\mu\nu}^{\text{CFT}}\rangle\;, \label{eqombiglaugh}\eeq
 where $\langle T_{\mu\nu}^{\text{CFT}}\rangle\equiv-\frac{2}{\sqrt{-g}}\frac{\delta I_{\text{CFT}}}{\delta g^{\mu\nu}}$ is the expectation value of the stress-energy tensor $T_{\mu\nu}^{\text{CFT}}$ with respect to some unspecified quantum state $|\Psi\rangle$.

 The conformal matter thus backreacts on the classical solution. To study the problem of backreaction consistently, we work in the large-$c$ limit\footnote{Since we have also maintained Newton's constant $G_{2}$, the proper semi-classical limit is $G_{2}\to0$, $c\to\infty$ while keeping $cG_{2}$ fixed, where $c\gg1$ keeps the 1-loop corrections to the dilaton suppressed compared to the CFT. Dimensional reduction tells us our semi-classical approximation is only valid in the regime $\phi_{0}/G_{2}\gg \phi_r/G_{2}\gg c\gg1$ \cite{Pedraza:2021cvx}. } such that $I_{\text{CFT}}$ is given by the non-local 1-loop Polyakov action $I_{\text{Poly}}$ \cite{Polyakov:1981rd}.
 This 1-loop action 
 can be put into a localized form by introducing a massless auxiliary scalar field $\chi$, modelling the 2D CFT, such that
 \beq I_{\text{Poly}}=-\frac{c}{24\pi}\int_{\mathcal{M}}\hspace{-2mm}d^{2}x\sqrt{-g}\left[(\nabla\chi)^{2}+\chi R\right]-\frac{c}{12\pi}\int_{\partial\mathcal{M}}\hspace{-3mm}dt\sqrt{-h}\chi K\;.\label{eq:Polyghyloc}\eeq
 The boundary contribution we have included is a GHY term such that the localized 1-loop action has a well-posed variational problem. The equation of motion for $\chi$ is 
 \beq 2\Box\chi=R\;,\label{eq:chieom}\eeq
 whose formal solution $\chi=\frac{1}{2}\int d^{2}y\sqrt{-g(y)}G(x,y)R(y)$ puts the local action (\ref{eq:Polyghyloc}) into its original non-local form.
 From the action (\ref{eq:Polyghyloc}), the semi-classical gravitational field equations are given by \eqref{eqombiglaugh} where $\langle T_{\mu\nu}^{\text{CFT}}\rangle$ is now replaced by  
 \beq \langle T_{\mu\nu}^{\chi}\rangle=\frac{c}{12\pi}\left[(g_{\mu\nu}\Box-\nabla_{\mu}\nabla_{\nu})\chi+(\nabla_{\mu}\chi)(\nabla_{\nu}\chi)-\frac{1}{2}g_{\mu\nu}(\nabla\chi)^{2}\right]\;.\label{eq:Tchi}\eeq
 Using the equation of motion for $\chi$ (\ref{eq:chieom}), it is easy to show $\langle T^{\chi}_{\mu\nu}\rangle$ has the well-known conformal anomaly 
 \beq g^{\mu\nu}\langle T^{\chi}_{\mu\nu}\rangle=\frac{c}{24\pi}R\;.\label{eq:anom}\eeq
 In \cite{Christensen:1977jc} it was recognized that in two dimensions the conformal anomaly captures all 1-loop quantum effects and the full backreaction.

 Crucially, since the Polyakov action (\ref{eq:Polyghyloc}) does not directly couple to the dilaton $\phi$, it does not alter the dilaton equation of motion (\ref{eq:dilaeom}) and the background geometry remains exact $\text{dS}_{2}$. The classical solution of $\phi$ will be modified due to backreaction, but in the case of interest, $\phi$ will only be shifted by a constant proportional to $cG_{2}$, as in the $\text{AdS}_{2}$ model. 
 
 To proceed with the semi-classical analysis, we must specify the vacuum state of the quantum matter. We accomplish this as follows. First, we work in the conformal gauge, $d\ell^2=-e^{2\rho(y^{+},y^{-})}dy^{+}dy^{-}$,
 where $(y^{+},y^{-})$ are some null conformal coordinates. One finds the solution for the auxiliary field $\chi$ to be 
 \beq \chi=-\rho+\xi(y^{+},y^{-})\;,\quad \text{with} \quad \Box\xi=0 \quad \Rightarrow \quad \xi(y^{+},y^{-})=\xi_{+}(y^{+})+\xi_{-}(y^{-})\;.\label{eq:chigensol}\eeq
 The  $\xi_{\pm}(y^{\pm})$ constitute functions $t_{\pm}(y^{\pm})$,
 \beq t_{+}(y^{+})=\partial^{2}_{+}\xi_{+}-(\partial_{+}\xi_{+})^{2}\;,\quad t_{-}(y^{-})=\partial_{-}^{2}\xi_{-}-(\partial_{-}\xi_{-})^{2}\;,\label{eq:tpmfluxes}\eeq
 which characterize the normal-ordered stress-tensor 
 \beq \langle\Psi|:T_{\pm\pm}^{\chi}(y^\pm):|\Psi\rangle=-\frac{c}{12\pi}t_{\pm}(y^{\pm})\;.\eeq
 From the definition of normal ordering, $:T_{\pm\pm}^{\chi}:\equiv T^{\chi}_{\pm\pm}(y^{\pm})-\langle0_{y}|T^{\chi}_{\pm\pm}(y^{\pm})|0_{y}\rangle$, the vacuum state $|0_{y}\rangle$ with respect to the positive frequency modes in coordinates $y^{\pm}$ is the state obeying
 \beq \langle 0_{y}|:T^{\chi}_{\pm\pm}(y^{\pm}):|0_{y}\rangle=0\;\;\Leftrightarrow\;\; t_{\pm}(y^{\pm})=0\;.\label{eq:expTtpm}\eeq
Moreover, the transformation properties of $\rho$ and $\xi$ reveal that the normal-ordered stress tensor obeys an anomalous transformation law under a conformal transformation $y^{\pm}\to x^{\pm}(y^{\pm})$, such that
 \beq \langle 0_{y}|:T^{\chi}_{\pm\pm}(x^\pm):|0_{y}\rangle=-\frac{c}{24\pi}\{y^{\pm},x^{\pm}\}\;,\label{eq:anomtransnormst} \eeq
 and 
 \beq \langle 0_{y}|T_{\pm\pm}^{\chi}(x^{\pm})|0_{y}\rangle=\langle 0_{x}|T_{\pm\pm}^{\chi}(x^{\pm})|0_{x}\rangle-\frac{c}{24\pi}\{y^{\pm},x^{\pm}\}\;.\label{eq:vactrans}\eeq
 Here $|0_{x}\rangle$ is the vacuum defined with respect to the positive frequency modes in the $x^{\pm}$ coordinate system, and $\{y^{\pm},x^{\pm}\}$ denotes the Schwarzian derivative
 \beq \{y^{\pm},x^{\pm}\}\equiv\frac{(y^{\pm})'''}{(y^{\pm})'}-\frac{3}{2}\left(\frac{(y^{\pm})''}{(y^{\pm})'}\right)^{2}\;,\quad (y^{\pm})'\equiv\frac{dy^{\pm}}{dx^{\pm}}\;.\eeq
 The central lesson of \eqref{eq:anomtransnormst} is that observers in different coordinates will experience the same vacuum differently.

 \subsection*{Static and Bunch--Davies vacua}
 
 Let us now be more explicit and consider two vacuum states of interest, the static (S) and Bunch--Davies (BD) vacua. We do this by choosing two conformally related sets of null coordinates. The \emph{static vacuum} $|\text{S}\rangle$ characterizes the state of a static observer confined to the static patch of $\text{dS}_{2}$, \emph{i.e.}, $(v,u)$ coordinates (\ref{eq:statnullmet}). The static vacuum is conformally related to the Rindler vacuum in Minkowski space, and   analogous to the Boulware vacuum of a  black hole. The \emph{Bunch--Davies vacuum} $|\text{BD}\rangle$ characterizes the state of an observer in Kruskal-like coordinates $(V,U)$, given in (\ref{eq:kruskalmet}). It is conformally related to the global Minkowski vacuum state, and analogous to the Hartle-Hawking state of a black hole in thermal equilibrium with its Hawking radiation.

 By the normal-ordered relation (\ref{eq:expTtpm}), we have
 \beq \langle \text{S}|:T_{vv}^{\chi}:|\text{S}\rangle=\langle \text{S}|:T_{uu}^{\chi}:|\text{S}\rangle=0\quad \Leftrightarrow\quad t_{v}(v)=t_{u}(u)=0\;,\eeq
 and 
 \beq \langle \text{BD}|:T_{VV}^{\chi}:|\text{BD}\rangle=\langle \text{BD}|:T_{UU}^{\chi}:|\text{BD}\rangle=0\quad \Leftrightarrow\quad t_{V}(V)=t_{U}(U)=0\;.\eeq
 Moreover, it is straightforward to show that for the static vacuum,
 \beq \langle \text{S}|:T^{\chi}_{VV}:|\text{S}\rangle=\langle \text{S}|T^{\chi}_{VV}|\text{S}\rangle=-\frac{c}{48\pi V^{2}}\;,\quad \langle \text{S}|:T^{\chi}_{UU}:|\text{S}\rangle=\langle \text{S}|T^{\chi}_{\text{UU}}|\text{S}\rangle=-\frac{c}{48\pi U^{2}}\;,\label{eq:statvacUUVV}\eeq
 \beq \langle \text{S}|T^{\chi}_{uu}|\text{S}\rangle=\langle \text{S}|T^{\chi}_{vv}|\text{S}\rangle=-\frac{c}{48\pi L^{2}}\;,\label{eq:statvacvvuu}\eeq
and for the Bunch--Davies vacuum,
 \beq \langle \text{BD}|:T^{\chi}_{uu}:|\text{BD}\rangle=\langle \text{BD}|:T^{\chi}_{vv}:|\text{BD}\rangle=\frac{c\pi}{12}T^{2}_{\text{GH}}\;,\label{eq:BDvacuuvv}\eeq
 \beq \langle \text{BD}|T^{\chi}_{VV}|\text{BD}\rangle=\langle \text{BD}|T^{\chi}_{UU}|\text{BD}\rangle=\langle \text{BD}|T^{\chi}_{uu}|\text{BD}\rangle=\langle \text{BD}|T^{\chi}_{vv}|\text{BD}\rangle=0\;.\eeq
 To summarize, from (\ref{eq:statvacUUVV}) we see the static vacuum state expectation value of the renormalized stress-tensor in Kruskal coordinates  becomes singular on the past and future cosmological horizons ($V=0$ and $U=0$). In static null coordinates $(v,u)$, we find a negative energy density, (\ref{eq:statvacvvuu}). This behavior is analogous to the Casimir energy of the Boulware state in an eternal black hole background. Alternatively, an observer in the static patch will see the Bunch--Davies vacuum   as a thermal state at the Gibbons--Hawking temperature, cf. (\ref{eq:BDvacuuvv}). More precisely, a static observer detects a left and right flux of particles at the same temperature $T_{\text{GH}}$, such that the static patch of $\text{dS}_{2}$ is a thermal system at temperature $T_{\text{GH}}$, and $|\text{BD}\rangle$ restricted to the static patch is a thermal equilibrium state. 
 Indeed, the   Bunch--Davies state can be written as a thermofield double state with respect to energy eigenstates $|E_{i}\rangle_{L,R}$ characterizing left and right static patches 
 \beq 
 | \text{BD} \rangle = \frac{1}{Z} \sum_i e^{-\beta E_i/2} |E_{i}\rangle_{L} |E_{i}\rangle_{R}\,.
 \eeq
 Tracing out the degrees of freedom of, say, the left static patch, the reduced density matrix is a thermal Gibbs state
 \beq \rho_{\text{BD}}^{R}=\text{tr}_{L}\rho_{\text{BD}}=\frac{1}{Z}\sum_{i}e^{-\beta E_{i}}|E_{i}\rangle_{R}\langle E_{i}|_R\;.\label{eq:GibbsBDstate}\eeq
 In the following we will only  work with the Bunch--Davies vacuum state precisely because of its thermal nature. 
 
 \subsection*{Wald entropy is generalized entropy}
 
Since it has a temperature, it is natural to assign a thermodynamic entropy to the cosmological horizon. Given the semi-classical JT action, we do this by following the Noether charge method and computing the Wald entropy \cite{Wald:1993nt}, including quantum backreaction. One finds
 \beq S_{\text{Wald}}=\frac{1}{4G_{2}}(\phi_{0}+\phi_H) -\frac{c}{6}\chi_H  \;,\label{eq:SWalddS2}\eeq
where the backreacted solutions for $\phi$ and $\chi$ are evaluated on the horizon. With respect to the Bunch--Davies vacuum, a state in thermal equilibrium, we justifiably interpret the Wald entropy as a thermodynamic entropy.\footnote{The static vacuum, found by taking the $\beta\to\infty$ limit of the   reduced state (\ref{eq:GibbsBDstate}), is a factorized pure state and is not thermal, such that the Wald entropy with respect to the backreacted solutions found in the static vacuum is not a thermal entropy.}    
 
 The first term in the Wald entropy (\ref{eq:SWalddS2}) is the usual ``area'' law in the Gibbons--Hawking entropy formula for de Sitter space. Assuming the conformal matter is in the BD vacuum, the backreacted $\phi$ is the classical solution shifted by an unimportant constant proportional to $cG_{2}$. The second term is purely due to the 1-loop Polyakov action, entirely encoding the entropy due to the CFT represented by $\chi$. In fact, as recently argued in \cite{Pedraza:2021cvx}, this second term is \emph{exactly} equal to the von Neumann entropy $S_{\text{vN}}$ of a 2D CFT restricted to a single interval $[(y^{+}_{1},y^{-}_{1}),(y^{+}_{2},y^{-}_{2})]$ on a two-dimensional background $d\ell^2=-e^{2\rho(y^{+},y^{-})}dy^{+}dy^{-}$ (cf. \cite{Fiola:1994ir})
 \beq S_{\text{vN}}=\frac{c}{6}\log\left[\frac{1}{\delta_{1}\delta_{2}}(y^{+}_{2}-y^{+}_{1})(y^{-}_{2}-y^{-}_{1})e^{\rho(y^{+}_{1},y^{-}_{1})}e^{\rho(y^{+}_{2},y^{-}_{2})}\right]\;.\label{eq:vnentgen}\eeq
 Here $\delta_{1,2}$ are independent UV regulators which resolve divergences arising from evaluating $S_{\text{vN}}$ at the endpoints of the interval. With respect to the Bunch--Davies vacuum, we can show explicitly that the general backreacted solution for $\chi$ in global coordinates $(V,U)$ (\ref{eq:kruskalmet}) takes precisely the form of $S_{\text{vN}}$   
 \beq \label{eq:vnentchikrusk}
 \begin{aligned}
  -\frac{c}{6}\chi &=\frac{c}{12}\log\left[\frac{16}{\left(1-\frac{U_{1}V_{1}}{L^{2}}\right)^{2}\left(1-\frac{U_{2}V_{2}}{L^{2}}\right)^{2}}\right]+\frac{c}{12}\log\left[\frac{1}{\delta_{1}^{2}\delta_{2}^{2}}(U_{2}-U_{1})^{2}(V_{2}-V_{1})^{2}\right]\\
  &=S_{\text{vN}}^{\text{BD}}\;.
 \end{aligned}
 \eeq
 Here the CFT is restricted to the interval $[(U_{1},V_{1}),(U_{2},V_{2})]$. The entropy is generically time-dependent, despite the spacetime being static. This form of the von Neumann entropy is for a single interval inside the shaded regions in Figure \ref{fig:dS2fullred}; it does not give the entropy for an interval with endpoints in different hyperbolic patches, as could be the case for the full reduction model. This scenario is dealt with by performing the continuation (\ref{eq:continuedcoordUVapp}) on one of the endpoints. 
 
 While we explicitly computed (\ref{eq:vnentchikrusk}), the result $S_{\text{vN}}=-\frac{c}{6}\chi$ holds for any 2D gravity theory coupled to a large $c$ CFT and with respect to any vacuum state \cite{Pedraza:2021cvx}. This primarily follows from the fact that generically $\chi=-\rho+\xi$, as in (\ref{eq:chigensol}), and imposing $\chi$ obeys Dirichlet boundary conditions.\footnote{Technically, $\chi$ formally diverges logarithmically  at the location where the Dirichlet boundary condition is imposed. The divergence is regularized via a cutoff, such that $\chi$ is equal to a constant which we set to zero.\label{fn:dirichletbc}}
 Thus, the general solution for $\chi$ is proportional to the von Neumann entropy (\ref{eq:vnentgen}) of a 2D CFT in vacuum reduced to a single interval in a curved background. The semi-classical Wald entropy (\ref{eq:SWalddS2}), then, is exactly equal to the generalized entropy\footnote{Note the von Neumann entropy depends on a cutoff $\delta$ such that, via (\ref{eq:SWaldSgen}), the generalized entropy depends on a UV cutoff. However, $S_{\text{gen}}$ is expected to be a UV finite quantity \cite{Susskind:1994sm}, independent of the cutoff (the regularized terms in the gravitational and matter sectors cancel). Likewise, the $S_{\text{gen}}$ here can be made UV finite by introducing a renormalized Newton's constant $G_{2}$, as done in higher dimensions, \emph{e.g.}, \cite{Jacobson:1994iw}. We thank Ted Jacobson for emphasizing this point.}  
 \beq S_{\text{Wald}}=S_{\text{gen}}\;.\label{eq:SWaldSgen}\eeq
 Relating $S_{\text{Wald}}$ to $S_{\text{gen}}$ was previously hinted at but not realized in \cite{Myers:1994sg} in the case of 2D flat space; the observation (\ref{eq:vnentchikrusk}) has seemingly only been recognized in \cite{Pedraza:2021cvx}. It is worth pointing out that normally the Wald entropy represents only the gravitational contribution to the generalized entropy, while the matter entropy is solely due to the von Neumann entropy of the quantum fields living on the background. In the context of two-dimensional gravity, however, the entire effect of conformal matter living on the background is encoded in the 1-loop Polyakov action, for which we may apply the Wald formalism to compute the entropy. We do not expect this observation to be true in higher dimensions as the trace anomaly does not provide complete information of the matter fields.

 \subsection*{A comment on CFTs in the half vs.\ full reduction models}
 
 As emphasized in Section \ref{sec:red2DdS}, the half reduction model of de Sitter JT gravity leads to a $\text{dS}_{2}$ geometry that is restricted, due to the fact that the dilaton $\Phi\geq0$. No such restriction occurs in the full reduction model. Here the quantum matter is described by a two-dimensional CFT in both versions of JT gravity. This is not a natural viewpoint for the half reduction because the CFT does not see the full space: the half reduction effectively restricts the CFT from a cylinder to the half plane. Therefore, it is more natural to describe the quantum matter as a CFT in the full reduction model, while in the half reduction one should probably  consider a boundary CFT.

 \subsubsection{Semi-classical thermodynamics}
 
 Briefly, let us now derive the semi-classical extension of the quasi-local Euler relation and first law of thermodynamics in dS$_2$. We again use Noether charge techniques, following~\cite{Pedraza:2021cvx}. In principle, we could have performed an on-shell Euclidean action analysis when $\chi$ is static, as we did in the classical case. However, for a time-dependent $\chi$   this approach is conceptually and computationally challenging, and a covariant analysis is desired. 
 
 \subsubsection*{Semi-classical quasi-local Euler relation}
 
 We use the integral identity (\ref{eq:Smarrid}), where now we include the Noether charge and current associated with the 1-loop Polyakov action.  We begin with the right-hand side of (\ref{eq:Smarrid}).  The Noether charge $Q_{\xi}^{\chi}$ for the auxiliary field $\chi$ is 
 \beq Q_{\xi}^{\chi}=\frac{c}{24\pi}\epsilon_{\mu\nu}[\chi\nabla^{\mu}\xi^{\nu}+2\xi^{\mu}\nabla^{\nu}\chi] \;.\eeq
 Evaluating this at the bifurcate point $\mathcal{H}$ of the Killing horizon $H$ yields
 \beq \oint_{\mathcal{H}}Q_{\xi}^{\chi}=\frac{\kappa c}{12\pi}\chi_{\mathcal H}\;,\label{eq:QxichiH}\eeq
 where $\chi_{\mathcal{H}}=\chi_{H}$. Clearly this is equal to the temperature times the semi-classical correction to the Wald entropy due to $\chi$ (\ref{eq:SWalddS2}). 
 Meanwhile, the   Noether charge $Q_{\xi}^{\chi}$ associated to $\chi$  evaluated at $\mathcal{S}$ is
 \beq Q^{\chi}_{\xi}|_{\mathcal{S}}= \frac{c}{12\pi}\left(-Nn^{\mu}a_{\mu}\chi  +Nn^{\nu}\nabla_{\nu}\chi\right)|_{\mathcal S}\;.\label{eq:QcxiatB}\eeq
Note $\chi_{\mathcal{S}}$ is not   the same as $\chi_B$ when $\chi$ is time dependent. Consider now the left-hand side of the relation (\ref{eq:Smarrid}), where the only new contribution arises from $\chi$. The associated Noether current 1-form $j_{\xi}^{\chi}$ on shell is given by 
 \beq j_{\xi}^{\chi}|_{\Sigma}=-\xi\cdot L_{\text{Poly}}=\frac{c}{12\pi L^{2}}\left[\chi+\frac{L^{2}}{2}(\nabla\chi)^{2}\right] \xi\cdot\epsilon \;,\eeq
 where we used $\theta_\chi(\psi,\mathcal{L}_{\xi}\psi)|_{\Sigma}=0$, since $\chi$ is static at $\Sigma$ and $\mathcal L_\xi g_{\mu \nu}=0$, and the Polyakov Lagrangian 2-form $L_{\text{Poly}}$ is 
 \beq L_{\text{Poly}}=-\frac{c}{24\pi}\epsilon[\chi R+(\nabla\chi)^{2}]\;.\eeq
 Thus, analogous to (\ref{eq:LHSsmarr}), we may express the left-hand side of (\ref{eq:Smarrid}) in terms of a semi-classical ``Killing volume'' $\Theta^{\chi}_{\xi}$:
 \beq \int_{\Sigma}j_{\xi}^{\chi}=\frac{c\Lambda}{12\pi}\Theta_{\xi}^{\chi}\;,\quad \Theta_{\xi}^{\chi}\equiv\int_{\Sigma}\left(\chi+\frac{L^{2}}{2}(\nabla\chi)^{2}\right)\xi\cdot\epsilon\;.\label{eq:intjchi}\eeq
  Adding the semi-classical corrections (\ref{eq:QxichiH}), (\ref{eq:QcxiatB}), and (\ref{eq:intjchi}) to the classical Euler relation (\ref{eq:classELrel}) yields the semi-classical quasi-local Euler relation for de Sitter JT gravity:
 \beq E=TS_{\text{gen},H}-\sigma\left((\phi_{0}+\phi_{B})-\frac{2G_{2}c}{3}\chi_{\mathcal S}\right)-\frac{\phi_{0}\Lambda}{8\pi G_{2}N}\Theta_{\xi}+\frac{c\Lambda}{12\pi N}\Theta_{\xi}^{\chi}\;.\label{eq:semiclassEuler}\eeq
 Here $E$ refers to the sum of classical and semi-classical contributions to the energy, namely, 
 \beq E= -\frac{1}{8\pi G_{2}}n^{\nu}\nabla_{\nu}\phi+\frac{c}{12\pi}n^{\nu}\nabla_{\nu}\chi= E_{\phi}+E_{\chi}\;. \label{eq:semiclassicalenergyy}\eeq
 The semi-classical quasi-local Euler relation (\ref{eq:semiclassEuler}) may be split into three equations. The first two involve $\phi_r$ and $\phi_0$, respectively, and  are equivalent to the classical  expressions in Eq.~(\ref{eq:eulerblabha}). The third equation is proportional to $c$ and is given by
\beq E_\chi=T S_{\chi_{H}}+ \frac{2G_{2}c}{3}\sigma\chi_{\mathcal S} +\frac{c\Lambda}{12\pi N}\Theta^{\chi}_{\xi}\;.\eeq
 Further, in the limit the thermodynamic systems become the full static patch, in the full reduction  we have that the quasi-local Smarr formula (\ref{eq:semiclassEuler}) becomes the semi-classical Euler relation
 \beq 0=T_{\text{GH}}S_{\text{gen,h}}+T_{\text{GH}}S_{\text{gen,c}}-\frac{\phi_{0}\Lambda}{8\pi G_{2}N}\Theta_{\xi}+\frac{c\Lambda}{12\pi N}\Theta_{\xi}^{\chi}\;.\label{eq:semiclassicalsmarrrrr}\eeq

 \subsubsection*{Semi-classical quasi-local first law}
 
 The semi-classical first law follows from the variational identity (\ref{eq:varSmarr}). Unlike the classical case, the left-hand side of (\ref{eq:varSmarr}) is generally non-zero solely due to the presence of the auxiliary field $\chi$. This is because the symplectic current 1-form with respect to the Polyakov action   is (see Eq. (4.88) of \cite{Pedraza:2021cvx})
 \beq \omega_{\chi}(\psi,\delta\psi,\mathcal{L}_{\xi}\psi)|_{\Sigma}=-\frac{c}{24\pi}\epsilon_{\mu}\left[(g^{\mu\beta}g^{\alpha\nu}-g^{\mu\nu}g^{\alpha\beta})\nabla_{\nu}(\mathcal{L}_{\xi}\chi)\delta g_{\alpha\beta}-2\nabla^{\mu}(\mathcal{L}_{\xi}\chi)\delta\chi\right]\Big|_{\Sigma}\;,\eeq
 where we used $\mathcal L_\xi g_{\mu \nu}=0$ and $\mathcal{L}_{\xi}\chi|_{\Sigma}=0$. This is non-zero because generally $\nabla_{\mu}(\mathcal{L}_{\xi}\chi)|_{\Sigma}\neq0$.\footnote{The contribution will in fact vanish in the special case the radius associated with endpoint $(V_{2},U_{2})$ lies near the cosmological horizon, \emph{i.e.}, $U_{2}V_{2}=0$. } 
 Moreover, we cannot explicitly evaluate the integral of the symplectic current over $\Sigma$. Thus, we express the left-hand side of (\ref{eq:varSmarr}) formally, via Hamilton’s equations, as the variation of the Hamiltonian   associated to the $\chi$ field, generating evolution along the flow of $\xi$,
 \beq \delta H_{\xi}^{\chi}=\int_{\Sigma}\omega_{\chi}(\psi,\delta\psi,\mathcal{L}_{\xi}\psi)\;.\label{eq:delHchi}\eeq
 Moving to the right-hand side of (\ref{eq:varSmarr}), the integral at the bifurcation point of the Killing horizon is 
 \beq \oint_{\mathcal H}[\delta Q^{\chi}_{\xi}-\xi\cdot\theta_\chi(\psi,\delta\psi)]=\frac{\kappa c}{12\pi}\delta\chi_{\mathcal H}\;.\label{eq:ointHchi}\eeq
 The integral at $\mathcal S$   due to $\chi$ is given by the difference of 
 \beq \delta Q_{\xi}^{\chi}|_{\mathcal S}=-\frac{c}{12\pi}\delta\left( N n^{\mu}a_{\mu}\chi_{\mathcal S} \right) +\delta (N E_{\chi} ) \;,\eeq
 and $\xi\cdot\theta_{\chi}(\psi,\delta\psi)|_{\mathcal S}$, where 
 \beq
 \begin{split}
  \theta_\chi(\psi,\delta\psi)|_{B}&=-\frac{c\epsilon_{B}}{12\pi}\left[\left(n^{\mu}a_{\mu}+\frac{12\pi}{c}E_{\chi}\right)\delta\chi_{B}+\frac{12\pi}{c}\frac{E_{\chi}}{N}\delta N\right]+\frac{c}{12\pi}\delta(\epsilon_{B}\chi_B n^{\mu}a_{\mu})\\
  &=-\epsilon_{B}E_{\chi}\left(\delta\chi_{B}+\frac{\delta N}{N}\right)+\frac{c}{12\pi}\epsilon_{B}\chi_{B}\left(\delta(n^{\mu}a_{\mu})+n^{\mu}a_{\mu}\frac{\delta N}{N}\right)\;,
 \end{split}
 \eeq
 where we used $n^{\mu}\nabla_{\mu}\chi=12\pi E_{\chi}/c$. With $\xi\cdot\epsilon_{B}=-N$, we find
 \beq\oint_{\mathcal S}[\delta Q^{\chi}_{\xi}-\xi\cdot\theta(\psi,\delta\psi)]=N\delta E_{\chi}-N E_\chi \delta\chi_{\mathcal S}-\frac{c}{12\pi}N n^{\mu}a_{\mu}\delta\chi_{\mathcal S}\;.\label{eq:ointBchi}\eeq
 Substituting (\ref{eq:delHchi}), (\ref{eq:ointHchi}), and (\ref{eq:ointBchi}) into (\ref{eq:varSmarr}) and adding the result to the classical quasi-local first law (\ref{eq:quasilocalclasscov}), we arrive to
 \beq \delta(E_{\phi}+E_{\chi})=T\delta S_{\text{gen},H} -\sigma\left(\delta\phi_{\mathcal S}-\frac{2G_{2}c}{3}\delta\chi_{\mathcal S}\right)+   E_\chi\delta\chi_{\mathcal S}+\frac{1}{N}\delta H^{\chi}_{\xi}\;.\label{eq:quasilocalsc}\eeq
This is the semi-classical quasi-local first law. The second and third term on the right side may   be more neatly expressed as   $-\sigma_\phi  \delta \phi_{\mathcal S} - \sigma_\chi   \delta \chi_{\mathcal S} $, with the ``dilaton surface pressure'' defined as $\sigma_\phi   = n^\mu a_\mu/8\pi G_2$ and the ``conformal matter surface pressure''  $\sigma_\chi = - \frac{c}{12 \pi}  n^\mu a_\mu - E_\chi$. 
Moreover, multiplying both sides of the quasi-local first law by $N$ and taking the limit where the thermodynamic systems become the full static patch, leads to the global first law in the full reduction model,
 \beq 0=T_{\text{GH}}\delta S_{\text{gen,h}}+T_{\text{GH}}\delta S_{\text{gen,c}}+\delta H_{\xi}^{\chi}\;,\label{eq:globalfirstlawscfull}\eeq
 where we used $NE_{\chi}\to0$ as $r_{B}\to\pm L$. This first law suggests that the total semi-classical entropy of the static patch is given by the sum of the generalized entropy associated to the black hole horizon and the one associated to the cosmological horizon, \emph{i.e.},
 \begin{equation}  
     S_{\text{tot}}=S_{\text{gen,h}}+S_{\text{gen,c}}\,,
 \end{equation}
 which is the semi-classical generalization of the standard Nariai entropy \eqref{eq:entropynariai}.
 Finally, in the half reduction model the global first law follows from the limit $r_B \to 0$ and $N\to 1$, yielding
 \beq \label{eq:halfreductionfullstaticsemicl}
 \delta (E_\phi + E_\chi) = T_{\text{GH}} \delta S_{\text{gen},H} + E_\chi  \delta \chi_{\mathcal S} + \delta H_\xi^\chi\,,
 \eeq
 where $E_\chi = E_\chi (r_B=0)=\pm\frac{c}{12\pi L}$, where the positive (negative) sign refers to the cosmological (black hole) system. This is the semi-classical extension of the first law \eqref{eq:global1stlawhalfredffff}.

\subsection{Stationarity of generalized entropy in the microcanonical ensemble}

We can use the quasi-local first laws (\ref{eq:quasilocalclasscov}) and (\ref{eq:quasilocalsc}) to define different thermal ensembles and find the  associated  equilibrium conditions. Recall from ordinary thermodynamics that the stationarity   of the Helmholtz free energy $F=E-TS$ at a fixed temperature~$T$ and volume $V$ follows from the first law $d E = Td S - p d V$, since $d F =-SdT - p d V $ vanishes at fixed $(T,V)$.  Importantly, the stationarity of the free energy $F$ in the canonical ensemble is equivalent to the stationarity of the entropy $S$ in the microcanonical ensemble. This is because $dF|_{T,V}= dE - T dS$ and $dS|_{E,V}= dS -\beta dE $, so $dF |_{T,V} = - T dS|_{E,V}$ which means that $dF |_{T,V} =0$ is equivalent to  $dS|_{E,V}=0$. The last equilibrium condition states that the microcanonical entropy is extremized at fixed energy and volume. Below we will derive a similar statement for the generalized entropy in  semi-classical JT gravity.

The  quasi-local Helmholtz  free energy $F $ \eqref{eq:freeencosmo} in classical JT gravity   is defined as
\beq F=E-TS_{H} \;,\label{eq:quasilocfreeclass}\eeq
whose stationarity   follows from an application of the first law (\ref{eq:quasilocalclasscov}), 
\beq \delta F \big|_{T,\phi_B}=-S_{H}\delta T-\sigma \delta \phi_{B} =0\;.\label{eq:quasilocfreeclassstat}\eeq
Compared to the discussion of standard thermodynamics above, here the pressure is replaced by surface pressure $\sigma$ and the volume by the dilaton $\phi_B$.
When we include semi-classical   corrections the classical entropy is replaced by the generalized entropy and the classical quasi-local energy is replaced by the semi-classical energy \eqref{eq:semiclassicalenergyy}, such that the free energy becomes 
\beq F_{\text{semi-cl}}=E_\phi + E_\chi -T S_{\text{gen},H} \;, 
\eeq
which is stationary at fixed $(T,\phi_B, \chi_{\mathcal S}, H_\xi^\chi)$ due to the semi-classical first law (\ref{eq:quasilocalsc}). 

The  stationarity condition of the Helmholtz free energy characterizes the canonical ensemble. 
The canonical ensemble may be transformed into the microcanonical ensemble by an appropriate Legendre transformation of the free energy. In particular, under a (negative) Legendre transform  of $\beta F $ with respect to $\beta$,    the classical entropy $S_{H}$ is recognized as the thermodynamic potential of the microcanonical ensemble, 
\beq S_{H}=-\beta (  F -  E )\;.\eeq
It follows from the classical first law (\ref{eq:quasilocalclasscov}) that $S_{H}$ is stationary at fixed    $E$ and $\phi_B$,
\beq \delta S_{H}\big|_{E,\phi_B}=0\;.\eeq
Likewise, when semi-classical corrections are included,   the generalized entropy $S_{\text{gen}}$ is identified with the microcanonical entropy, 
\beq
 S_{\text{gen},H}  = - \beta (F_{\text{semi-cl}} - E_\phi - E_\chi)\,,
\eeq
and obeys the stationarity condition 
\beq \delta  S_{\text{gen},H} \big|_{(E_\phi, E_\chi,\phi_B, \chi_\mathcal{S}, H_{\xi}^{\chi})}=0\;. \label{eq:stationarityentropysc}\eeq
We may interpret this as the microcanonical equilibrium condition for semi-classical de Sitter JT gravity. It holds both for the black hole system as well as for the cosmological system, in the sense that $H$ can represent both the black hole horizon and  the cosmological horizon in~dS$_2$. If the thermodynamic systems become  the full static patch, less variables need to be kept fixed in the microcanonical ensemble: in the full reduction model the sum $S_{\text{gen,h}}+S_{\text{gen,c}}$ is stationary at fixed $H_\xi^\chi$ in the static patch, as follows from \eqref{eq:globalfirstlawscfull}, while in the half reduction model $S_{\text{gen},H}$ is stationary at fixed $(E_\phi, E_\chi, \chi_{\mathcal S}, H_\xi^\chi)$, according to \eqref{eq:halfreductionfullstaticsemicl}.  A similar relation as \eqref{eq:stationarityentropysc} was uncovered for semi-classical JT gravity in AdS in \cite{Pedraza:2021cvx}. 

It is worth recalling  that quantum extremal surfaces are defined as codimension-2 surfaces which extremize the generalized entropy; this is the essential content of the QES prescription (\ref{eq:QESformula}). Thus, when backreaction effects are taken into account, the semi-classical first law in the microcanonical ensemble may be regarded as the first law of thermodynamics of quantum extremal surfaces in $\text{dS}_{2}$. This observation motivates us to explore the connection between the QES formula and microcanonical semi-classical thermodynamics in the next section.


 \section{Islands from the microcanonical action} \label{sec:islandsfrommicact}

 In the previous section we have established two key insights about semi-classical JT gravity in de Sitter space: (i)  the semi-classical Wald entropy is equal to the generalized entropy (\ref{eq:SWaldSgen}), and (ii) $S_{\text{gen}}$ is the microcanonical entropy   and is stationary in the microcanonical ensemble.
 Following \cite{Pedraza:2021cvx,Pedraza:2021ssc}, we may combine these two observations and provide a first principles derivation of the extremization condition appearing in the QES formula (\ref{eq:QESformula}) via a Euclidean microcanonical gravitational path integral. More precisely, at leading order in a saddle-point approximation, the Euclidean microcanonical action $I^{\text{mc}}_{\text E}$ is equal to (minus) the generalized entropy, where the extremization of $S_{\text{gen}}$ follows from minimizing $I^{\text{mc}}_{\text E}$. An important distinction is that in the previous section we studied the thermodynamics of Killing horizons, while here we consider the thermodynamics of finite causal diamonds which have a conformal Killing horizon in dS$_2$. This is because we are interested in the entropy of entanglement wedges, which take the form of   causal diamonds. Time evolution in the former quasi-local set-up is generated by the standard time translation Killing vector $\partial_{t}$, while in the latter diamond context it is generated by a conformal Killing vector (see Sec. \ref{sec:cdds2yeah}).

It is worth emphasizing our approach does not rely on an underlying holographic duality, such as AdS/CFT or dS/CFT. We also will not need to invoke the replica trick, as done in \cite{Almheiri:2019qdq,Penington:2019kki,Goto:2020wnk}, since we are working with an eternal background with a $U(1)$ Killing symmetry. Thus, we will not find replica wormhole geometries. Moreover, while the arguments below hold for two-dimensional models, we will provide a derivation of the island formula for de Sitter JT gravity, which thus far has  been assumed to hold in the literature. 
 
 \subsection{Microcanonical action}

 
 Recall from ordinary thermodynamics that a system may be described using various ensembles depending on which thermodynamic data is held fixed. For example, the canonical partition function $Z(\beta)$ characterizes a system of fixed size and temperature $T=\beta^{-1}$, defining the canonical ensemble. Meanwhile, when the total energy $E_{0}$ is fixed, the system is best described using the microcanonical partition function, \emph{i.e.}, the density of states $W(E_{0})$. One may relate the canonical and microcanonical ensembles via an appropriate Legendre transform of the thermodynamic potentials, as described above.
 
 It is well known, moreover, that the canonical partition function may be cast as a Euclidean path integral, \emph{i.e.}, a functional integral over field configurations $\psi$ with fixed boundary data, weighted by the (canonical) Euclidean action $I_{\text E}^{\text{can}}$ defining the theory, all at fixed temperature, 
 \beq Z(\beta)=\int \mathcal{D}\psi \hspace{1mm} e^{-I_{\text E}^{\text{can}}[\psi]}\;.\label{eq:canZgen}\eeq
 Here $\mathcal{D}\psi$ denotes the functional integration measure over dynamical fields $\psi$, and, as is standard practice with thermal path integrals, the Euclidean time variable is periodic in $\beta$. In a saddle-point approximation we have $Z\approx e^{-I^{\text{can}}_{\text E}[\psi_{0}]}$, where $\psi_{0}$ are solutions to the semi-classical field equations.

 It is not immediately clear whether the density of states $W(E_{0})$ can likewise be cast in terms of a path integral. This is because, for a theory without gravity, the total energy of matter fields permeates all space and is not fixed by only specifying boundary data.
 However, as recognized by Brown and York \cite{Brown:1992bq} (see also \cite{Brown:1989fa}), when gravity is included, the total energy of the system is entirely given by the behavior of gravitational field variables at the boundary. 
 This makes it possible to express $W(E_{0})$ as a path integral over field configurations at a fixed energy, weighted by the Euclidean microcanonical action $I_{\text E}^{\text{mc}}$,
 \beq W(E_{0})=\int \mathcal{D}\psi \hspace{1mm} e^{-I_{\text E}^{\text{mc}}[\psi]}\approx e^{-I^{\text{mc}}_{\text E}[\psi_{0}]}\;.\label{eq:mcdengen}\eeq
 The form of the microcanonical action can be deduced, at least to leading order, in a saddle-point approximation, since the canonical and microcanonical actions are related via a standard Legendre transform. To see this, recall the canonical and microcanonical partition functions are connected by a Laplace integral transform 
    \beq Z(\beta)=\int dE_0 W(E_0)e^{-\beta E_{0}}\;.\eeq
  In a stationary phase approximation and in the (near) thermodynamic limit, the canonical partition function  is given by  $\log Z(\beta)\approx \log W (E_0)-\beta E_{0}$. Identifying the canonical free energy $-\beta F(\beta)=\log Z(\beta)$ and microcanonical entropy $S_{\text{mc}}(E_{0})=\log W(E_{0})$, this relation is recognized as the   Legendre transform $-\beta F=S_{\text{mc}}-\beta E_{0}$. Expressing $Z(\beta)$ in terms of a path integral as in (\ref{eq:canZgen}) with $\log Z(\beta)\approx-I_{\text E}^{\text{can}}$, to leading order one finds a transformation between the microcanonical and canonical actions, $I^{\text{mc}}_{\text E}=I_{\text E}^{\text{can}}-\beta E_{0}$. 
  
Formally, for a gravity theory on a Euclidean manifold $\mathcal{M}_{\text E}$ with a timelike Killing symmetry, generated by $\xi=\partial_t$, the off-shell  Euclidean microcanonical action is given by a Legendre-like transform of the (canonical) Euclidean action involving the Noether charge $Q_{\xi}$~\cite{Iyer:1995kg}
 \beq I^{\text{mc}}_{\text E}=-i\left(\int_{\mathcal{M}_{\text E}}\hspace{-2mm}L-\int_{\partial \mathcal{M}_{\text E}}\hspace{-3mm}dt\wedge Q_{\xi}\right)\;.\label{eq:ImcoffshellBHs}\eeq
 Here,   $L$ is the Lagrangian form in Euclidean signature. This version of the action is found by explicitly comparing its variation $\delta I^{\text{mc}}_{\text E}$ to the variation of the microcanonical action developed in \cite{Brown:1992bq}. 
  In the context of an eternal black hole  in an arbitrary diffeomorphism invariant theory, one finds the on-shell microcanonical action is equal to the Wald entropy
 \beq I^{\text{mc}}_{\text E}=-S_{\text{Wald}}\;.\label{eq:ImcSwald}\eeq
 This on-shell relation  can be understood as a path integral derivation of the Wald entropy functional for stationary black holes in the microcanonical ensemble. 
 
 Another, seemingly less well-known, path integral method for deriving the entropy of a bifurcate Killing horizon in an arbitrary theory is known as the Hilbert action surface term method, developed by Ba\~{n}ados-Teitelboim-Zanelli (BTZ) \cite{Banados:1993qp}. In this approach, as detailed in \cite{Brown:1995su}, the on-shell microcanonical action is equal to the Gibbons--Hawking--York surface term evaluated on the boundary of an infinitesimal disk $D_{\epsilon}$ of radius $\epsilon$ orthogonal to punctures in the Euclidean spacetime, corresponding to the bifurcate Killing horizon in Lorentzian signature. Hence,    the Wald entropy may be written as the GHY surface term evaluated on infinitesimal boundaries surrounding the analytic continuation of the bifurcate horizon.\footnote{It is worth emphasizing that the horizon entropy does not follow from inserting a GHY boundary term near the horizon in   the standard (canonical) Gibbons--Hawking path-integral method. For example, in asymptotically flat backgrounds, the on-shell canonical Euclidean action is given by the GHY term evaluated at infinity, while in   Euclidean dS  there is   no boundary term and the entropy is computed using the bulk action.} Providing more details below, we will use the BTZ prescription to compute the on-shell microcanonical action; an equivalence between this method \cite{Banados:1993qp} and the Noether charge formalism was established in \cite{Iyer:1995kg} (see also Appendix C in \cite{Pedraza:2021ssc}). 
 
 Lastly, as eluded to in the introduction, the Bekenstein--Hawking entropy formula applies to surfaces other than black hole horizons. As such, the respective off-shell and on-shell relations (\ref{eq:ImcoffshellBHs}) and (\ref{eq:ImcSwald}), as well as the BTZ method may be generalized to other spacetimes with horizons. In the next two subsections we will introduce  causal diamonds in $\text{dS}_{2}$ and apply the microcanonical action to this geometric setup.


 \subsection{Causal diamonds in $\text{dS}_{2}$}
 \label{sec:cdds2yeah}
 
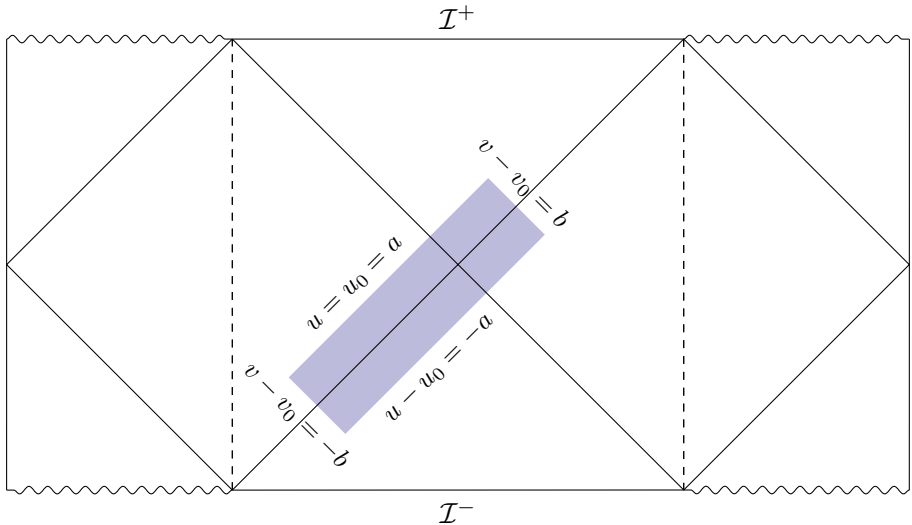
\begin{figure}[t!]
\centering
\begin{tikzpicture}[scale=1.5]
	\pgfmathsetmacro\myunit{4}
	\path (0.5,1) --++(-45:0.71) node[pos=.4, below, sloped] {{\small $v - v_0 = -b$}} --++(45:2.5) node[pos=.4, below, sloped] {{\small $u - u_0 = -a$}} --++(135:0.71) node[pos=.65, above, sloped] {{\small $v - v_0 = b$}} -- cycle node[pos=.6, above, sloped] {{\small $u = u_0 = a$}};
	\fill[fill=Periwinkle, fill opacity=0.5] (0.5,1) --++(-45:0.71) --++(45:2.5) --++(135:0.71) -- cycle;
	\draw[dashed]	(0,0)			coordinate (a)
		--++(90:\myunit)	coordinate (b);
	\draw (b) --++(0:\myunit)		coordinate (c)
							node[pos=.5, above] {$\mathcal{I}^+$};
	\draw[dashed] (c) --++(-90:\myunit)	coordinate (d);
    \draw (a) -- (-2,2) -- (b);
	\draw (d) -- (a) 		node[pos=.5, below] {$\mathcal{I}^-$};
	\draw (b) -- (d)  -- (6,2) -- (c) -- (a);
    \draw[dashed] (a) -- (b);
    \draw[dashed] (c) -- (d);
    \draw[decorate, decoration={snake, amplitude=0.5mm, segment length=2.5mm}] (a) -- (-2,0) coordinate (e);
    \draw (e) -- (-2,4) coordinate (f);
    \draw[decorate, decoration={snake, amplitude=0.5mm, segment length=2.5mm}] (f) -- (b); 
    \draw[decorate, decoration={snake, amplitude=0.5mm, segment length=2.5mm}] (c) -- (6,4) coordinate (g); 
    \draw (g) -- (6,0) coordinate (h);
    \draw[decorate, decoration={snake, amplitude=0.5mm, segment length=2.5mm}] (h) -- (d);
\end{tikzpicture}
\caption{A (rectangular) Lorentzian causal diamond in two-dimensional de Sitter space in the full reduction model.
\label{fig:Lordiamond}}
\end{figure}

 We are interested in evaluating the microcanonical action on the entanglement wedge 
 of an interval $\Sigma$ in $\text{dS}_{2}$, \emph{i.e.}, the domain of dependence  of any achronal surface with boundary $\partial \Sigma$. The entanglement wedge is given by a finite, rectangular causal diamond, the intersection of the past and future domains of dependence of $\Sigma$. In a generic two-dimensional spacetime in the conformal gauge $d\ell^2=-e^{2\rho}dudv$, the causal diamond consists of the intersection of the regions $[u-u_0 = -a,u-u_0= a]$ and $[v-v_0 = -b, v-v_0=b]$ for constants $a,b,u_{0},v_{0}$. Positive length scales $a$ and $b$ define the null boundaries of the diamond, $(u-u_{0}=\pm a,\,v-v_{0}=\pm b)$. A square diamond is one with $a=b$. The maximal spatial slice $\Sigma$ in the diamond is given by $u-u_0 =  - \sqrt{(v-v_0)^2 +a^2 - b^2}$,  and the line between the future and past vertices is given by $u-u_0 =   \sqrt{(v-v_0)^2 +a^2 - b^2}$ (see Appendix A in \cite{Pedraza:2021ssc}).  An illustration of a (Lorentzian) causal diamond  is given in Figure \ref{fig:Lordiamond}.
 

 Such a causal diamond has a conformal isometry generated by a conformal Killing vector~$\zeta$, obeying the conformal Killing equation in two dimensions  $2\nabla_{(\mu}\zeta_{\nu)}=g_{\mu\nu}(\nabla\cdot \zeta)$  \cite{Jacobson:2015hqa,Jacobson:2018ahi,Visser:2019muv}. Specifically, when we put the diamond into two-dimensional de Sitter space, and we require that $\zeta$ is proportional to   $\partial_t$ in static  coordinates in the maximal diamond limit $a,b \to \infty$, then the  conformal Killing vector takes the unique form (see    Appendix~\ref{app:diamondcoordinates})
\beq \zeta=A_{a}(u-u_{0})\partial_{u}+A_{b}(v-v_{0})\partial_{v}\;,\label{eq:ckvdS2}\eeq
with 
\beq A_{a}(y)=\frac{L\kappa_{a}}{\sinh(a/L)}\left[\cosh(a/L)-\cosh(y/L)\right]\;,\eeq 
and similarly for $A_{b}(y)$. Here $\kappa_{a}$ and $\kappa_{b}$ are surface gravities associated with length scales $a$ and $b$. On the null boundaries of the diamond we have $\zeta^{2}=0$, thus they are conformal Killing horizons generated by $\zeta$. The surface gravities are constant and positive (negative) along the future (past) horizon.


We can cover the causal diamond with inextendible ``diamond universe'' coordinates $(s,x)$ adapted to the flow of $\zeta$ \cite{Jacobson:2018ahi}. Here $s$ is the conformal Killing time, satisfying $\zeta \cdot ds =1$, with range $s\in[-\infty,\infty]$, while $x$ is a spatial coordinate $x\in[-\infty,\infty]$. In these coordinates, the two-dimensional line element is
 \beq d\ell^2=C^{2}(s,x)(-ds^{2}+dx^{2}).\label{eq:diamonduni}\eeq
The conformal factor $C^{2}$ is explicitly derived in Appendix \ref{app:diamondcoordinates} for diamonds in dS$_2$. In these coordinates the conformal Killing vector is simply the generator of the conformal Killing time, $\zeta=\partial_{s}$. The null boundaries of the horizon   are located at $x=\pm\infty$, where  the diamond line element (\ref{eq:diamonduni}) approximates to 
  \beq d\ell^2 \approx  4L^{2}\kappa_a \kappa_b e^{\mp (\kappa_a + \kappa_b) x}(-ds^{2}+dx^{2}).\label{eq:nearhorizonCD}\eeq
We recognize this as the flat Rindler metric $d\ell^2 = -\kappa^{2}\varrho^{2}ds^{2}+d\varrho^{2}$, with  radial coordinate  $\varrho\equiv  4 L \sqrt{\kappa_a \kappa_b} (\kappa_a + \kappa_b)^{-1} e^{\mp (\kappa_a + \kappa_b)x/2}$ and surface gravity $ \kappa=\frac{1}{2}(\kappa_{a}+\kappa_{b})=\mp C^{-1}\partial_{x}C |_{x\to\pm\infty}$. Thence, $\zeta = \partial_s$ approaches an approximate boost Killing vector near  $x=\pm\infty$. 

\begin{figure}[t]
 \begin{center}
 \includegraphics[width=8cm]{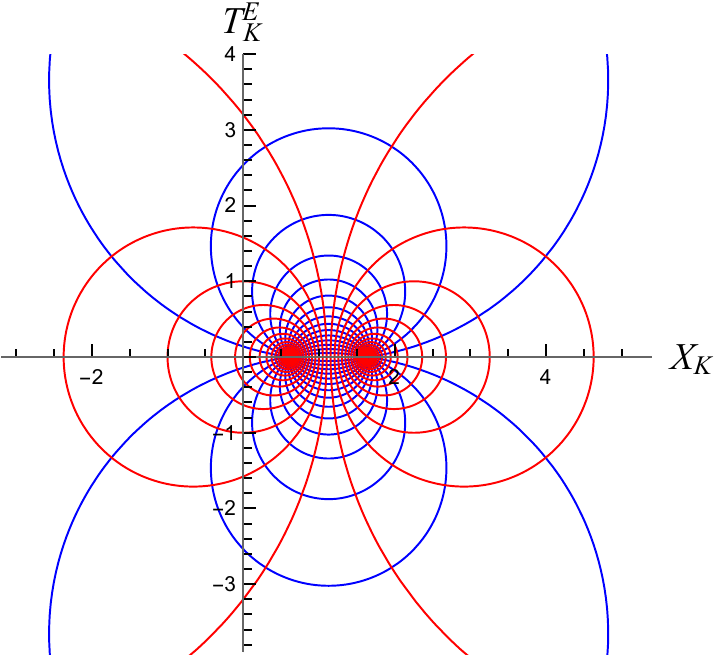}
 \end{center}
\caption{Euclidean $\text{dS}_{2}$ diamond spacetime in Kruskal coordinates $(T^{\text{E}}_{\text{K}},X_{\text{K}})$ (\ref{eq:EucKruskdiacoord}). Lines of constant $x$ (red) and lines of constant $s_{\text{E}}$ (blue) are  at equal  intervals of 0.2. High contour density corresponds to the two horizon punctures at $x=\pm\infty$. We have set $a=b=1/2$, $L=\kappa_{a}=\kappa_{b}=1$ and $u_0\neq v_0\neq 0$. For the square diamond, the punctures at $x\to\pm\infty$ are mapped to the points $(T_{\text{K}}^{\text{E}},X_{\text{K}})=(0,L e^{\pm a/L})$.
\label{fig:Eucdiamond}}
\end{figure}

The Euclidean continuation of the diamond universe coordinates follows from Wick rotating the conformal Killing time $s\to -is_{\text{E}}$. Similar to causal diamonds in four dimensions \cite{Banks:2020tox}, the Euclidean continuation of the finite diamond covers nearly the entire space of Euclidean $\text{dS}_{2}$; only the bifurcation points $x\to\pm\infty$ are missing (see Appendix \ref{app:diamondcoordinates}). Thus, the null boundaries  are mapped to punctures in the Euclidean spacetime, and correspond to a conical singularity $\varrho=0$ in the Rindler metric (\ref{eq:nearhorizonCD}). To remove the conical singularity, we must periodically identify the Euclidean time coordinate, $s_{\text{E}}\sim s_{\text{E}}+ 2\pi/\kappa$. As such, when we restrict  the  Bunch--Davies state to the causal diamond, the diamond has a natural temperature $T_{\text{CD}}=\kappa/2\pi$. 
Thus, the Euclidean causal diamond in dS$_2$ may be represented by a two-sphere with two punctures corresponding to the horizons of the Lorentzian diamond. We illustrate the Euclidean diamond spacetime in Figure \ref{fig:Eucdiamond}.


 \subsection{Generalized entropy from the microcanonical action}
 
We now have all of the ingredients to find the microcanonical action of $\text{dS}_{2}$ causal diamonds in semi-classical JT gravity. We start from the microcanonical density of states $W(E_{0})$ (\ref{eq:mcdengen}) in the saddle point approximation, where $\psi_{0}=\{g_{\mu\nu},\phi,\chi\}$ are the solutions to the semi-classical JT equations. Following \cite{Pedraza:2021ssc}, the off-shell Euclidean microcanonical action for causal diamonds is defined as
\beq
\begin{split}
I^{\text{mc}}_{\text E}&\equiv-i \left[\int_{  \mathcal{M}_{\text E}^{\text{CD}}}\hspace{-2mm}  L-\int_{\mathcal{M}_{\text E}^{\text{CD}}}\hspace{-2mm}ds\wedge \theta(\psi,\mathcal{L}_{\zeta}\psi)\right].
\end{split}
\label{eq:Imicrogen}\eeq
This differs from the microcanonical action for black holes \eqref{eq:ImcoffshellBHs}, since causal diamonds have no asymptotic region like black holes and they admit a conformal isometry instead of a true isometry. Here $\theta$ is the symplectic potential 1-form,  with $\theta(\psi, \mathcal L_\zeta \psi)$ non-vanishing since $\zeta$ is a conformal Killing vector rather than an exact Killing vector. Writing $L = ds \wedge \zeta \cdot L$, we see the two terms between brackets combine into an integral over  the   Noether current 1-form $j_{\zeta}\equiv\theta(\psi,\mathcal{L}_{\zeta}\psi)-\zeta\cdot L$ associated with diffeomorphisms generated by~$\zeta$. Using the on-shell identity $j_{\zeta}=dQ_{\zeta}$, with $Q_{\zeta}$ the Noether charge 0-form, and applying Stokes' theorem we find the  on-shell Euclidean microcanonical action for diamonds is equal to 
\begin{equation}
     I^{\text{mc}}_{\text E} = \int_{\partial \mathcal{M}_{\text E}^{\text{CD}}} \hspace{-2mm}ds_{E} \wedge Q_\zeta=\frac{2\pi}{\kappa}\oint_{\partial \Sigma} Q_\zeta =   - S_{\text{Wald}} \big |_{ \partial \Sigma}\;.
\label{eq:microaction2}\end{equation} 
To arrive to the second equality we used the fact that $\partial \mathcal{M}_{\text E}^{\text{CD}}$ has topology $S^1 \times \partial \Sigma$, such that the Noether charge restricted to $\partial \Sigma$ is independent of Euclidean time $s_{E}$ since the dilaton $\phi$ and auxiliary field $ \chi$ are constant in the limit $x \to \pm\infty.$ This allows us to integrate out the Euclidean time. The last equality follows from the definition of the Wald entropy, with $S_{\text{Wald}}= \frac{1}{4G_{2}}(\phi_{0}+\phi)-\frac{c}{6}\chi$. Thus, the on-shell microcanonical action of Euclidean $\text{dS}_{2}$ causal diamonds is equal to minus the Wald entropy.

Equivalently, the on-shell microcanonical action (\ref{eq:microaction2}) is given by a GHY boundary term inserted at the bifurcation points $\{\partial\Sigma: x =\pm \infty\}$ \cite{Banks:2020tox,Pedraza:2021ssc}. To see this, note  the Hamiltonian $H_{\zeta}$ for a theory, which fixes the induced metric of the boundary $\partial \mathcal{M}$ of a (Lorentzian) manifold $\mathcal{M}$, is  given by an integral over the codimension-2 slices $\mathcal{C}_{s}$ where $\Sigma_{s}$ orthogonally intersects $\partial \mathcal{M}$ \cite{Pedraza:2021cvx,Iyer:1995kg},
\beq H_{\zeta}=\oint_{\mathcal{C}_{s}}(Q_{\zeta}-\zeta\cdot b)=\oint_{\mathcal{C}_{s}}\epsilon_{\partial\Sigma} N\varepsilon\;.\label{eq:HzetaCs}\eeq
Here $b$ is the GHY boundary term 1-form (\ref{eq:Cdefn}),  $\varepsilon$ is the quasi-local energy density (\ref{eq:quasilocalen}), and $N=-\zeta^\mu  u_\mu$ is the lapse. Importantly, at the bifurcation points $\partial\Sigma$ the lapse $N=0$ such that $H_{\zeta}=0$ on $\partial\Sigma$. Now, let $(\partial \Sigma)_{\epsilon}$ denote a 1-parameter family of surfaces in $\Sigma_{s_{\text{E}}}$ obeying $\lim_{\epsilon\to0}(\partial \Sigma)_{\epsilon}\to \partial \Sigma$. Using $H_{\zeta}=0$ in (\ref{eq:HzetaCs}), it follows
\beq \lim_{\epsilon\to0}\int_{(\partial \Sigma)_{\epsilon}}Q_{\zeta}=\lim_{\epsilon\to0}\int_{(\partial \Sigma)_{\epsilon}}\zeta\cdot b ~,\eeq
leading to, for the case of semi-classical JT gravity,
\beq I^{\text{mc}}_{\text E}=-\lim_{\epsilon\to0}\int_{\partial D_{\epsilon}\times \partial \Sigma}\hspace{-4mm} ds_{E}\sqrt{\gamma}K\left[\frac{(\phi_{0}+\phi)}{8\pi G_{2}}-\frac{c\chi}{12\pi}\right]\;.\label{eq:micactionCDGHY}\eeq
Here $\sqrt{\gamma} = C$ is the induced metric on constant $s_{E}$ slices and the trace of the extrinsic curvature  of these slices is $K= \mp C^{-2}\partial_x C$. Since the fields $\phi$, $\chi$  are independent of $s_{\text E}$ and $\sqrt{\gamma}K\to\kappa$ in the limit $x\to\pm\infty$, the integral over $s_{\text E}$ is trivial, and the right-hand side is equal to minus the Wald entropy. This establishes the equivalence between the BTZ \cite{Banados:1993qp,Banks:2020tox} and Noether charge \cite{Iyer:1995kg} methods for the case of causal diamonds.

From either (\ref{eq:microaction2}) or (\ref{eq:micactionCDGHY}), since $S_{\text{Wald}}=S_{\text{gen}}$ (\ref{eq:SWaldSgen}), we see the on-shell microcanonical action in semi-classical JT gravity is given by the generalized entropy
\beq I^{\text{mc}}_{\text E}=-S_{\text{gen}} \big |_{\partial \Sigma}\;. \label{eq:actionentropy}\eeq
The density of states is thus $W(E_{0})\approx e^{S_{\text{gen}}}$, identifying $S_{\text{gen}}$ as the microcanonical entropy. As a microcanonical entropy, $S_{\text{gen}}$ is maximized at a fixed energy. Therefore, the microcanonical action is minimized at fixed energy $E_{0}$. We may formally determine the energy $E_{0}$ by computing the variation of $I_{E}^{\text{mc}}$ over the full Euclidean causal diamond. Specifically, 
\beq \delta I_{\text{E}}^{\text{mc}}=\int_{\mathcal{M}_{\text{E}}^{\text{CD}}}\hspace{-2mm}ds_{\text{E}}\wedge \omega(\psi,\delta\psi,\mathcal{L}_{\zeta}\psi)=\int_{S^{1}}\hspace{-1mm}ds_{\text{E}}\delta H_{\zeta}\;,\eeq
where $\delta H_{\zeta}=\int_{\Sigma_{s_{\text{E}}}}\hspace{-1mm} \omega(\psi,\delta\psi,\mathcal{L}_{\zeta}\psi)$ is the variation of the Hamiltonian generating the evolution along $\zeta$. This shows   $I_{\text{E}}^{\text{mc}}$ is stationary at fixed energy $E_{0}=\pm H_{\zeta}+\text{const}$. We set the constant to zero, and the sign is determined by imposing consistency with the first law of causal diamonds, $\frac{\kappa}{2\pi}\delta S_{\text{Wald}}=-\delta H_{\zeta}$ \cite{Jacobson:2018ahi,Pedraza:2021ssc}. Hence, the energy to be fixed is $E_{0}=-H_{\zeta}$. 

Furthermore, minimizing the microcanonical action with respect to the background is equivalent to extremizing $S_{\text{gen}}$ with respect to the shape and location of $\partial\Sigma$. This is consistent with the extremization prescription in the QES formula. One subtle difference with the QES formula is that we derived the extremization of $S_{\text{gen}}$ in Euclidean signature, whereas the QES formula is usually stated in Lorentzian signature. We have thus derived the generalized entropy in de Sitter JT gravity and its extremization from a Euclidean action principle.

Lastly, note that here the Euclidean time $s_{\text{E}}=is$ is imposed to be periodic, $s_{\text{E}}\sim s_{\text{E}}+ \frac{2\pi}{\kappa}$, in order to remove the conical singularities at $x =\pm \infty$. This is a regularity condition at the horizon which happens to be consistent with our choice of   vacuum state. Thus, while we work in the microcanonical ensemble, the vacuum state of matter remains in the Bunch--Davies vacuum, which is a thermal state when restricted to the causal diamond  at a fixed, positive temperature $T_{\text{CD}}=\kappa/2\pi$.

\subsection{Islands in the full reduction}

Quantum extremal surfaces arise from extremizing the generalized entropy  \eqref{eq:SWalddS2}, where the von Neumann entropy \eqref{eq:vnentchikrusk} is of a single interval with endpoints $[(U_1, V_1), (U_2,V_2)]$. Equivalently, we search for QESs by minimizing the  microcanonical action of a causal diamond in $\text{dS}_{2}$, where the bifurcation points $\partial\Sigma$ of the diamond are identified with the endpoints of the interval. In our computation, we will keep one endpoint of the interval fixed, and vary the position of the other endpoint. When looking for QESs, it is important to distinguish between the $\text{dS}_{2}$ geometry which arises from the half or full spherical reduction. In the full reduction, one may consider an interval with one endpoint in one hyperbolic patch, and another endpoint in a different hyperbolic patch. The authors of \cite{Hartman:2020khs} showed non-pathological quantum extremal islands only arise in this scenario, thus implying islands do \emph{not} arise in the half reduction model. Our calculations below are consistent with the results in \cite{Hartman:2020khs}. 

\subsection*{QESs in half reduction}

Let us look for quantum extremal surfaces, and, consequently, islands, in the $\text{dS}_{2}$ geometry found via half reduction (Figure \ref{fig:dS2halfred}). This follows from extremizing the generalized entropy. The matter entanglement entropy is given by the von Neumann entropy of the conformal matter, $\chi$, in the Bunch--Davies vacuum restricted to an interval with both endpoints in the half reduction dS$_2$ space. One can consider a similar set-up for the $\text{dS}_{2}$ geometry from full reduction, and therefore our discussion here applies equally to that case as well (hence $\phi_{0}$ is not set to zero).

The total generalized entropy \eqref{eq:SWalddS2} is
\beq S_{\text{gen}}=\frac{1}{4G}\left(\phi_{0}+\phi_{r}\left(\frac{1+\frac{U_{1}V_{1}}{L^{2}}}{1-\frac{U_{1}V_{1}}{L^{2}}}\right)+\frac{Gc}{3}\right)+\frac{c}{12}\log\left[\frac{16}{\delta_{1}^{2}\delta_{2}^{2}}\frac{(U_{2}-U_{1})^{2}(V_{2}-V_{1})^{2}}{\left(1-\frac{U_{1}V_{1}}{L^{2}}\right)^{2}\left(1-\frac{U_{2}V_{2}}{L^{2}}\right)^{2}}\right]\;.\eeq
Here we will keep the endpoint $(U_{2},V_{2})$ fixed while varying the first endpoint $(U_{1},V_{1})$. Doing so, we find two possible locations for a QES. The first is at
\beq V_{1}\approx \frac{L^{2}\epsilon}{3U_{2}}+\mathcal{O}(\epsilon^{2})\;,\quad U_{1}\approx\frac{L^{2}\epsilon}{3V_{2}}+\mathcal{O}(\epsilon^{2})\;,\label{eq:QEShalf1}\eeq
and the second is at
\beq V_{1}\approx V_{2}-\frac{(L^{2}-2U_{2}V_{2})}{3U_{2}}\epsilon+\mathcal{O}(\epsilon^{2})\;,\quad U_{1}\approx U_{2}-\frac{(L^{2}-2U_{2}V_{2})}{3V_{2}}\epsilon+\mathcal{O}(\epsilon^{2})\;,\label{eq:QEShalf2}\eeq
where $\epsilon\equiv \frac{G_{2}c}{\phi_{r}}\ll1$. In the classical limit\footnote{When Planck's constant $\hbar$ is restored one has $\epsilon=G_{2}\hbar c/\phi_{r}$, and the $\epsilon\to0$ limit corresponds to $\hbar\to0$.} $\epsilon\to0$, the first solution reduces to the cosmological horizon ($U_1=V_1=0$) for any choice of $(U_{2},V_{2})$. The second solution places the two endpoints very near each other, coinciding as $\epsilon\to0$. We thus reject the second solution, and find that the QES (\ref{eq:QEShalf1}) lies near the cosmological horizon, 
\beq r_{\text{QES}}\approx L-\frac{2L}{9}\left(\frac{L+r_{2}}{L-r_{2}}\right)\epsilon^{2}\;,\label{eq:QESlochalf}\eeq
where $r_{2}$ denotes the radial coordinate associated with endpoint $(U_{2},V_{2})$. The location of this QES coincides with the one attained in \cite{Aalsma:2021bit}. Further, note  that for  $U_{2}V_{2}=L^{2}$ or $r_2 \to \infty$, then $S_{\text{gen}}(r=L)>S_{\text{gen}}(r=r_{\text{QES}})$, consistent with the QES formula.\footnote{Moreover, as $r_{2}\to\infty$, the QES position (\ref{eq:QESlochalf}) simplifies to $r_{\text{QES}}\approx L (1 + \frac{2}{9} \epsilon^2)$, identical to the  location of a QES in AdS$_2$ \cite{Pedraza:2021cvx}. This is  an approximation to the exact value $r_{\text{QES}}(r_2 \to \infty)=\frac{2}{3} L \epsilon \sqrt{1 + \frac{9}{4 \epsilon^2}}$ \cite{Pedraza:2021ssc}.} Further, the generalized entropy for the second solution (\ref{eq:QEShalf2}) is parametrically larger than $S_{\text{gen}}$ evaluated at the QES in (\ref{eq:QEShalf1}), which is  another reason to ignore the second solution.

The island formula is an application of the QES formula (\ref{eq:QESformula}), which may be used to compute the von Neumann entropy associated with radiation emitted from a horizon. Formally, one computes the von Neumann entropy associated to the entanglement wedge of radiation, namely, the causal development of the codimension-1 slice $\Sigma_{\partial I}=\Sigma_{\text{rad}}\cup I$. One imagines collecting the radiation in a weakly gravitating region $\Sigma_{\text{rad}}$, here placed near future infinity $\mathcal{I}^{+}$, as the dilaton diverges near there, a herald for weak gravity. The boundary of the island $\partial I$ corresponds to the location of the QES. For $\Sigma_{\text{rad}}$ near $\mathcal{I}^{+}$ ($r_{2}\to\infty$) we see the QES (\ref{eq:QESlochalf}) is located just outside of the cosmological horizon, and hence the island is timelike separated from the radiation region $\Sigma_{\text{rad}}$.\footnote{Further note that the neglected QES (\ref{eq:QEShalf2}) is also timelike separated from $\Sigma_{\text{rad}}$.} While computing the entanglement entropy of an interval between timelike separated points is not unreasonable, such entropies have been shown to lead to bag-of-gold and strong subadditivity paradoxes \cite{Anous:2020lka,Chen:2020tes}. Moreover, our derivation of the QES prescription only applies for an interval between spacelike separated points, and therefore, as in  \cite{Hartman:2020khs}, we neglect such scenarios. Consequently, there are no non-trivial islands spacelike separated from $\Sigma_\text{rad}$ to consider.

\subsection*{QESs in full reduction}

We now turn to the $\text{dS}_{2}$ geometry found via full reduction, where we place the endpoint $(U_{1},V_{1})$ inside the hyperbolic patch coinciding with the black hole interior, whilst fixing the endpoint $(U_{2},V_{2})$ in the neighboring hyperbolic patch (future blue region in Figure \ref{fig:dS2fullred}).
To move the point $(U_{1},V_{1})$ into the other hyperbolic patch we employ the continuation (\ref{eq:continuedcoordUVapp}), such that the generalized entropy is now
\beq S_{\text{gen}}=\frac{1}{4G}\left(\phi_{0}-\phi_{r}\left(\frac{1+\frac{U_{1}V_{1}}{L^{2}}}{1-\frac{U_{1}V_{1}}{L^{2}}}\right)+\frac{Gc}{3}\right)+\frac{c}{12}\log\left[\frac{16L^{4}}{\delta_{1}^{2}\delta_{2}^{2}}\frac{(L^{2}+U_{1}U_{2})^{2}(L^{2}+V_{1}V_{2})^{2}}{(L^{2}-U_{1}V_{1})^{2}(L^{2}-U_{2}V_{2})^{2}}\right]\;,\eeq
where we point out the relative minus sign in front of $\phi_{r}$ in the ``area'' term. It is worth noting that the quantum state of matter is in the vacuum with respect to global coordinates of the full space $(\sigma,\varphi)$ (\ref{eq:globalcoordapp}). This vacuum state is still the Bunch--Davies vacuum, which follows from the fact the continuation (\ref{eq:continuedcoordUVapp}) leaves the line element invariant. 

\begin{figure}[t!]
\centering
\begin{tikzpicture}[scale=1.8]
    \draw (0,2) coordinate (a) -- (0,0) coordinate (b) -- (2,2) coordinate (c) -- (4,0) coordinate (m) -- (6,2) coordinate (d) -- (8,0) coordinate (e) -- (8,2) coordinate (f);
    \draw[decorate, decoration={snake, amplitude=0.5mm, segment length=2.5mm}] (a) -- (c); 
    \draw[decorate, decoration={snake, amplitude=0.5mm, segment length=2.5mm}] (d) -- (f); 
    \draw[dashed] (c) --++(-90:2);
    \draw[dashed] (d) --++(-90:2);
    \draw (c) --++(0:0.5) coordinate (L) node[pos=1, above] {$P_L$};
    \draw (d) --++(180:0.5) coordinate (R) node[pos=1, above] {$P_R$};
    \draw (L) -- (R) node[pos=.5, above] {$\Sigma_{\rm rad}$};
    \draw[very thick, red] (L) -- (R);
    \draw[thick, red] (0,1.7) --++(0:1.25) node[pos=.5,above] {{\small\color{black} $I$}} node[pos=.9, below] {\color{black} $Q_L$} coordinate (QL);
    \draw[thick, red] (8, 1.7) --++(180:1.25) node[pos=.5, above] {{\small\color{black} $I$}} node[pos=.9, below] {\color{black}$Q_R$} coordinate (QR);
    \draw[thick, Purple] (QL) -- (L); 
    \draw[thick, Purple] (QR) -- (R);
    \path[name path = QL1] (L) --++(225:2);
    \path[name path = QL2] (QL) --++(-45:2);
    \path[name path = sing1] (L) --++(180:2);
    \path[name path = sing2] (QL) --++(45:1);
    \fill[Periwinkle, fill opacity=0.5, name intersections={of=QL1 and QL2, by={intL}}, name intersections={of=sing1 and sing2, by={intsing1}}] (QL) -- (intL)-- (L) -- (intsing1) -- (QL);
    \path[name path = QR1] (R) --++(-45:2);
    \path[name path = QR2] (QR) --++(-135:2);
    \path[name path = sing3] (R) --++(0:2);
    \path[name path = sing4] (QR) --++(135:1);
    \fill[Periwinkle, fill opacity=0.5, name intersections={of=QR1 and QR2, by={intR}}, name intersections={of=sing3 and sing4, by={intsing2}}] (QR) -- (intR)-- (R) -- (intsing2) -- (QR);
    \filldraw[black] (QR) circle (0.02cm);
    \filldraw[black] (QL) circle (0.02cm);
    \filldraw[black] (L) circle (0.02cm);
    \filldraw[black] (R) circle (0.02cm);
\end{tikzpicture}
\caption{Islands in two-dimensional de Sitter in the full reduction. When an island $I$ is included, the entanglement wedge of radiation is the causal development of $\Sigma_{\text{rad}}\cup I$. Since the global vacuum state is pure, one instead computes the entanglement entropy of the complement $(\Sigma_{\text{rad}}\cup I)^{c}$, the two intervals $[Q_{L},P_{L}]\cup[P_{R},Q_{R}]$ (purple). The entanglement wedge of the complement is given by two rectangular causal diamonds (blue).}
\label{fig:islandsds2full} 
\end{figure}
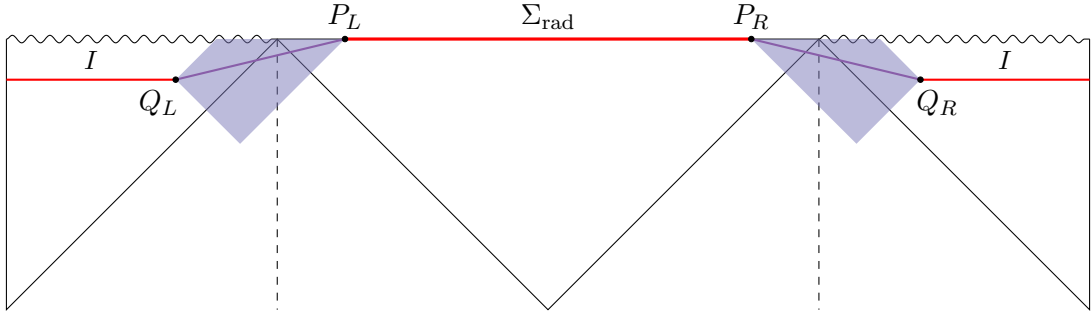

Varying with respect to endpoint $(U_{1},V_{1})$ while keeping $(U_{2},V_{2})$ fixed, we find two possible locations for a QES. The first one is
\beq V_{1}\approx \frac{U_{2}\epsilon}{3}+\mathcal{O}(\epsilon^{2})\;,\quad U_{1}\approx \frac{V_{2}\epsilon}{3}+\mathcal{O}(\epsilon^{2})\;,\label{eq:QESfull1}\eeq
and the second is
\beq V_{1}\approx -\frac{L^{2}}{V_{2}}+\left(\frac{2L^{2}}{3V_{2}}-\frac{U_{2}}{3}\right)\epsilon+\mathcal{O}(\epsilon^{2})\;,\quad U_{1}\approx -\frac{L^{2}}{U_{2}}+\left(\frac{2L^{2}}{3U_{2}}-\frac{V_{2}}{3}\right)\epsilon+\mathcal{O}(\epsilon^{2})\;.
\label{eq:QESfull2}\eeq
When the point $(U_{2},V_{2})$ lives near $\mathcal{I}^{+}$, it is straightforward to show the second solution (\ref{eq:QESfull2}) is timelike separated from the radiation region, and for the reasons described above, we neglect such a solution. The first solution (\ref{eq:QESfull1}) is located near the black hole singularity, and thus the associated island is spacelike separated from $\Sigma_{\text{rad}}$ (see Figure \ref{fig:islandsds2full}). In static patch coordinates, the QES (\ref{eq:QESfull1}) is
\beq r_{\text{QES}}\approx -L+\frac{2L}{9}\left(\frac{L-r_{2}}{L+r_{2}}\right)\epsilon^{2}\;.\eeq
This QES is the same one uncovered in \cite{Hartman:2020khs}. In the classical limit, the QES lies at the black hole horizon $r=-L$, while for $\epsilon\neq0$ and $r_{2}\to\infty$, the value of the dilaton at this QES is
\beq \phi(r_{\text{QES}})=-\frac{\phi_{r}}{L}r_{\text{QES}}+\frac{Gc}{3}=-\phi_{r}\sqrt{1+\left(\frac{2\epsilon}{3}\right)^{2}}+\frac{Gc}{3}\;.\eeq
 As in the half reduction model, $S_{\text{gen}}(r=r_{\text{QES}})<S_{\text{gen}}(r=-L)$, consistent with the QES formula.

\begin{figure}[t!]
\centering
\begin{tikzpicture}[scale=1.8]
    \draw (0,2) coordinate (a) -- (0,0) coordinate (b) -- (2,2) coordinate (c) -- (4,0) coordinate (m) -- (6,2) coordinate (d) -- (8,0) coordinate (e) -- (8,2) coordinate (f);
    \draw[decorate, decoration={snake, amplitude=0.5mm, segment length=2.5mm}] (a) -- (c); 
    \draw[decorate, decoration={snake, amplitude=0.5mm, segment length=2.5mm}] (d) -- (f); 
    \draw[dashed] (c) --++(-90:2);
    \draw[dashed] (d) --++(-90:2);
    \draw (c) --++(0:1) coordinate (L) node[pos=1, above] {$P_L$};
    \draw (d) --++(180:1) coordinate (R) node[pos=1, above] {$P_R$};
    \fill[Periwinkle, opacity=0.5] (L) --++(-45:1.41) -- (R);
    \draw (L) -- (R) node[pos=.5, above] {$\Sigma_{\rm rad}$};
    \draw[very thick, red] (L) -- (R);
    \filldraw[black] (L) circle (0.02cm);
    \filldraw[black] (R) circle (0.02cm);
\end{tikzpicture}
\caption{Causal diamond associated with computing semi-classical entanglement entropy of $\Sigma_{\text{rad}}$ in the absence of an island.} 
\label{fig:islandds2fullhawk} 
\end{figure}
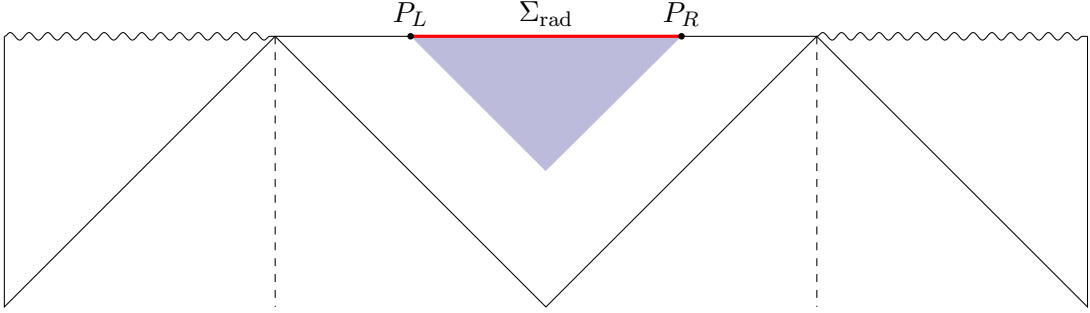

Note that the radiation, modeled by $\chi$, is in the Bunch--Davies vacuum, a pure state. Hence, $S_{\text{vN}}(\Sigma_{\partial I})=S_{\text{vN}}(\Sigma_{\partial I}^{c})$, where $\Sigma_{\partial I}^{c}=(\Sigma_{\text{rad}}\cup I)^{c}$ is the complement of $\Sigma_{\partial I}=\Sigma_{\text{rad}}\cup I$, and in practice we therefore compute $S_{\text{vN}}(\Sigma_{\partial I}^{c})$ using the island formula. Thus we consider the microcanonical action of the complement of the entanglement wedge of radiation:   the union of the domain of dependence of achronal surfaces $\Sigma$ and $\Sigma'$ with boundaries $\partial\Sigma=\mathcal{B}\cup P_{L}$ and $\partial\Sigma'=P_{R}\cup\mathcal{B}'$, respectively. Upon extremizing the microcanonical action, the point $\mathcal{B}$ becomes the QES $Q_{L}$, and similarly $\mathcal{B}'=Q_{R}$, such that we compute the entropy of the two intervals $[Q_{L},P_{L}]\cup[P_{R},Q_{R}]$, and the island is $I=[Q_{L},Q_{R}]$ with boundary $\partial I=Q_{R}\cup Q_{L}$ (Figure \ref{fig:islandsds2full}). We thus evaluate the on-shell microcanonical action for two identical causal diamonds with edges $\partial\Sigma$ and $\partial\Sigma'$. In the previous subsection we showed the on-shell (Euclidean) microcanonical action of a causal diamond in semiclassical JT gravity is  equal to minus the generalized entropy of the diamond. Even though the causal diamonds in Figure \ref{fig:islandsds2full} include the black hole singularity, this does not pose a problem to evaluating the microcanonical action of the diamonds. The action is on-shell given by a boundary term, i.e. ($\beta$ times) the Noether charge of the  edge $\partial \Sigma$,   which is not close to the black hole singularity.

In an appropriate OPE limit, the von Neumann entropy of the CFT factorizes, such that we may treat each causal diamond separately. Consequently, the total island entropy is, to leading order in $U_{2}V_{2}\approx L^{2}$,
\beq  \label{eq:entropyqes}
\begin{aligned} 
S_{\text{vN}}(\Sigma_{\partial I})&=S_{\text{gen}}(Q_{L})+S_{\text{gen}}(Q_{R})\\
&\approx \frac{2}{4G}\left(\phi_{0}+\phi(r_{\text{QES}})\right)+\frac{c}{3}\log\left[\frac{4L^{4}}{\delta_{1}\delta_{2}(L^{2}-U_{2}V_{2})}\right]+\frac{2c\epsilon}{9}\;,
\end{aligned}
\eeq
consistent with the result found in \cite{Hartman:2020khs}. Here we used $S_{\text{gen}}|_{\partial\Sigma}=S_{\text{gen}}(Q_{L})$, and similarly for $S_{\text{gen}}|_{\partial\Sigma'}$, since   $P_{R,L}$ belong to the weakly gravitating region, such that we ignore the contribution to the dilaton, and where $\chi=0$, due to the Dirichlet boundary condition imposed on $\chi$ (see footnote \ref{fn:dirichletbc}) \cite{Pedraza:2021cvx}. 


Note that the entropy (\ref{eq:entropyqes}) is constant with respect to time $t$ or equivalently ``length'' $X=\frac{L}{2}\log(V/U)$ in coordinates (\ref{eq:nenewwecoord}). This is analogous to the behavior of the entropy of radiation emitted by an eternal $\text{AdS}_{2}$ black hole in the island phase \cite{Almheiri:2019yqk}. When the island is taken to be the empty set, however, one neglects the dilaton and the von Neumann entropy of radiation is given by the semi-classical entanglement entropy, which is found to grow linearly in $X$ at large $X$ \cite{Hartman:2020khs}. The linear growth also follows from extremizing the microcanonical action, where now there is only a single (square) causal diamond with edges $\partial\Sigma=P_{L}\cup P_{R}$ (Figure~\ref{fig:islandds2fullhawk}). Specifically, similar to \cite{Pedraza:2021ssc},  extremization of $S_{\text{vN}}^{\text{sc}}$ for a single interval contained in the expanding hyperbolic patch yields  $U_{L}=V_R$ and $V_{L}=U_{R}$ (such that $T_{R}=T_{L}\equiv T$ and $X_{R}=-X_{L}\equiv X$), and $X \gg L$ and $T \ll L$. Substituting these conditions into 
the semi-classical entanglement entropy gives
\beq S_{\text{vN}}^{\text{sc}}(\Sigma_{\text{rad}})=\frac{c}{3}\log\left(\frac{2L^{2}(V_{R}-U_{R})}{\delta (L^{2}-U_{R}V_{R})}\right)=\frac{c}{3} \log \left ( \frac{2L}{\delta} \frac{\sinh X}{\sinh (-T)}\right)\approx \frac{c}{3} \frac{X}{L}+ ...\;,\label{eq:svnrad}\eeq
where we expanded for $X \gg L$ to obtain the   linear growth in $X$, analogous to the ``Hawking phase'' for black hole radiation, and  in agreement with \cite{Hartman:2020khs}. There exists a critical length, the ``Page length'' $X_{\text{P}}$ at which the island entropy (\ref{eq:entropyqes}) equals the semi-classical entanglement entropy (\ref{eq:svnrad}):
\beq (V_{P}-U_{P})\approx\frac{2L^{2}}{\delta_{1}}e^{\frac{6}{4Gc}[\phi_{0}+\phi(r_{\text{QES}})]}\;,\;\;
\text{or}
\;\;\; X_{\text{P}}\approx L\text{arcsinh}\left(\frac{L}{\delta_{1}}e^{\frac{6}{4Gc}[\phi_{0}+\phi(r_{\text{QES}})]}\right)\;.\eeq
A global minimization of entropies (\ref{eq:svnrad}) and (\ref{eq:entropyqes}) reveals a transition occurring at this length, analogous to the transition seen in the Page curve for eternal $\text{AdS}_{2}$ black holes \cite{Almheiri:2019yqk}.

\section{Discussion}\label{sec:conclusion}

In this article we explored thermodynamic and microscopic aspects of two-dimensional de Sitter space using semi-classical de Sitter JT gravity. Specifically, we extended the quasi-local analysis of York to the case of $\text{dS}_{2}$ by introducing an auxiliary timelike boundary in the static patch that interpolates between the black hole and cosmological horizons. With this timelike boundary we were able to properly define conserved charges, namely the energy, and uncovered a quasi-local first law of thermodynamics. Backreaction due to quantum matter is fully incorporated via the 1-loop Polyakov action, leading to a semi-classical extension of the quasi-local first law of thermodynamics, where the classical Gibbons--Hawking entropy is replaced by the generalized entropy. Crucial to this extension was the observation that in two dimensions the semi-classical Wald entropy is \emph{exactly} equal to the generalized entropy, where the semi-classical contribution arises from expressing the Polyakov action in a localized form. Including semi-classical backreaction, the first law of horizon thermodynamics was modified such that the classical entropy is replaced by the generalized entropy. This is expected to be a feature for systems which include backreaction; indeed, the same modification appears in the three-dimensional semi-classical Schwarzschild-de Sitter black hole \cite{Emparan:2022ijy}.

Further, in the microcanonical ensemble, we found that the generalized entropy is equal to the microcanonical entropy, whose stationarity condition   implies extremizing the generalized entropy, similar to recent results for eternal $\text{AdS}_{2}$ black holes \cite{Pedraza:2021cvx}. This observation suggests a first principles derivation of the QES formula \cite{Pedraza:2021ssc} in $U(1)$ symmetric backgrounds (alternative to previous derivations invoking the replica trick) which we have extended to the case of de Sitter JT gravity. Thus, we provided evidence that the QES and island prescriptions hold beyond $\text{AdS}_{2}$ systems. The crucial new insight is that the on-shell microcanonical action of (Euclidean) causal diamonds computes the generalized entropy, whose extremization follows from the minimization of the action. This leads to the appearance of quantum extremal islands in the full reduction model of JT gravity, consistent with \cite{Hartman:2020khs}, where the island lives near the singularity of the black hole.

There are a number of exciting prospects of our work, which we have only eluded to thus far. Let us discuss them now in some detail. 

\paragraph{De Sitter holography.} The Gibbons--Hawking entropy formula suggests the microscopic description of de Sitter space obeys the holographic principle. That is, the  putative  dual quantum theory accounting for the underlying microscopics of dS lives on a holographic screen. Evidence is mounting that dS holography is strikingly different from AdS/CFT holography (cf. \cite{Leuven:2018ejp}). For example, the number of degrees of freedom increases in the IR direction of the microscopic theory, indicating the dual quantum theory description of dS is unlikely to be a local quantum field theory. Additional evidence that the underlying microscopics of de Sitter space is not well characterized by a local quantum field theory has been given via matrix model descriptions of $\text{dS}_{2}$ \cite{Anninos:2021eit}. Further, the UV/IR connection for dS appears to be inverted: long distances (IR) in the bulk correspond to low energies (IR) in the microscopic theory. This is consistent with the worldline holography proposed in \cite{Anninos:2011af} (see also \cite{Anninos:2017hhn}) where the UV theory lives on a surface near the origin $r=0$, in contrast with AdS/CFT where the UV description lies on the conformal boundary. 

The differences between dS holography and AdS/CFT are further exemplified in the way entanglement entropy of the dual quantum mechanical theory is computed using bulk quantities. The Ryu--Takayanagi entropy formula says that the entanglement entropy of a CFT state restricted to a boundary subregion is equal to the area of the (bulk AdS) extremal surface anchored at the endpoints of the boundary subregion. In contrast, it was recently proposed that in de Sitter space the extremal surface whose area computes the entanglement entropy is anchored between the two stretched
 horizons   where the holographic degrees of freedom reside \cite{Susskind:2021dfc,Susskind:2021esx,Shaghoulian:2021cef,Shaghoulian:2022fop} (see also \cite{Sanches:2016sxy,Nomura:2017fyh}). Thus, the UV boundary of AdS is replaced by the IR boundary of the static patch, compatible with the aforementioned worldline holography. 

A toy quantum mechanical model which exhibits these features of static patch holography has been conjectured to be the SYK model in the ``hyperfast'' limit \cite{Susskind:2021esx}. This is because the holographic degrees of freedom of the cosmological horizon are hyperfast scramblers, scrambling on a time scale equal to the de Sitter radius, implying the complexity growth is hyperfast. 
In the SYK model, the hyperfast scrambling property is a consequence of taking the infinite temperature limit of SYK, such that the temperature is greater or equal to the fundamental energy scale of SYK. 

The transition from low to high temperature in the dual bulk (AdS JT gravity) picture suggests a connection with the quasi-local thermodynamics studied here. At low temperature, the boundary bends slightly inward toward the horizon \cite{Maldacena:2016upp}, while at high temperature the boundary nearly coincides with the horizon, such that the holographic boundary degrees of freedom become horizon quasinormal modes. The timelike screen one introduces to study quasi-local thermodynamics interpolates between the UV $(r=0)$ surface and the IR static patch boundary (stretched horizon), and shifting its position
may be capturing this low to high temperature transition of SYK. It would be interesting to pursue this connection further and see whether the quasi-local thermodynamics of dS JT gravity provides insights into hyperfast SYK, and vice versa.

Furthermore, the   semi-classical thermodynamics may deepen our understanding of entanglement entropy in de Sitter space. For example, according to the proposals of \cite{Susskind:2021dfc,Shaghoulian:2021cef,Shaghoulian:2022fop}, the Gibbons--Hawking entropy of pure dS is identified with the entanglement entropy between modes living on the left and right horizons, $S_{\text{ent}}=S_{\text{GH}}$. Similarly, the entanglement entropy between left and right sides in a Schwarzschild-de Sitter background   is given by the sum of the gravitational entropies of the black hole and cosmological horizon, $S_{\text{ent}}=S_{\text h}+S_{\text c}$. Therefore, thermodynamic relations directly translate into relations for the entanglement entropy. Our global first law for dS$_2$ \eqref{eq:globalfirstlawscfull} suggests that when quantum matter is included, the total semi-classical entropy of SdS is $S_{\text{ent}}=S_{\text{gen,h}}+S_{\text{gen,c}}$. Moreover, semi-classical corrections affect  the probability of creating a black hole in de Sitter, which in semi-classical gravity should be $P \sim \exp(-\Delta S)$ with entropy deficit $\Delta S= S_{\text{gen,dS}} - S_{\text{gen,h}} - S_{\text{gen,c}}$, where the last term is the generalized entropy of  pure dS. This semi-classical modification of the entropy deficit also features in the three-dimensional semi-classical Schwarzschild-de Sitter black hole \cite{Emparan:2022ijy}. 

\paragraph{Quasi-local thermodynamics and $T\bar{T}$ deformations in $\text{dS}_{2}$.} It is well known that finite cutoff holography in $\text{AdS}_{3}$ is dual to  $T\bar{T}$ deformations of a holographic CFT, where $T\bar{T}$ is related to the trace of the quasi-local Brown-York stress tensor. Further, AdS JT gravity with a finite cutoff is precisely described by a Schwarzian theory deformed by the one-dimensonal analog of $T\bar{T}$ \cite{Iliesiu:2020zld}, providing evidence that $T\bar{T}$ deformations in a holographic theory correspond to moving the conformal boundary to a finite radial distance in bulk AdS. In three (bulk) dimensions, these deformations were generalized to $T\bar{T}+\Lambda_{2}$ deformations of the CFT to reconstruct patches in three-dimensional de Sitter space \cite{Gorbenko:2018oov}, and were recently used to provide a microstate counting interpretation of the Gibbons--Hawking entropy of global de Sitter space \cite{Coleman:2021nor}. In particular, one constructs microstates of the patch containing the cosmological horizon from the (dressed) microstates of the BTZ black hole at a particular energy level. This ``cosmic horizon patch" is defined as the region between the cosmological horizon and a timelike boundary $B$. This picture suggests, holographically, that the $T\bar{T}$ deformation corresponds to the movement of a holographic screen in the bulk. The quasi-local thermodynamics studied here may shed light on the one-dimensional analog of $T\bar{T}$ deformations to the dual quantum mechanical theory.

\paragraph{Microcanonical action and multiverse models.} We found that the von Neumann entropy of radiation collected at $\mathcal{I}^{+}$ in global $\text{dS}_{2}$ during the island phase is equivalent to evaluating the microcanonical action on two finite causal diamonds (Figure \ref{fig:islandsds2full}). Recently, JT de Sitter multiverses, where global $\text{dS}_{2}$ is extended,  have been used as toy models to study false vacuum decay in inflationary universes with multiple vacua. For sufficiently large radiation subregions $\Sigma_{\text{rad}}$, the fine grained entropy of radiation, captured using the island rule, leads to an analogous Page-like transition \cite{Aguilar-Gutierrez:2021bns}. Crucial to this analysis is the assumption the island formula holds for the extended $\text{dS}_{2}$ multiverse. However, it is not immediately clear how reasonable this assumption is as a Euclidean description of JT multiverses is currently lacking, as is the replicated manifold necessary for the replica trick.\footnote{See, however, \cite{Balasubramanian:2020coy}, where the replica method is used to compute the entanglement entropy between two disjoint universes, one gravitating and one non-gravitating.}

An appealing feature of our first principles derivation of the QES prescription is that it naturally applies to these multiverse models. We find an island develops and covers nearly the entire $\text{dS}_{2}$ multiverse, consistent with \cite{Aguilar-Gutierrez:2021bns}. To carry out the analysis explicitly, one must consider causal diamonds in extended $\text{dS}_{2}^{n}$, which globally has the same line element and static dilaton solution as in (\ref{eq:globalcoordmetapp}), except where the coordinate $\varphi\in(0,2\pi n)$ for integer $n$. The Euclidean diamond universe in $\text{dS}_{2}^{n}$, which the microcanonical action is evaluated over, is given by the Euclidean continuation of $\text{dS}_{2}^{n}$, modulo the horizon bifurcation points. The calculation then proceeds as before (\ref{eq:micactionCDGHY}), where the microcanonical action is equal to the Gibbons--Hawking--York boundary term evaluated on the boundary of an infinitesimal disk surrounding the punctures. This is reminiscent of the observation made in \cite{Aguilar-Gutierrez:2021bns}: an island always forms for sufficiently large $\Sigma_{\text{rad}}$, independent of the global geometry beyond $\Sigma_{\text{rad}}$. This suggests the global spacetime does not capture fundamental degrees of freedom independent from those in $\Sigma_{\text{rad}}$, \emph{i.e.}, there is a redundancy in the rest of the global spacetime. 

\paragraph{Dynamical backgrounds and beyond two dimensions.} Our derivation of the QES prescription only applies to two-dimensional static backgrounds. It is natural to wonder whether we can extend this work to higher-dimensional and dynamical settings.\footnote{For example, one can consider the quantum matter to be in the Unruh--de Sitter state \cite{Aalsma:2019rpt}, in which case the horizon evaporates. In the latter case, it appears that catastrophic backreaction occurs at the Page time, preventing any observer from seeing a unitary Page curve \cite{Aalsma:2021bit,Kames-King:2021etp}.} The derivation  $I_{\text{E}}^{\text{mc}}=-S_{\text{Wald}}$ for causal diamonds is actually valid for any diffeomorphism invariant theory in any dimension \cite{Pedraza:2021ssc}. It is unclear, however, whether the connection $S_{\text{Wald}}=S_{\text{gen}}$ holds beyond two dimensions; in fact it seems unlikely since here we used the fact that the Polyakov action is 1-loop exact in two dimensions. Nonetheless, it would be interesting to see what one could glean from the Wald entropy associated with the anomaly induced action for four-dimensional general relativity. Finally, our results reasonably apply in equilibrium settings, and therefore we cannot comment on Page curves in dynamical backgrounds, including the $\text{dS}_{2}$ scenarios described in \cite{Sybesma:2020fxg,Aalsma:2021bit,Kames-King:2021etp,Teresi:2021qff}. It would be interesting to see whether one can express the dynamical black hole entropy proposal of \cite{Iyer:1994ys} as a ``microcanonical" action, and apply it to diamonds.  This could also prove useful to extend the de Sitter holographic entropy proposals in \cite{Susskind:2021dfc,Susskind:2021esx,Shaghoulian:2021cef,Shaghoulian:2022fop} to dynamical setups.  We leave this for future exploration.

\noindent\section*{Acknowledgments}

We would like to thank Batoul Banihashemi, Ted Jacobson, Edward Morvan,  Juan  Pedraza, Watse Sybesma, and Jan Pieter van der Schaar for helpful and illuminating discussions.  AS is supported by the Simons Foundation through \emph{It from Qubit: Simons Collaboration on Quantum fields, gravity, and information}, and EPSRC. MV is supported by the Republic and canton of Geneva and the Swiss National Science Foundation, through Project Grants No. 200020-182513 and No. 51NF40-141869 The Mathematics of Physics (SwissMAP). EV and EPV are supported by the Spinoza grant and the Delta ITP consortium, a program of the NWO funded by the Dutch Ministry of Education, Culture and Science (OCW), and the work of EV is part of the research programme of the Foundation for Fundamental Research on Matter (FOM), which is financially supported by NWO.

\appendix

\section{Reductions and coordinate systems} \label{app:reductionscoords}

Here we provide details of the half and full spherical reductions of the $d$-dimensional Einstein--Hilbert action (\ref{eq:ddimaction}), leading to two different models of de Sitter JT gravity. 

\section*{Half reduction}

For the half reduction, consider the metric ansatz
\beq d\ell^{2}=\hat{g}_{MN}dX^{M}dX^{N}=g_{\mu\nu}(x)dx^{\mu}dx^{\nu}+L^{2}_{d}\Phi^{2/(d-2)}(x)d\Omega_{d-2}^{2}\;.\label{eq:dimansatzapp}\eeq
Here $M,N=0,1,...,d-1$, $\mu,\nu=0,1$, and $\Phi(x)$ is the dilaton.  A standard calculation using Cartan's structure equations (see \emph{e.g.} \cite{Grumiller:2001ea,Narayan:2020pyj}) shows the $d$-dimensional Ricci scalar is\footnote{Our conventions differ slightly from \cite{Grumiller:2001ea}. Namely, our metric Ansatz takes a more convenient form and we work in the ``mostly plus'' Lorentzian signature.}
\beq \hat{R}=R+\frac{(d-3)(d-2)}{L^{2}_{d}\Phi^{2/(d-2)}}+\frac{(d-3)}{(d-2)}\frac{1}{\Phi^{2}}(\nabla\Phi)^{2}-\frac{2}{\Phi}\Box\Phi\;,\label{eq:Riccscalred}\eeq
where $R$ is the Ricci scalar and $\nabla$ the covariant derivative with respect to the two-dimensional metric $g_{\mu\nu}$. It is also straightforward to show the $(d-1)$-dimensional extrinsic curvature $\hat{K}$ reduces to 
\beq \hat{K}=K+\frac{1}{\Phi}n^{\mu}\nabla_{\mu}\Phi\;,\label{eq:Kred}\eeq
where $K$ is the extrinsic curvature of the 1-dimensional boundary and $n^{\mu}$ is the normal vector used to define the $(d-1)$-dimensional boundary metric. We further have
\beq \int_{\hat{M}}\hspace{-2mm}d^{d}X\sqrt{-\hat{g}}=L_{d}^{(d-2)}\Omega_{d-2}\int_{\mathcal{M}}\hspace{-2mm}d^{2}x\sqrt{-g}\,\Phi\label{eq:detred}\;,\eeq
where $\mathcal{M}$ is the two-dimensional Lorentzian manifold endowed with metric $g_{\mu\nu}$. Substituting the dimensionally reduced scalar curvatures (\ref{eq:Riccscalred}) and (\ref{eq:Kred}), and (\ref{eq:detred}) into the $d$-dimensional Einstein--Hilbert action results in the following two-dimensional dilaton theory of gravity,
\beq
\begin{split}
I_{d}&=\frac{1}{16\pi G_{2}}\int_{\mathcal{M}}\hspace{-2mm}d^{2}x\sqrt{-g}\biggr(\Phi R-2\Lambda\Phi+\frac{(d-3)(d-2)}{L^{2}_{d}}\Phi^{\frac{(d-4)}{(d-2)}}+\frac{(d-3)}{(d-2)}\frac{(\nabla\Phi)^{2}}{\Phi}\biggr)\\
&+\frac{1}{8\pi G_{2}}\int_{\partial\mathcal{M}}\hspace{-3mm}dy\sqrt{-h}\Phi K\;,
\end{split}
\label{eq:redactgen}\eeq
where we performed an integration by parts on the $\Box\Phi$ term in the two-dimensional ``bulk'' integral, which cancels against an identical  term in the GHY integral, and we defined the two-dimensional Newton's constant as
\begin{equation}
    \frac{1}{G_2} \equiv \frac{L_d^{(d-2)}\Omega_{d-2}}{G_d}\,.
\end{equation} 
We notice a dramatic simplification when $d=3$:
\beq I_{\text{JT}}=\frac{1}{16\pi G_{2}}\int_{\mathcal{M}}\hspace{-2mm}d^{2}x\sqrt{-g}\,\Phi\!\left(R-\frac{2}{L_{3}^{2}}\right)+\frac{1}{8\pi G_{2}}\int_{\partial\mathcal{M}}\hspace{-3mm}dy\sqrt{-h}\Phi K\;,\label{eq:JTactv1app}\eeq
where   $1/G_{2}=2\pi L_{3}/G_{3}$. Importantly, here the dilaton $\Phi$ only takes on positive values.

\section*{Full reduction}

When $d=3$ the dimensionally reduced action (\ref{eq:redactgen}) greatly simplifies since the potential and kinetic terms drop out. For $d>3$ this is no longer the case. For $d>3$ we can remove the kinetic term via an appropriate Weyl rescaling of the two-dimensional metric, such that a spherical reduction of the near-Nariai solution in  $d$-dimensions leads to another form of de Sitter JT  gravity.\footnote{We thank Watse Sybesma for discussions on this and for sharing notes on the spherical reduction in $d=4$.} 
This is accomplished by first modifying the metric ansatz (\ref{eq:dimansatzapp}) to
\beq ds_{d}^{2}=\hat{g}_{MN}dX^{M}dX^{N}=g_{\mu\nu}(x)dx^{\mu}dx^{\nu}+r_{\text{N}}^{2}\Phi^{2/(d-2)}(x)d\Omega_{d-2}^{2}\;,\label{eq:dimansatz2}\eeq
which reduces to the Nariai geometry  for $\Phi=1$.  Going through the same steps as above, we arrive at the same form of the action (\ref{eq:redactgen}), except   with $L_{d}\to r_{\text{N}}$ and where we have now identified the dimensionless two-dimensional Newton's constant $G_{2}$ as
\beq \frac{1}{G_{2}}\equiv\frac{\Omega_{d-2}r_{\text{N}}^{d-2}}{G_{d}}\;.\eeq
Next, recall how the Ricci scalar and trace of the extrinsic curvature  transform under the Weyl rescaling $\bar{g}_{\mu\nu}=\omega^{2}g_{\mu\nu}$ in a two-dimensional spacetime\footnote{To see how the conformal transformation of $\bar{K}$ arises, first note the induced metric transforms as $\bar{h}_{\mu\nu}=\omega^{2}g_{\mu\nu}$, such that $\bar{n}_{\mu}=\omega n_{\mu}$. Then, 
$\bar{\nabla}_{\mu}\bar{n}_{\nu}=\omega \nabla_{\mu}n_{\nu}-(n_{\nu}\partial_{\mu}\omega-n_{\rho}g_{\mu\nu}g^{\rho\delta}\partial_{\delta}\omega)$
where we used $\bar{\Gamma}^{\rho}_{\;\mu\nu}=\Gamma^{\rho}_{\;\mu\nu}+\omega^{-1}(\delta^{\rho}_{\mu}\partial_{\nu}\omega+\delta^{\rho}_{\nu}\partial_{\mu}\omega-g_{\mu\nu}g^{\rho\delta}\partial_{\delta}\omega)$. Using $\bar{K}=\bar{g}^{\mu\nu}\bar{\nabla}_{\mu}\bar{n}_{\nu}$, the transformation rule follows.}
\beq \bar{R}=\omega^{-2}R-2\omega^{-3}\Box\omega+2\omega^{-4}(\nabla\omega)^{2}\;,\quad \bar{K}=\omega^{-1}K+\omega^{-2}n_{\mu}\nabla^{\mu}\omega\;.\eeq
Then, rescaling $g_{\mu\nu}\to \omega^{2}g_{\mu\nu}$, the reduced action (\ref{eq:redactgen}) becomes
\beq
\begin{aligned}
I_{d}&=\frac{1}{16\pi G_{2}}\int_{\mathcal{M}}\hspace{-2mm}d^{2}x\sqrt{-g}\biggr[\Phi R-2\omega^{-1}\Phi\Box\omega+2\omega^{-2}\Phi(\nabla\omega)^{2}+\frac{(d-3)(d-2)\omega^{2}}{r^{2}_{\text{N}}}\Phi^{\frac{(d-4)}{(d-2)}}\\
&-2\Lambda\Phi\omega^{2}+\frac{(d-3)}{(d-2)}\frac{1}{\Phi}(\nabla\Phi)^{2}\biggr]+\frac{1}{8\pi G_{2}}\int_{\partial\mathcal{M}}\hspace{-3mm}dy\sqrt{-h}(\Phi K+\omega^{-1}\Phi n_{\mu}\nabla^{\mu}\omega)\;.
\end{aligned}
\eeq
By an integration by parts, the $\Box\omega$ term is partially cancelled by the GHY integral and after simplifying we are left with\footnote{We thank Marija Toma\v{s}evi\'c for pointing out a typo in an intermediate step in our derivation. The final result, however, is unchanged from previous versions.}
\beq 
\begin{split}
I_{d}&=\frac{1}{16\pi G_{2}}\int \hspace{-1mm}d^{2}x\sqrt{-g}\biggr[\Phi R+2\omega^{-1}(\nabla^{\mu}\Phi)(\nabla_{\mu}\omega)+\frac{(d-3)}{(d-2)}\frac{1}{\Phi}(\nabla\Phi)^{2}-2\Lambda\Phi\omega^{2}\\
&+\frac{(d-3)(d-2)\omega^{2}}{r_{\text{N}}^{2}}\Phi^{\frac{(d-4)}{(d-2)}}\biggr]+\frac{1}{8\pi G_{2}}\int_{\partial\mathcal{M}}\hspace{-3mm}dy\sqrt{-h}\Phi K\;.
\end{split}
\eeq
We can eliminate the kinetic term by choosing $\omega=\alpha\Phi^{\beta}$ for $\beta=-(d-3)/2(d-2)$ and $\alpha$ some constant, which we judiciously  choose to be $\alpha=1/\sqrt{d-1}$. Then we have
\beq I_{d}=\frac{1}{16\pi G_{2}}\int_{\mathcal{M}}\hspace{-2mm}d^{2}x\sqrt{-g}[\Phi R+U(\Phi)]+\frac{1}{8\pi G_{2}}\int_{\partial\mathcal{M}}\hspace{-3mm}dy\sqrt{-h}\Phi K\;,\label{eq:redactnariai}\eeq
where we used $r_{\text{N}}=\sqrt{\frac{d-3}{d-1}}L_{d}$ in $\Lambda$. We also introduced the dilaton potential,
\beq U(\Phi)=\frac{(d-2)}{L_{d}^{2}}\left(\Phi^{-1/(d-2)}-\Phi^{1/(d-2)}\right)\label{eq:dilatonpotential2}\;.\eeq
which clearly obeys
\beq U(\Phi=1)=0\;,\quad \frac{dU}{d\Phi}\biggr|_{\Phi=1}=-\frac{2}{L_{d}^{2}}\;.\eeq
Then, expanding the reduced action (\ref{eq:redactnariai}) about $\Phi\approx \phi_{0}+\phi$ for $\phi_{0}=1$, we find to leading order  
\beq 
\begin{split}
I_{\text{JT}}&=\frac{1}{16\pi G_{2}}\int_{\mathcal{M}}\hspace{-2mm}d^{2}x\sqrt{-g}\left((\phi_{0}+\phi)R- \frac{2}{L_{d}^{2}} \phi\right)+\frac{1}{8\pi G_{2}}\int_{\partial\mathcal{M}}\hspace{-3mm}dy\sqrt{-h}(\phi_{0}+\phi)K\,,
\end{split}
\label{eq:JTactv2app}\eeq
Notice then $\phi_{0}$ is proportional to the entropy \eqref{eq:entropynariai} of the Nariai black hole:
\beq \frac{\phi_{0}}{4G_{2}}=\frac{\Omega_{d-2}r_{\text{N}}^{d-2}}{4G_{d}}=\frac{1}{2}S_{\text{N}}\;.\eeq
Thus, analogous to the case of JT gravity in AdS, $\phi$ represents deviations from the Nariai solution.

\section*{Coordinate systems}

Here we summarize various useful coordinates to describe two-dimensional de Sitter space (see also \cite{Maldacena:2019cbz,Kames-King:2021etp,Moitra:2022glw}).

\paragraph{Static patch.} In static patch coordinates $(t,r)$, the $\text{dS}_{2}$ line element and static dilaton take the form:
\beq d\ell^2=-\left(1-\frac{r^{2}}{L^{2}}\right)dt^{2}+\left(1-\frac{r^{2}}{L^{2}}\right)^{-1}dr^{2}\;,\quad \phi(r)=\phi_{r}\frac{r}{L} \;.\label{eq:statpatchapp}\eeq
Let $(v,u)$ denote advanced and retarded null coordinates for the static patch (\ref{eq:statpatchapp}), respectively defined by
 \beq v=t+r_{\ast}\;,\quad u=t-r_{\ast}\;,\eeq
with $r_{\ast}$ being the tortoise coordinate, 
 \beq r_{\ast}\equiv\int^{r}_0\frac{dr'}{1-\frac{r'^{2}}{L^{2}}}=L\text{arctanh}(r/L)\;.\eeq
 In the full reduction the ranges are $r \in [-L,L]$ and $r_{\ast}\in[-\infty,\infty]$ in the static patch, where $r_{\ast}=\infty$ $(r=L)$ is the location of the cosmological horizon    and $r_{\ast}=-\infty$ $(r=-L)$ is the location of the black hole horizon. In the half reduction the ranges are $r \in [0,L]$ and $r_{\ast}\in[0,\infty]$, where  $r_{\ast}=\infty$ and $r=L$ correspond to the location of the cosmological horizon. 
 
 In these null coordinates the static patch line element (\ref{eq:statpatchapp}) and dilaton become
 \beq d\ell^2=-\text{sech}^{2}\left(\frac{v-u}{2L}\right)dvdu\,,\;\quad \phi(u,v)=\phi_{r}\tanh\left(\frac{v-u}{2L}\right)\;.\label{eq:statnullmetapp}\eeq
 
\paragraph{Global conformal coordinates.} The full space of $\text{dS}_{2}$ is covered by global conformal coordinates $(\sigma,\varphi)$,  
 \beq \tan \varphi =\frac{1}{r}\sqrt{L^{2}-r^{2}}\cosh(t/L)\;,\quad \tan \sigma =\frac{1}{L}\sqrt{L^{2}-r^{2}}\sinh(t/L)\;,\label{eq:globalcoordapp}\eeq
 with line element and dilaton:
 \beq d\ell^2=\frac{L^2}{\cos^{2} \sigma }(-d\sigma^{2}+d\varphi^{2})\;,\quad \phi(\sigma, \varphi) =\phi_{r}\frac{\cos\varphi}{\cos\sigma}\;.\label{eq:globalcoordmetapp}\eeq
Here the ranges are $\sigma \in(-\pi/2,\pi/2)$, where future/past infinity corresponds  to $\sigma=\pm\pi  /2$, and  $\varphi \in (0,2\pi)$ for the full reduction model. In the half reduction model the metric takes the same form, however, the dilaton is given by $\phi=\phi_{r}\frac{\sin\varphi}{\cos\sigma}$, with $\varphi \in(0,\pi)$. 
 
\paragraph{Global coordinates.} In standard global coordinates the line element and dilaton are
\begin{equation} \label{eq:global2}
    d\ell^2 = - d\tau^2 + L^2 \cosh^2 (\tau/L) d \varphi^2\,, \qquad \phi (\tau, \varphi)= \phi_r \cos \varphi \cosh (\tau /L) \,,
\end{equation}
where $\tau \in(-\infty , \infty)$ and the range of $\varphi$  is the same as for the global conformal coordinates.
The  global coordinates $(\tau, \varphi)$ are related to the static patch coordinates ($t,r$) by 
\begin{equation}
    r = L \cosh (\tau /L) \sin \varphi \,,\qquad \sinh (t/L) = \frac{\sinh (\tau/ L)}{\sqrt{1 - \cosh^2 (\tau/ L) \sin^2 \varphi}}\,,
\end{equation}
and to global   conformal coordinates $(\sigma, \varphi)$ by
\begin{equation}
    \tan  (\sigma / 2) = \tanh (\tau / 2L)\,.
\end{equation}

\paragraph{Kruskal coordinates.} We introduce global Kruskal-like coordinates $(V,U)$ to cover the full two-dimensional geometry in the half reduction model
 \beq V=Le^{v/L}=Le^{t/L}\sqrt{\frac{L-r}{L+r}}\;,\quad U=-Le^{-u/L}=-Le^{-t/L}\sqrt{\frac{L-r}{L+r}}\;. \label{eq:kruskalnullcoordapp}\eeq
 The line element and dilaton become
 \beq d\ell^2=-\frac{4L^4}{\left(L^2-UV\right)^{2}}dVdU\;,\quad \phi(U,V)=\phi_{r}\left(\frac{L^2+UV}{L^2-UV}\right)\;.\label{eq:kruskalmetapp}\eeq
 In these coordinates, $UV=-L^{2}$ corresponds to the location of the poles $r=0$, while $UV=+L^{2}$ corresponds to the past and future conformal boundary $\mathcal{I}^{\pm}$. Moreover, the past (future) cosmological horizon is located at $V=0$ ($U=0$). 
 
 It is also useful to express Kruskal coordinates $(V,U)$ as
 \beq V=Le^{(T+X)/L}\;,\quad U=Le^{(T-X)/L}\;,\label{eq:nenewwecoord}\eeq
 such that the line element (\ref{eq:kruskalmetapp}) and static dilaton solution are
 \beq d\ell^2=\frac{1}{\sinh^{2}(T/L)}[-dT^{2}+dX^{2}]\;,\quad \phi(T)=-\phi_{r}\coth(T/L)\;.\label{eq:XTlineeleapp}\eeq
 In these coordinates $\mathcal{I}^{\pm}$ is located at $T=0$. The region to the future of the cosmological horizon is defined by $X\in\mathbb{R}$ and $T<0$, such that for $\phi_{r}>0$ the dilaton is strictly positive, diverging to $+\infty$ near $\mathcal{I}^{+}$. Coordinates $(T,X)$ are related to coordinates $(\sigma,\varphi)$ (\ref{eq:globalcoordapp})
 \beq \sigma=L\arctan \left[-\frac{\cosh(X/L)}{\sinh(T/L)}\right]\;,\quad \varphi=L\arctan \left[\frac{\sinh(X/L)}{\cosh(T/L)}\right]\;,\label{eq:coordglobTX}\eeq
 and to static coordinates by 
 \beq \coth(T/L)=-\frac{r}{L}\;,\quad \tanh(X/L)=\coth(t/L)\;.\eeq

\paragraph{Interior region in full reduction.}  To describe physics in the interior region containing the black hole singularity in the full reduction model, one must analytically extend the Kruskal coordinates to move between each hyperbolic patch \cite{Hartman:2020khs}. The standard branch of the $\arctan$ function  in (\ref{eq:coordglobTX}) only covers the hyperbolic region in the exterior of the black hole (blue region in Figure \ref{fig:dS2fullred}). The hyperbolic patch in the interior of the black hole (white region in Figure \ref{fig:dS2fullred}) is attained by shifting $\varphi/L\to\varphi/L+\pi$. This amounts to performing the continuation $(T,X)\to(-T+i\pi L,-X)$, such that 
 \beq V\to-L e^{-(T+X)/L}=-\frac{L^{2}}{V}\;,\quad U\to-L e^{-(T-X)/L}=-\frac{L^{2}}{U}\;,\label{eq:continuedcoordUVapp}\eeq
 which leaves the  line elements (\ref{eq:kruskalmetapp}) and (\ref{eq:XTlineeleapp}) invariant, but alters the sign of the dilaton, \emph{i.e.}, $\phi = - \phi_r \frac{L^2 + U V }{L^2- U V}= \phi_r \coth (T/L)$.
 Equivalently, the continuation of static patch coordinates is $(t,r)\to(-t+i\pi L,-r)$, or $(v,u)\to(-v+i\pi L,-u+i\pi L)$.

\subsection*{Euclidean   two-dimensional de Sitter space}

As is well known,  the Euclidean continuation of two-dimensional de Sitter space is a two-sphere $S^2.$ In static patch coordinates \eqref{eq:statpatchapp} this can be seen by   analytically continuing   $t /L\to i \phi $ and introducing a polar coordinate via $r = L \cos \theta $. This leads to the line element of a round two-sphere
\begin{equation} \label{eq:euclideandsstatic}
    d\ell^2 = L^2 (d\theta^2 + \sin^2 \theta d \phi^2 )\,,
\end{equation}
with $\theta \in (0,\pi)$ and $\phi \sim \phi + 2\pi$. The periodicity in the Euclidean time $\phi$ follows from the requirement that the Euclidean geometry is regular at the poles.  Note the cosmological horizon $r=L$ resides at $\theta = 0$, whereas the black hole horizon is located at $\theta = \pi$. 

Alternatively, in global coordinates one can Euclideanize de Sitter space  by taking $\tau /L\to i (\vartheta - \pi/2) $, such that the line element becomes  
\begin{equation}
    d\ell^2 = L^2 (d \vartheta^2 + \sin^2 \vartheta d \varphi^2)\,,
\end{equation}
where $\vartheta \in (0,\pi)$ and $\varphi \sim \varphi + 2\pi$. So both static patch and global coordinates describe the geometry of a two-sphere in Euclidean signature.

\section{Nariai geometry in general dimensions} \label{app:nariaigeom}

The Schwarzschild-de Sitter (SdS) black hole in $d$ spacetime dimensions in static coordinates   has the line element
\beq d\ell^2=-f(r)dt^{2}+f^{-1}(r)dr^{2}+r^{2}d\Omega^{2}_{d-2}\;,\qquad f(r)=1-\frac{r^{2}}{L_{d}^{2}}-\frac{16\pi G_{d}M}{(d-2)\Omega_{d-2}r^{d-3}}\;,\label{eq:SdSmetapp}\eeq
where $M$ is the mass parameter of the black hole and $\Omega_{d-2}=2\pi^{(d-1)/2}/\Gamma[(d-1)/2]$ is the volume of the unit $(d-2)$-sphere. For $d>3$, the blackening factor $f(r)$ will have two positive roots corresponding to the locations of the black hole and cosmological horizons, $r_{\text{h}}$ and $r_{\text{c}}$, respectively, with $r_{\text{h}}\le r_{\text{c}}$. Using $f(r_{\text{h}})=f(r_{\text{c}})=0$, we can express the dS radius $L_{d}$ and black hole mass parameter $M$ as
\beq L_{d}^{2}=\frac{r_{\text{c}}^{d-1}-r_{\text{h}}^{d-1}}{r_{\text{c}}^{d-3}-r_{\text{h}}^{d-3}}\;,\qquad \frac{16\pi G_{d}M}{(d-2)\Omega_{d-2}}=\frac{r^{d-3}_{\text{h}}r_{\text c}^{d-1}-r^{d-1}_{\text{h}}r_{\text c}^{d-3}}{r_{\text{c}}^{d-1}-r_{\text{h}}^{d-1} }\;.\label{eq:paramsrhrcapp}\eeq
The Nariai solution is the special case of the SdS black hole when $r_{\text{h}}=r_{\text{c}}\equiv r_{\text{N}}$, for which the mass of the resulting black hole $M_{\text{N}}$ forms an upper bound on the mass parameter~$M$, avoiding a naked singularity. The Nariai radius  and mass may be found using $f(r_{\text{N}})=f'(r_{\text{N}})=0$, yielding
\beq r_{\text{N}}=\sqrt{\frac{d-3}{d-1}}L_{d}\;,\qquad M_{\text{N}}=\frac{d-2}{d-1}\frac{\Omega_{d-2}}{8\pi G_{d}}r_{\text{N}}^{d-3}\;.\label{eq:Nariailimapp}\eeq
In the Nariai limit, the static coordinates $(t,r)$ are insufficient since $f(r)\to0$ between the two horizons. We may nonetheless take the  near horizon limit where we zoom into the region between the two horizons. 

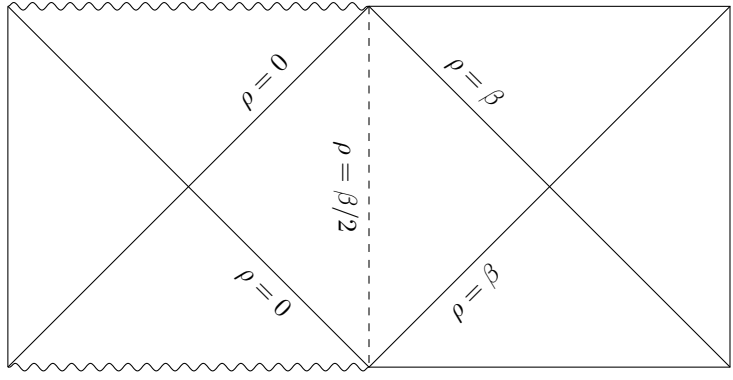
\begin{figure}[t]
\centering
\begin{tikzpicture}[scale=1.2]
    \draw (0,0) coordinate (a) -- (0,4) coordinate (b); 
    \draw[decorate, decoration={snake, amplitude=0.5mm, segment length=2.5mm}] (b) -- (4,4) coordinate (c) ;
    \draw (c) -- (8,4) coordinate (d) -- (8,0) coordinate (e) -- (4,0) coordinate (f);
    \draw[decorate, decoration={snake, amplitude=0.5mm, segment length=2.5mm}] (f) -- (a); 
    \draw (a) -- (c) node[pos=.75, above, sloped] {{\small $\rho = 0$}} -- (e) node[pos=.25, above, sloped] {{\small $\rho = \beta$}}; 
    \draw (b) -- (f) node[pos=.75, below, sloped] {{\small $\rho = 0$}} -- (d) node[pos=.25, below, sloped] {{\small $\rho = \beta$}};
    \draw[dashed] (c) -- (f) node[pos=.5, below, sloped] {{\small $\rho = \beta/2$}};
\end{tikzpicture}
\caption{Penrose diagram of the Nariai black hole. The black hole and cosmological horizons are located at $\rho=0$ and $\rho=\beta$, respectively, and are in thermal equilibrium. Clearly there is a finite proper distance between the two horizons.}
\label{fig:ds2penapp} 
\end{figure}

To find the  Nariai geometry in $d$ dimensions we follow \cite{Anninos:2012qw,Visser:2019muv} (see also \cite{Maldacena:2019cbz,Moitra:2022glw,Cardoso:2004uz}). First note that using (\ref{eq:paramsrhrcapp}) the function $f(r)$ factorizes as
\beq 
\begin{split}
f(r)&=\frac{1}{L_{d}^{2}r^{d-3}}\left(L_{d}^{2}r^{d-3}-r^{d-1}-\frac{16\pi G_{d}ML_{d}^{2}}{(d-2)\Omega_{d-2}}\right)=\frac{1}{L_{d}^{2}r^{d-3}}(r-r_{\text{h}})(r_{\text{c}}-r)\mathcal{P}(r)\;,
\end{split}
\label{eq:frfact}\eeq
where $\mathcal{P}(r)$ is a polynomial in $r$ that is invariant under the exchange $r_{\text{c}}\leftrightarrow r_{\text{h}}$. For example, 
\beq \mathcal{P}_{d=4}(r)=r+r_{\text{h}}+r_{\text{c}}\;,\quad \mathcal{P}_{d=5}(r)=(r+r_{\text{h}})(r+r_{\text{c}})\;.\eeq
Next, introduce dimensionful coordinates $(\tau,\rho)$ and parameter $\beta$
\beq \tau=\tilde{\epsilon} t\;,\quad \rho=\frac{r-r_{\text{h}}}{\tilde{\epsilon}}\;,\quad \beta=\frac{r_{\text{c}}-r_{\text{h}}}{\tilde{\epsilon}}\;.\eeq
Substitute $r=\tilde{\epsilon}\rho+r_{\text{h}}$ into (\ref{eq:frfact}) and expand around $\tilde{\epsilon}=0$, leading to
\beq
\begin{aligned}
f(\rho)\approx&-\frac{2r_{\text{h}}}{L_{d}^{2}}\rho\tilde{\epsilon}+\frac{(d-3)\beta\rho\tilde{\epsilon}^{2} }{L_{d}^{2}}\frac{r_{\text{c}}^{d}r_{\text{h}}^{2}(r_{\text{c}}+r_{\text{h}})}{r_{\text{c}}^{d}r_{\text{h}}^{3}-r_{\text{h}}^{d}r_{\text{c}}^{3}}\\
& -\frac{\rho^{2}\tilde{\epsilon}^{2}}{L_{d}^{2}}\left(1-\frac{(d-3)(d-2)r_{\text{c}}^{d}r_{\text{h}}(r^{2}_{\text{c}}-r^{2}_{\text{h}})}{2(r_{\text{h}}^{d}r_{\text{c}}^{3}-r_{\text{c}}^{d}r_{\text{h}}^{3})}\right)+\mathcal{O}(\tilde{\epsilon}^{3})\;.
\end{aligned}
\eeq
\noindent Carefully taking the limit $r_{\text{c}}\to r_{\text{h}}$ and $\epsilon\to0$ while keeping $\beta$ fixed, it is straightforward to show\footnote{Taking the simultaneous limit $r_{\text{c}}\to r_{\text{h}}$ and $\tilde{\epsilon}\to0$ is delicate in $d>5$. First send $r_{\text{c}}\to r_{\text{N}}+\delta$ and $r_{\text{h}}\to r_{\text{N}}-\delta$ for small $\delta$ and then take the limit $\delta\to\tilde{\epsilon}$ using L'H\^{o}pital's rule.}
\beq \lim_{\overset{r_{\text{c}}\to r_{\text{h}}}{\tilde{\epsilon}\to0}}\frac{f(\rho)}{\tilde{\epsilon}^{2}}=\frac{\rho(\beta-\rho)}{\hat L_{d}^{2}}\;, \qquad \text{with} \qquad \hat{L}_{d}=\frac{L_{d}}{\sqrt{d-1}},
\eeq
from which the line element (\ref{eq:SdSmetapp}) becomes
\beq d\ell^2=-\frac{\rho(\beta-\rho)}{\hat L_d^2}d\tau^{2}+\frac{\hat L_d^2 d\rho^{2}}{\rho(\beta-\rho)}+r_{\text{N}}^{2}d\Omega_{d-2}^{2}\;.\eeq
The original black hole horizon now lives at $\rho=0$ while the cosmological horizon is at $\rho=\beta$ (see Figure \ref{fig:ds2penapp} for an illustration of the Penrose diagram).
The Nariai geometry is $\text{dS}_{2}\times S^{d-2}$, which is more easily seen by introducing the coordinates
\beq \tilde{\tau}=\frac{\beta}{2 \hat L_d }{\tau}\;,\quad \tilde{\rho}=\frac{2 \hat L_d}{\beta}\left(\rho-\frac{\beta}{2}\right)\;,\eeq
such that
\beq d\ell^2=-\left(1-\frac{\tilde{\rho}^{2}}{\hat{L}_{d}^{2}}\right)d\tilde{\tau}^{2}+\left(1-\frac{\tilde{\rho}^{2}}{\hat{L}_{d}^{2}}\right)^{-1}d\tilde{\rho}^{2}+r_{\text{N}}^{2}d\Omega^{2}_{d-2}\;.\label{eq:Nariaimetapp}\eeq
The curvature radii of the two-dimensional de Sitter space and the sphere $S^{d-2}$, given by $\hat L_d = L_d / \sqrt{d-1}$ and $r_{\text N} = L_d \sqrt{\frac{d-3}{d-1}}$  respectively,  are thus generically different, but they    coincide for $d=4$.  With respect to the metric (\ref{eq:Nariaimetapp}), it is easy to see there is a finite proper distance between the two horizons
\beq  \ell=2\int_{0}^{\hat{L}_{d}}\frac{d\tilde{\rho}}{\sqrt{1-\frac{\tilde{\rho}^{2}}{\hat{L}_{d}^{2}}}}=\pi \hat{L}_{d}\;.\eeq
Moreover, in the Nariai limit  the black hole and cosmological horizons are in thermal equilibrium at a temperature 
\begin{figure}[t]
\begin{center}
\includegraphics[width=10cm]{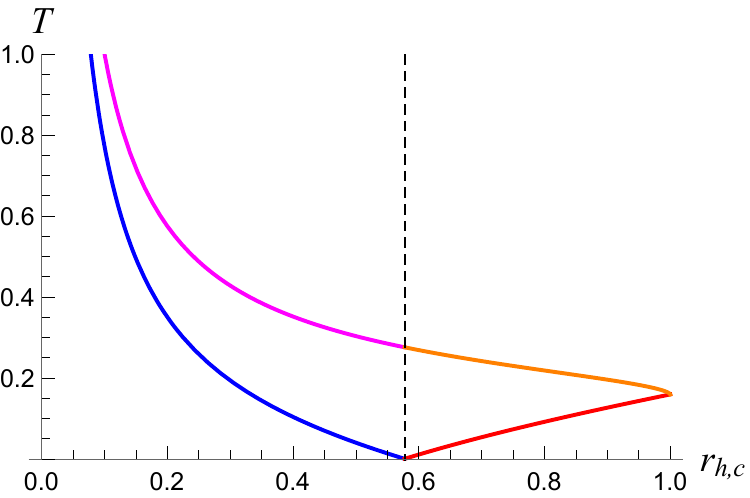}
\end{center}
\caption{Temperatures of SdS vs. horizon radii $r_{\text{h,c}}$, where we have set $d=4$ and $L_4=1$. The blue and red curves correspond to the temperatures $T_{\text{h}}=\kappa_{\text{h}}/2\pi$ and $T_{\text{c}}=\kappa_{\text{c}}/2\pi$, respectively, where the surface gravities are defined with respect to the timelike Killing vector $\xi = \partial_t$. The pink and orange curves show the Bousso-Hawking temperatures $\tilde T_{\text{h}}=\tilde \kappa_{\text{h}}/2\pi$ and $\tilde T_{\text{c}}=\tilde \kappa_{\text{c}}/2\pi$, respectively, where  $\tilde \kappa_{\text{h,c}}=  \kappa_{\text{h,c}}/\sqrt{f(r_0)}$ given in \eqref{eq:newsurfacegravities}. Whereas the former temperatures $T_{\text{h,c}}$  vanish  for the Nariai black hole with horizon radius $r_{\text N} = L_4 / \sqrt{3} $ (the dashed line), the latter   $\tilde T_{\text{h,c}}$ are finite  and equal to  $T_{\text{N}}=\sqrt{3}/2\pi L_4 $ in the Nariai limit.}
\label{fig:tempsSdS} 
\end{figure}
\beq T_{\text{N}}=\frac{1}{2\pi\hat{L}_{d}}\;.\label{eq:tempnariai2}\eeq
This temperature can be derived in two-dimensional de Sitter space, for instance,  by removing the conical singularity in the Euclidean static patch geometry. It can also be obtained from the higher-dimensional perspective by taking the Nariai limit of the temperatures of the black hole and cosmological horizons. Interestingly, if we normalize the timelike Killing vector of SdS as $\xi = \partial_t$, then the temperature of  the Nariai black hole vanishes. Namely, for this normalization the (positive) surface gravities of the black hole and cosmological horizon can be found to be
\begin{equation}
\begin{aligned}
    \kappa_{\text h} &= \frac{(d-3) - (d-1) r_{\text h}^2/L^2_d}{2 r_{\text h}}= \frac{r_{\text N}^2 - r_{\text h}^2}{2 r_{\text h} \hat L_d^2}\, , \\
    \kappa_{\text c} &= \frac{-(d-3) + (d-1) r_{\text c}^2/L^2_d}{2 r_{\text c}}= \frac{r_{\text c}^2 - r_{\text N}^2}{2 r_{\text c} \hat L_d^2}\, .
\end{aligned}
\end{equation}
We  indeed observe that $\kappa_{\text {h,c}}$ vanishes  if $r_{\text h} = r_{\text c} = r_{\text N}.$
However, Bousso and Hawking \cite{Bousso:1996au} argued that $\xi = \partial_t$ is not the correct normalization of the timelike Killing vector of SdS. This is only true for pure de Sitter space, where this choice corresponds to setting $\xi^2 =-1$ at the origin $r=0$,  and for the asymptotically flat  Schwarzschild geometry   where this normalization yields $\xi^2=-1$ at spatial infinity $r=\infty$. These locations in pure de Sitter and Schwarzschild have in common that   an observer can stay in place without accelerating. In other words, these radii are maxima of the blackening factor  $f(r)$ in the respective spacetimes. Alternatively, for 
 Schwarzschild-de Sitter space the function $f(r)$ attains its maximum when  
\begin{equation}
    f'(r_0)=0 \quad \rightarrow \quad r_0^{d-1} = \frac{d-3}{d-2} \frac{8 \pi G_d M L_d^2}{\Omega_{d-2}} 
   =\frac{d-3}{2} r_{\text{h,c}}^{d-3} (L^2_d - r_{\text{h,c}}^2) \, .
\end{equation}
The sphere of radius $r_0$ is the place where the cosmological expansion and the black hole attraction cancel each other exactly. 
The function $f(r)$ at this radius is given by
\begin{equation}
    f(r_0)=1- \frac{d-1}{d-3} \frac{r_0^2}{L_{d}^2} = 1- \frac{r_0^2}{r_{\text N}^2} \,. 
\end{equation}
The idea by Bousso and Hawking is now to normalize the Killing vector on the geodesic at fixed radius $r_0$, such that $\xi = \frac{1}{\sqrt{f(r_0)}} \partial_t$. This has a similar effect on  the surface gravities of the cosmological and black hole horizon  
\begin{equation} \label{eq:newsurfacegravities}
    \tilde \kappa_{\text{h,c}} = \frac{\kappa_{\text{h,c}}}{\sqrt{f(r_0)}} =\frac{\frac{r_{\text N}}{2 r_{\text{h,c}} \hat L_d^2}\big |  r_{\text N}^2 -r_{\text{h,c}}^2  \big |}   {\sqrt{r_{\text{N}}^2-     \left [\frac{d-3}{2} r_{\text{h,c}}^{d-3} ( \frac{d-1}{d-3} r_{\text N}^2 - r_{\text{h,c}}^2) \right]^{2/(d-1)}}} ~.
\end{equation}
Expanding the denominator  on the right   near $r_{\text{h,c}} = r_{\text N}$ gives $\sqrt{r_{\text N}^2 - r_0^2} \approx \sqrt{d-3} |r_{\text{N}} - r_{\text h}|  $.
Thus, in the Nariai limit   we find
\begin{equation} \label{eq:nariaisurfacegravity}
\tilde \kappa_{\text N} =    \lim_{r_{\text{h,c}}\to r_{\text{N}}} \tilde \kappa_{\text{h,c}}=   \frac{1}{\hat L_d}=\frac{\sqrt{d-1}}{L_d} \qquad \rightarrow \qquad T_{\text{N}} = \frac{\sqrt{d-1}}{2 \pi   L_d}\,. 
\end{equation}
This agrees with the expected temperature \eqref{eq:tempnariai2} in two-dimensional de Sitter space.


\section{Noether charge formalism for 2D dilaton gravity} \label{app:noetherchargesumm}

To keep this article self contained, here we summarize elements of the Noether charge formalism \cite{Wald:1993nt,Iyer:1994ys} in the context of a wide class of two-dimensional dilaton theories of gravity. For a more thorough analysis, see Appendix C of \cite{Pedraza:2021cvx}. 

\subsection*{Lagrangian formalism}

Let $\psi = (g_{\mu\nu}, \Phi)$ denote a collection of dynamical fields, where $g_{\mu\nu}$ is an arbitrary background metric of a $(1+1)$-dimensional Lorenztian spacetime $M$ and $\Phi$ represents any   scalar field on $M$. Consider the following covariant Lagrangian  2-form  $L$
\begin{equation} \label{eq:genericdilatonlagr}
	L = L_{0}\epsilon\left [  R Z(\Phi) +   U(\Phi)( \nabla \Phi)^2 - V(\Phi)\right ]\;,
\end{equation}
where $\epsilon$ is the spacetime volume form,  $L_{0}$ is some coupling constant, and $R$ is the Ricci scalar. This is the most general Lagrangian for two-dimensional Einstein gravity coupled non-minimally to a dynamical scalar field, and includes JT gravity and the localized form of the 1-loop Polyakov action. For spacetimes $M$ with boundary $\partial M$, one should also add a boundary Gibbons--Hawking--York 1-form $b$
\begin{equation}
b =-2L_{0}\epsilon_{\partial  M} Z(\Phi)  K\;.
\label{eq:Cdefn}\end{equation}
Here $\epsilon_{\partial M}$ is the volume form on $\partial M$, $K$ is the trace of the extrinsic curvature $K_{\mu \nu} = \frac{1}{2}\mathcal L_n \gamma_{\mu \nu}$  of the timelike boundary, and $\gamma_{\mu \nu} = - n_\mu n_\nu + g_{\mu \nu}$ is the induced metric on $\partial M$, with $n_\mu$ the (outward pointing) unit normal to $\partial M$. 

The symplectic potential 1-form $\theta(\psi,\delta\psi)$ and symplectic current 1-form $\omega(\psi,\delta_{1}\psi,\delta_{2}\psi)\equiv\delta_{1}\theta(\psi,\delta_{2}\psi)-\delta_{2}\theta(\psi,\delta_{1}\psi)$ are, respectively, 
\begin{equation}
\begin{aligned}
	 \theta &= L_{0}\epsilon_\mu\Big[ Z(\Phi)  (g^{\mu \beta} g^{\alpha \nu} - g^{\mu \nu} g^{\alpha \beta})\nabla_\nu \delta g_{\alpha \beta} + (g^{\alpha \beta} \nabla^\mu Z(\Phi) \\
	 &- g^{\beta \mu} \nabla^\alpha  Z(\Phi))\delta g_{\alpha \beta} + 2 U(\Phi) \nabla^\mu \Phi \delta \Phi \Big]\;,
\end{aligned}
\label{eq:sympotgen}
\end{equation}
and
\beq
\begin{aligned}
	\omega&= L_{0} \epsilon_\mu \Big [ Z(\Phi)S^{\mu \alpha \beta \nu \rho \sigma}\delta_1 g_{\rho \sigma} \nabla_\nu \delta_2 g_{\alpha \beta} + \frac{1}{2} g^{\mu \beta} g^{\alpha \nu} g^{\rho \sigma} \delta_1 g_{\rho \sigma} \delta_2 g_{\alpha\beta} \nabla_\nu Z(\Phi) \\
	&+ (g^{\mu \beta}g^{\alpha\nu} - g^{\mu \nu} g^{\alpha \beta}) ( \delta_1( Z(\Phi))  \nabla_\nu \delta_2 g_{\alpha \beta} - \delta_1 (\nabla_\nu Z(\Phi) )\delta_2 g_{\alpha \beta}) \\
	& + U(\Phi)\nabla^\mu \Phi g^{\alpha\beta}\delta_1 g_{\alpha \beta}\delta_2 \Phi  + 2 \delta_1 (U(\Phi) \nabla^\mu \Phi) \delta_2 \Phi - [1 \leftrightarrow 2] \Big ]\;. 
\end{aligned}
\label{eq:sympcurrentgen}\eeq
Here $Z'(\Phi)=\frac{dZ}{d\Phi}$ and $S^{\mu\alpha\beta\nu\rho\sigma}$ is given by   
\beq S^{\mu\alpha\beta\nu\rho\sigma}=g^{\mu\rho}g^{\alpha\sigma}g^{\beta\nu}-\frac{1}{2}g^{\mu\nu}g^{\alpha\rho}g^{\beta\sigma}-\frac{1}{2}g^{\mu\alpha}g^{\beta\nu}g^{\rho\sigma}-\frac{1}{2}g^{\alpha\beta}g^{\mu\rho}g^{\sigma\nu}+\frac{1}{2}g^{\alpha\beta}g^{\mu\nu}g^{\rho\sigma}\;.\label{eq:tensorS}\eeq
Let $\zeta$ be an arbitrary smooth vector field on $M$ representing an infinitesimal generator of a diffeomorphism. The Noether current 1-form $j_{\zeta}$ associated with $\zeta$ and arbitrary field configuration $\psi$ is defined as $j_{\zeta}\equiv \theta (\psi, \mathcal{L}_{\zeta} \psi) - \zeta\cdot L$, with $\mathcal{L}_{\zeta}$ being the Lie derivative along $\zeta$. The associated Noether charge $0$-form $Q_{\zeta}$ is defined on-shell via $j_\zeta= d Q_\zeta$. Explicitly, with respect to the theory (\ref{eq:genericdilatonlagr})
\begin{equation}
	j_\zeta = \epsilon_\mu \left [2L_{0}\nabla_\nu (Z(\Phi) \nabla^{[\nu} \zeta^{\mu ]} + 2 \zeta^{[\nu } \nabla^{\mu]} Z(\Phi)) + 2 {E^\mu}_\nu \zeta^\nu\right]\;,
\label{eq:Noethercurrent}\end{equation}
 \begin{equation}
	Q_\zeta = -L_{0}\epsilon_{\mu \nu} \left [Z(\Phi) \nabla^\mu \zeta^\nu + 2 \zeta^\mu \nabla^\nu Z(\Phi) \right]\;,
\label{eq:Noethercharge}\end{equation}
where $\epsilon_{\mu\nu}$ is the volume form for the codimension-0 surface $\partial\Sigma$, which   is a   cross section of the spatial part of $\partial M$, and $E_{\mu \nu}$ are the metric equations of motion. On-shell, $\omega$, $Q_{\zeta}$ and $\theta$ obey the fundamental   variational identity
\begin{equation}
	\omega (\psi, \delta \psi, \mathcal L_\zeta \psi) =d \left[ \delta Q_\zeta - \zeta \cdot \theta (\psi, \delta \psi) \right]\;.
\label{eq:variationalid}\end{equation}

\subsection*{Hamiltonian formalism}

The Hamiltonian $H_{\zeta}$ generating time evolution along the flow of $\zeta$ follows from the variational identity (\ref{eq:variationalid}). Specifically, let $\Sigma$ be any Cauchy slice of $M$. Denote the induced metric on   $\Sigma$ by $h_{\mu \nu} = u_\mu u_\nu + g_{\mu \nu}$, with $u_\mu$ the (future-pointing) unit normal to $\Sigma$. Then, the variation $\delta H_{\zeta}$ and the Hamiltonian itself are given by \cite{Harlow:2019yfa,Pedraza:2021cvx}
\beq \delta H_{\zeta}=\oint_{\partial\Sigma}[\delta Q_{\zeta}-\zeta\cdot\delta b-\delta C(\psi,\mathcal{L}_{\zeta}\psi)]\;,\eeq
\beq H_{\zeta}=\int_{\partial\Sigma}[Q_{\zeta}-\zeta\cdot b- C(\psi,\mathcal{L}_{\zeta}\psi)]+ \text{cst} \;.\eeq
The constant  represents a standard ambiguity to the energy of any Hamiltonian system, which we will set to zero.  Moreover, $C$ is a local 0-form defined over the boundary $\partial M$ and is  covariant under diffeomorphisms preserving the location of the (spatial) boundary $B$ in $\partial M$, and obeys $\theta \big |_{B} = \delta b + d C$ if Dirichlet boundary conditions are imposed \cite{Harlow:2019yfa}. Explicitly, 
\beq C = c \cdot \epsilon_{\partial M}\;,\quad  c^\mu=-L_{0}Z(\Phi)  \gamma^{\mu \lambda}n^\nu \delta g_{\lambda \nu}\;.\eeq
For the 2D dilaton-gravity model (\ref{eq:genericdilatonlagr}), the terms on the right-hand side of $H_{\zeta}$ pulled back to $\partial\Sigma$ are \cite{Pedraza:2021cvx}
\beq
Q_\zeta \big |_{\partial \Sigma}=+L_{0}\epsilon_{\partial \Sigma} (u_\mu n_\nu - u_\nu n_\mu) \left [  Z(\Phi) \nabla^\mu \zeta^\nu + 2 \zeta^\mu \nabla^\nu Z(\Phi) \right]\;,\eeq
\beq \zeta\cdot b |_{\partial \Sigma}= -2L_{0}\epsilon_{\partial \Sigma} \zeta^\mu u_\mu Z(\Phi)  K\;,\eeq
\beq C(\psi,\mathcal L_\zeta \psi )|_{\partial \Sigma}= -L_{0}\epsilon_{\partial \Sigma}(u^\mu n^\nu + u^\nu n^\mu) Z(\Phi)   \nabla_\mu \zeta_\nu\;,\label{eq:CLiepsi}\eeq
where we used $\epsilon_{\mu\nu}|_{\partial\Sigma}=n_{\mu}u_{\nu}-n_{\nu}u_{\mu}$ and $(\epsilon_{\partial M})_{\mu}=+u_{\mu}\epsilon_{\partial\Sigma}$. It follows that the Hamiltonian for $2D$ dilaton gravity is  
\begin{equation}
H_\zeta=  \oint_{\partial \Sigma} \epsilon_{\partial\Sigma} N \varepsilon\;, 
\label{eq:quasilocaHam}\end{equation}
with $N = - \zeta^\mu u_\mu$ being the lapse function and $\varepsilon$ being the quasi-local energy density,
\beq \varepsilon \equiv u_\mu u_\nu \tau^{\mu \nu} =-2L_{0} n^\alpha \nabla_\alpha Z(\Phi)\;.\label{eq:quasilocalen}\eeq
Here $\tau_{\mu\nu}\equiv2L_{0}\gamma_{\mu \nu} n^\alpha \nabla_\alpha  Z(\Phi)$ is the Brown-York stress-energy tensor \cite{Brown:1992br}.


 \section{Conformal isometry and diamond universe coordinates}\label{app:diamondcoordinates}
 
Here we detail the geometry of rectangular causal diamonds in two-dimensional de Sitter space. To accomplish this it is necessary to   compute the conformal Killing vectors of $\text{dS}_{2}$, and for completeness we start by deriving the true Killing vectors of $\text{dS}_{2}$. Our approach largely follows Appendices A and B of \cite{Pedraza:2021ssc}  (see also \cite{Jacobson:2018ahi,Visser:2019muv}). 

\subsection*{Killing   vectors of $\text{dS}_{2}$}

A systematic way of deriving the Killing  vectors of $\text{dS}_{2}$  is to use the embedding formalism. Two-dimensional de Sitter space  can be embedded into three-dimensional Minkowski space $\mathbb{R}^{1,2}$ with   line element
\beq d\ell^2=-(dX^0)^{2}+(dX^{1})^{2}+(dX^{2})^2\,. \eeq
dS$_2$ is a hyperboloid in this embedding space, described by the equation
\beq -(X^0)^{2}+(X^{1})^{2}+(X^{2})^2=+L^{2}\;.\eeq
The embedding space induces a metric on the hyperboloid, recognized as the metric on $\text{dS}_{2}$. For example, $\text{dS}_{2}$ in static patch coordinates follows from the embedding coordinates
\beq X^0=\sqrt{L^{2}-r^{2}}\sinh(t/L)\;,\quad X^{1}=r\;,\quad X^2=\sqrt{L^{2}-r^{2}}\cosh(t/L)\;,\label{eq:embeddcoordtr}\eeq
when $r<L$, while for $r>L$ we have
\beq X^0=\sqrt{L^{2}-r^{2}}\sinh(t/L)\;,\quad X^{1}=r\;,\quad X^2=-\sqrt{L^{2}-r^{2}}\cosh(t/L)\;.\eeq
We will be primarily interested in the former case $r<L$, for which (\ref{eq:embeddcoordtr}) in terms of advanced/retarded coordinates $(v,u)$ is
\beq X^{0}= L\frac{\sinh\left(\frac{v+u}{2L}\right)}{\cosh\left(\frac{v-u}{2L}\right)}\;,\quad
X^{1}=  L\frac{\sinh\left(\frac{v-u}{2L}\right)}{\cosh\left(\frac{v-u}{2L}\right)} \;,\quad X^2=L\frac{\cosh\left(\frac{v+u}{2L}\right)}{\cosh\left(\frac{v-u}{2L}\right)}\;.\label{eq:embeddcoorvu}\eeq
The isometry group of Lorentzian $\text{dS}_{2}$ is $O(1,2)$ with $\frac{1}{2}(2)(2+1)=3$ Killing vectors. With respect to the embedding coordinates  the single rotation generator $J$ and the two boost generators $B_{1,2}$ of the isometry group are
\beq J=X^{1}\partial_{X^{2}}-X^{2}\partial_{X^{1}}\;,\quad B_{1}=X^1\partial_{X^{0}}+X^{0}\partial_{X^1}\;,\quad B_{2}=X^2\partial_{X^{0}}+X^{0}\partial_{X^2}\;,\label{eq:gengenerators}\eeq
obeying the algebra
\beq [J,B_{1}]=-B_{2}\;,\quad [J,B_{2}]=B_{1}\;,\quad [B_{1},B_{2}]=J\;.\eeq
Substituting the embedding coordinates (\ref{eq:embeddcoordtr}) into the generators (\ref{eq:gengenerators}) yields the true Killing vectors of $\text{dS}_{2}$ in static patch coordinates:
\beq
\begin{split}
&J=-\frac{rL}{\sqrt{L^{2}-r^{2}}}\sinh(t/L)\partial_{t}-\sqrt{L^{2}-r^{2}}\cosh(t/L)\partial_{r}\;,\\
&B_{1}=\frac{rL}{\sqrt{L^{2}-r^{2}}}\cosh(t/L)\partial_{t}+\sqrt{L^{2}-r^{2}}\sinh(t/L)\partial_{r}\;,\quad B_{2}=L\partial_{t}\;. 
\end{split}
\label{eq:generatorstr}\eeq
We see the boost $B_{2}$ generates   time translations in    the static patch.  Alternatively, the generators in null coordinates $(v,u)$ are\footnote{It is useful to know $\partial_{t}=\partial_{v}+\partial_{u}$ and $\partial_{r}=\cosh^{2}\left(\frac{v-u}{2L}\right)(\partial_{v}-\partial_{u})$.}
\beq 
\begin{split}
 &J=L[\cosh(u/L)\partial_{u}-\cosh(v/L)\partial_{v}]\;,\\
 &B_{1}=L[\sinh(v/L)\partial_{v}-\sinh(u/L)\partial_{u}]\;,\quad B_{2}=L(\partial_{v}+\partial_{u})\;. \label{eq:isometrygenerators}
\end{split}
\eeq

\subsection*{Conformal Killing vectors of causal diamonds in $\text{dS}_{2}$}

Consider a rectangular causal diamond in a generic two-dimensional spacetime in conformal gauge $d\ell^2=-e^{2\rho}dudv$, as described in Section \ref{sec:islandsfrommicact}. The conformal isometry of such a diamond is generated by \cite{Pedraza:2021ssc} 
\beq
    \zeta=A_a(u-u_0) \partial_u + A_b(v-v_0) \partial_v \, , \quad  A_j(y)= g_{j} (y)[h(j)- h(y)]\;,
\label{eq:appckv1}\eeq
 where $g$ and $h$ are even functions, and  $j$ is a length scale taking the values $a$ or $b$. The conformal Killing vector field $\zeta$ is constructed by demanding it respects the reflection symmetries across the maximal slice $\Sigma$ and the line intersecting past and future vertices when $a$ and $b$ are interchanged. Furthermore, we require that $\zeta$ maps the diamond onto itself, \emph{i.e.}, $\zeta$ must be tangent to the null generators of the null boundaries, imposing $A_{j}(\pm j)=0$. This implies that the past and future null boundaries are  conformal Killing horizons.  
 
 Generally, the length scales $a$ and $b$ are different, leading to two positive surface gravities, 
  \begin{equation} \label{eq:surfacegravity}
 	\kappa_{a} =   g_a(a) h'(a) \,,\quad \quad \kappa_{b} =g_b(b) h'(b)\,,
 \end{equation}
 defined via $\nabla_\mu \zeta^2 = - 2 \kappa \zeta_\mu$,   evaluated on the future null boundaries $u-u_0 =   a$ and $v-v_0=b$, respectively. The prime denotes the derivative with respect to $y$ (denoting $u$ and $v $ respectively). The surface gravities are constant and hence they satisfy the zeroth law for bifurcate conformal Killing horizons proven in Appendix C of \cite{Jacobson:2018ahi}.  
 
 Furthermore, for square causal diamonds (when $a=b$) there is only a single surface gravity $\kappa$, and the conformal Killing vector   becomes approximately a boost Killing vector near the bifurcation surface of the   horizon  
 \begin{equation}
 	\zeta  \approx  g(a) h'(a) \left [ \tilde v \partial_{\tilde v} - \tilde u \partial_{\tilde u}  \right] \qquad \text{at}  \qquad   \tilde u= \tilde v =0\,,
 \end{equation} 
 with $\tilde u \equiv u-u_0-a$ and $\tilde v \equiv v -v_0+a$. The vector field between brackets is the boost Killing vector in   null coordinates in flat space, and we regonize   the normalization  as  being  the  surface gravity in \eqref{eq:surfacegravity}.

We can restrict the form of $\zeta$ in \eqref{eq:appckv1} further by placing the diamond in two-dimensional de Sitter space and requiring that $\zeta$ become the generator of time translations when the causal diamond coincides with the static patch (the maximal diamond). 
First, we    express the conformal Killing vector in terms of Kruskal coordinates $(U,V)=(-L e^{-u/L},L e^{v/L})$,  and set $u_0=v_0=0$ and $a=b=-L \log (B/L)$   for simplicity,
\beq \label{eq:formofckvinkruskal}
\zeta=\frac{1}{L}\left [ A(V)V\partial_V - A(U)U\partial_U \right] \, , \qquad A(Y)=\frac{\kappa L  }{B^{-1}h'(B^{-1})}   \frac{g (Y)}{g(B)} \big [h(B) - h(Y)\big]        \,,
\eeq
where now the prime denotes the derivative with respect to $Y$,   denoting $U$ and $V$ respectively. Further, the functions $g$ and $h$ satisfy $g(Y)=g(Y^{-1})$ and $h(Y)=h(Y^{-1})$. 
Next, we require that in the maximal diamond limit ($a,b \to \infty$ or $B \to 0$) we have
\beq \label{eq:extrarequirckv}
\lim_{B \to 0} \zeta = \kappa (V \partial_V - U \partial_U)  = \kappa B_2\,,
\eeq
where $B_2$ is the time translation generator in \eqref{eq:isometrygenerators}. This implies
\beq \label{eq:restrictionsonh}
g(Y)=1, \qquad \lim_{B \to 0}\frac{B  h(B)}{ h'(B^{-1})}=1,  \qquad \lim_{B \to 0}\frac{C  }{h'(B^{-1})}   =0\,.
\eeq
Assuming the function $h(B)$ can be expanded as 
\beq
h(B)= \sum_{n=-\infty}^{n=\infty} a_n B^{n}  \, 
\eeq
with $a_{n}=a_{-n}$, since $h(B)=h(B^{-1})$.
The second condition in \eqref{eq:restrictionsonh} implies $a_n=0$ for $n>1$, hence only $a_0$ and $a_1$ are nonvanishing. Therefore, we arrive at the unique form for the conformal Killing vector by inserting $g(Y)=1$ and $h(Y)=a_1 Y^{-1} + a_0 + a_1 Y$ into \eqref{eq:formofckvinkruskal} 
\beq
\zeta= \frac{1}{L}\left [ A(V)V\partial_V - A(U)U\partial_U \right]  \, , \qquad A(Y)= \kappa L \frac{1 + B^2}{1-B^2}     \left [ 1-\frac{Y + Y^{-1}}{B + B^{-1}}\right]        \,.
\eeq
Transforming back to static null coordinates using \eqref{eq:kruskalnullcoord}   yields
\beq \label{eq:uniqueckvstatic1}
\zeta = A(u)\partial_u + A(v) \partial_v\, , \qquad A(y) =\frac{ \kappa L}{\sinh(a/L)}\left[\cosh(a/L)-\cosh(y/L)\right] \,.
\eeq
This is the expected result  since it   is identical  to the conformal Killing vector of a spherically symmetric causal diamond in $d$-dimensional de Sitter   space \cite{Jacobson:2018ahi}. In higher dimensions, however, the spherical symmetry of a  diamond restricts the form of the conformal Killing vector uniquely, whereas in $d=2$ we need the additional requirement \eqref{eq:extrarequirckv} to arrive at the same unique  form. 
For a  rectangular   diamond in $\text{dS}_{2}$ centered at $(u_0,v_0)$ this generalizes to  
\beq \zeta=A_{a}(u-u_{0})\partial_{u}+A_{b}(v-v_{0})\partial_{v}\;,\quad A_{a}(y)=\frac{\kappa_{a}L}{\sinh(a/L)}\left[\cosh(a/L)-\cosh(y/L)\right]\;, \label{eq:ckvdS2app}\eeq
and similarly for $A_{b}(y)$. It is straightforward to verify $\zeta$ is a conformal Killing vector, obeying the conformal Killing equation, $2\nabla_{(\mu}\zeta_{\nu)}=(\nabla\cdot\zeta)g_{\mu\nu}$. Incidentally, the expression (\ref{eq:ckvdS2app}) is equivalent to the form of the conformal Killing vector preserving a causal diamond in $\text{AdS}_{2}$ when we set $\mu=-1$ and $L\to iL$ in equation (A7) of \cite{Pedraza:2021ssc}. 
 To summarize, the maximal diamond in the de Sitter static patch (which is the static patch itself) admits a true  isometry, whereas finite causal diamonds  only admit a conformal  isometry generated by \eqref{eq:ckvdS2app}. 


\subsection*{Diamond universe coordinates}

We now introduce inextendible coordinates $(s,x)$  adapted to the flow of $\zeta$ that cover  the causal diamond \cite{Jacobson:2018ahi}. The coordinate $s$ is the conformal Killing  time  defined such that $\zeta \cdot ds =1$ and, for $a =b$, $s=0$ on the maximal slice $\Sigma$ and has the range $s \in [-\infty,\infty]$. Similarly, the spatial coordinate $x \in [-\infty, \infty]$ obeys $\zeta \cdot dx =0$ and $|dx| = |ds|$ and, for $a=b$, the origin $x=0$ is at $r_*=r_{*,0}\equiv \frac{1}{2}(v_0-u_0)$. From these conditions, the two-dimensional line element in so-called ``diamond universe'' coordinates is
\begin{equation} \label{eq:diamondmetric1}
    d\ell^2 = C^2 (s,x) (-ds^2 + dx^2)= -C^2 (\bar u, \bar v)d \bar ud\bar v\,,
\end{equation}
with null coordinates, $\bar u = s-x$ and $\bar v = s+x$, and $C^{2}$ is a conformal factor determined below.  In these coordinates,  the null boundaries of the  diamond are located  at $\bar u =\pm\infty$ ($u-u_0=\pm a$) and $\bar v = \pm \infty$ ($v-v_0 = \pm b$). 

From the line element (\ref{eq:diamondmetric1}), it is clear $\zeta = \partial_s=\partial_{\bar u} +\partial_{\bar v}$ is a conformal Killing vector, which  should be equivalent to the vector field (\ref{eq:ckvdS2app}) preserving the diamond in $\text{dS}_{2}$. Setting the expression (\ref{eq:ckvdS2app}) for $\zeta$    equal to $\partial_{\bar u} +\partial_{\bar v}$ yields the following transformation between null coordinates $(\bar{u},\bar{v})$ and $(u,v)$ \cite{Pedraza:2021ssc}
\begin{equation}
\begin{split}  \label{eq:transfdiamond2}
    e^{(u-u_0)/ L} = \frac{\cosh \left [ (a /L + \kappa_a \bar u)/2 \right]}{ \cosh \left [ (a/L - \kappa_a \bar u)/2\right]} \,,\quad   e^{( v-v_0)/ L} = \frac{\cosh \left [ (b /L + \kappa_b \bar v)/2 \right]}{ \cosh \left [ (b/L - \kappa_b \bar v)/2\right]}\,. 
\end{split}
\end{equation}
With this coordinate transformation, we uncover the conformal factor $C^{2}(\bar{u},\bar{v})$ by comparing line elements (\ref{eq:statnullmet}) and (\ref{eq:diamondmetric1}),
\beq C^{2}(\bar{u},\bar{v})=\frac{4\kappa_{a}\kappa_{b}L^{2}e^{\kappa_{a}\bar{u}+\kappa_{b}\bar{v}+2r_{\ast,0}/L}\left(e^{2b/L}-1\right)\left(e^{2a/L}-1\right)}{\left[\left(e^{b/L}+e^{\kappa_{b}\bar{v}}\right)\left(e^{a/L+\kappa_{a}\bar{u}}+1\right)+e^{2r_{\ast,0}/L}\left(e^{a/L}+e^{\kappa_{a}\bar{u}}\right)\left(e^{b/L+\kappa_{b}\bar{v}}+1\right)\right]^{2}}\;.\eeq
For the special case $ v_0 = u_0 $ and $a=b$, the      (square root of the)   conformal factor reduces to
\beq
C (s,x) = \frac{\kappa L \sinh (a/L)}{\cosh (a/L) \cosh (\kappa x) + \cosh (\kappa s) }\,. \label{eq:conformalsqaure}
\eeq
This is   the conformal factor for a  spherically symmetric causal diamond in higher-dimensional de Sitter space centered at $r_0=0$, see equation (B5) in \cite{Jacobson:2018ahi} (where we have set $\kappa=1$). Furthermore, for a maximal diamond we see that the conformal factor \eqref{eq:conformalsqaure}  becomes identical to the conformal factor of the static patch   itself given in (\ref{eq:statnullmet}), \emph{i.e.},  in the limit $a,b\to \infty$ and $v_0 \to u_0 $ we have $C \to \kappa L \text{sech} (\kappa x)$.

\subsection*{Euclidean continuation of causal diamonds}

Ultimately we are interested in the Euclideanized diamond, where we Wick rotate the diamond time $s\to -is_{\text{E}}$. In diamond universe coordinates, the line element (\ref{eq:diamondmetric1}) becomes
\beq d\ell^{2}=C^{2}(s_{\text{E}},x)(ds_{\text{E}}^{2}+dx^{2})\;,\eeq
where $s_{\text{E}}$ is periodic in $2\pi/\kappa$,   so as to remove the conical singularity in the Euclidean spacetime due to the horizon. To visualize the Euclidean diamond, it is convenient to introduce a set of Kruskal coordinates 
\beq T_{\text{K}}=\frac{1}{2}(V+U)\;,\quad X_{\text{K}}=\frac{1}{2}(V-U)\;,\eeq
with $U=-L e^{-u/L}$ and $V=L e^{v/L}$. For convenience, consider the case when $u_{0}=v_{0}=0$ and $a=b$, where the conformal factor $C(s,x)$ is given by (\ref{eq:conformalsqaure}). From the coordinate transformation (\ref{eq:transfdiamond2}), we have
\beq T_{\text{K}}=\frac{L\sinh(a/L)\sinh(\kappa s)}{\cosh(\kappa s)+\cosh(a/L-\kappa x)}\;,\quad X_{\text{K}}=\frac{L\left(\cosh(a/L)\cosh(\kappa s)+\cosh(\kappa x)\right)}{\cosh(\kappa s)+\cosh(a/L-\kappa x)}\;.\eeq
Upon Wick rotating the conformal Killing time $s$, we have
\beq 
\begin{aligned} 
T_{\text{K}}\to-i\frac{L\sinh(a/L)\sin(\kappa s_{\text{E}})}{\cos(\kappa s_{\text{E}})+\cosh(a/L-\kappa x)}\equiv -iT_{\text{K}}^{\text{E}}\;,\\
X_{\text{K}}\to\frac{L\left(\cosh(a/L)\cos(\kappa s_{\text{E}})+\cosh(\kappa x)\right)}{\cos(\kappa s_{\text{E}})+\cosh(a/L-\kappa x)}\;.\label{eq:EucKruskdiacoord}
\end{aligned}
\eeq
Remarkably, as realized in higher-dimensional  black hole geometries \cite{Banks:2005bm} and in $\text{AdS}_{2}$ \cite{Pedraza:2021ssc}, the Euclidean continuation of the finite causal diamond covers nearly all of Euclidean $\text{dS}_{2}$. The only difference is that the Euclidean diamond spacetime has two punctures corresponding to the horizons at $x\to\pm\infty$, as visualized in Figure \ref{fig:Eucdiamond}. For the square diamond, the punctures at $x\to\pm\infty$ are mapped to the points $(T_{\text{K}}^{\text{E}},X_{\text{K}})=(0,L e^{\pm a/L})$.

\bibliography{Diamond2d}

\end{document}